\documentclass[12pt]{article}

\usepackage{amsmath,amssymb,amsfonts,amsthm}
\usepackage{color}
\usepackage{graphicx}
\usepackage{epsf}
\def\R{\hbox{$\mit I$\kern-.33em$\mit R$}}
\def\supp{\mbox{supp}}

\newtheorem{theorem}{Theorem}[section]
\newtheorem{thm}[theorem]{Theorem}

\newtheorem{lem}[theorem]{Lemma}
\newtheorem{cor}[theorem]{Corollary}
\newtheorem{prop}[theorem]{Proposition}

\renewcommand{\theequation}{\thesection.\arabic{equation}}

\newcommand{\lam}{\lambda}

\newcommand{\f}{\frac}
\newcommand{\lt}{\left}
\newcommand{\rt}{\right}
\newcommand{\ep}{\epsilon}
\newcommand{\pa}{\partial}
\newcommand{\mb}{\mathbb}
\newcommand{\dsp}{\displaystyle}

\newcommand{\sg}{\sigma}

\newcommand{\al}{\alpha}
\newcommand{\gm}{\gamma}
\newcommand{\tl}{\tilde}
\newcommand{\mH}{\mathcal{H}}

\newcommand{\Id}{\left(\begin{array}{cc} 1 & 0\\0 & 1 \end{array}\right)}

\newcommand{\Iten}{\otimes\,I_2}

\newcommand{\abs}[1]{\lvert#1\rvert}

\newcommand{\norm}[1]{\lt\lvert \lt\lvert#1\rt\rvert \rt\rvert}

\begin{document}

\title{A Born-Oppenheimer Expansion in a Neighborhood of a Renner-Teller
Intersection}

\author{
Mark S. Herman\thanks{This research was supported in part by National Science
Foundation Grant DMS--0600944 while at Virginia Polytechnic Institute and State University, and also by the Institute for Mathematics and its Applications at the University of Minnesota, with funds provided by the National Science Foundation.}\\ Institute for Mathematics and its Applications, University of Minnesota\\ Minneapolis, Minnesota 55455-0134, U.S.A.}

\maketitle

\vskip 5mm \abstract{We perform a rigorous mathematical analysis of
the bending modes of a linear triatomic molecule that exhibits the
Renner-Teller effect. Assuming the potentials are smooth, we prove
that the wave functions and energy levels have asymptotic expansions
in powers of $\ep$, where $\ep^4$ is the ratio of an electron mass
to the mass of a nucleus.  To prove the validity of the expansion,
we must prove various properties of the leading order equations and
their solutions.  The leading order eigenvalue problem is analyzed
in terms of a parameter $\tl{b}$, which is equivalent to the
parameter originally used by Renner.  For $0<\tl{b}<1$, we prove
self-adjointness of the leading order Hamiltonian, that it has
purely discrete spectrum, and that its eigenfunctions and their
derivatives decay exponentially.  Perturbation theory and finite
difference calculations suggest that the ground bending vibrational
state is involved in a level crossing near $\tl{b}=0.925$.  We also
discuss the degeneracy of the eigenvalues.  Because of the crossing,
the ground state is degenerate for $0<\tl{b}<0.925$ and
non-degenerate for $0.925<\tl{b}<1$.}

\newpage

\section{Introduction and Background}
\setcounter{equation}{0} \setcounter{theorem}{0}

In their original paper, \cite{BornOpp}, Born and Oppenheimer let
$\ep^4$ be the ratio of the electron mass to the nuclear mass and
expanded the wave functions and eigenvalues of the time independent
Schr\"odinger equation in powers of $\ep$. We shall refer to such an
expansion as \emph{a Born-Oppenheimer expansion}. Since $\ep$ is
small, the first few orders of the expansions are thought to provide
reasonably accurate results for the bound states of the molecular
system.  Often only the lowest (or leading) order terms of the
expansions are even considered.

The focus of this paper is the Renner-Teller effect (also called the
Renner effect), which is later described in more detail.  In
short, a symmetry induced degeneracy exists in the electron states
at a particular nuclear configuration, but when the nuclei move away
from this configuration the degeneracy splits.  As a result one
must use more than one electronic state when attempting to
solve for the total wave function and energy using the
Born-Oppenheimer approximation. This effect was first predicted in
1933 by Herzberg and Teller \cite{Herz} and was analyzed one year
later by Renner \cite{Renner} in a simplified model.  We consider
the current paper as an extension of the mathematically rigorous
works related to the Born-Oppenheimer approximation, such as
\cite{CombesandFriends2,CombesandFriends1,HagSmooth,HagNonSmooth,Hagedorn/Toloza,Hag/Joye1,KleinandFriends},
to the model originally considered by Renner \cite{Renner}.  The main results are contained in theorem \ref{maintheorem}.  We show rigorously that a Born-Oppenheimer expansion exists to all
orders of $\ep$, with minimal mathematical assumptions.  We prove that under our hypotheses, the molecular energy and wave
function can be approximated by an asymptotic series in $\ep$ that
is truncated at arbitrary order.  The leading order equations we
obtain are unitarily equivalent to those found by Renner in
\cite{Renner}. This is the first rigorous derivation of the leading
order equations of which we are aware.  We feel it is especially
important to make contact with a rigorous Born-Oppenheimer expansion
here, since the Renner-Teller effect is not a straightforward
application of the Born-Oppenheimer approximation.  In their
extensive review of the subject \cite{nicereview}, Peri\'{c} and
Peyerimhoff give several interpretations of the origin of the
Renner-Teller effect, and in particular they state ``from the
quantum chemical standpoint, the R-T effect is a consequence of
violation of validity of the Born-Oppenheimer approximation.''  We
will see that in the Renner-Teller case there is a valid
Born-Oppenheimer expansion, but it differs significantly from the
usual Born-Oppenheimer approximation since the degeneracy cannot be
ignored.  It must be analyzed in terms of degenerate perturbation
theory.

In recent years there have been several mathematically rigorous
results justifying the validity of Born-Oppenheimer expansions under
various hypotheses.  The first rigorous proof related to the
Born-Oppenheimer approximation in a physically realistic model was
given by Combes, Duclos, and Seiler \cite{CombesandFriends2,CombesandFriends1}.  They proved the validity of the fourth order
approximation for the eigenvalue and the leading order approximation
for the eigenfunction.  A few years later, Hagedorn proved
\cite{HagSmooth} the existence of a Born-Oppenheimer expansion to
all orders using the method of multiple scales, assuming that the
potentials are smooth functions.  In particular, he proved that for
arbitrary $K$, there exist quasimode energies of the form
$E_K(\ep)\, =\, \sum_{k=0}^K\, \ep^k\, E^{(k)}$ and quasimodes of
the form $\Psi_K(\ep)\, =\, \sum_{k=0}^K\, \ep^k\,
\Psi_{\ep}^{(k)}$, that asymptotically approximate an exact
eigenvalue and eigenfunction below the essential spectrum of a
Hamiltonian $H(\ep)$, in the sense that
\begin{eqnarray}
\norm{H(\ep)\,\Psi_K(\ep)\, -\, E_K(\ep)\, \Psi_K(\ep)}\, \leq\,
C_K\, \ep^{K+1}.\label{eboundformula}
\end{eqnarray}

The first five orders of $E(\ep)$ were determined explicitly, and it
is discussed how one could proceed to any arbitrary order $K$. These
results were then extended to the case of Coulomb potentials for
diatomic molecules in \cite{HagNonSmooth} and to general polyatomic
molecules by Klein {\it et al.}\;\cite{KleinandFriends}. Here, we
will assume that the potentials are smooth, but we believe our
results can be extended in a similar manner to the case of Coulomb
potentials.

\section{Description of the Model and Statement of the Main Theorem}

\begin{figure}[htb]
\begin{centering}
\includegraphics[width=0.21\textwidth, height=.3 \textwidth]
{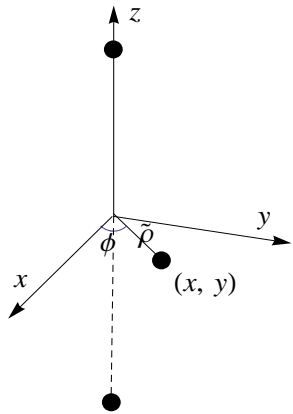}  \caption{The reference frame for the middle
nucleus.} \vskip .5cm \label{FigReferenceFrame}
\end{centering}
\end{figure}

Consider a triatomic molecule and fix the reference frame so that
when the molecule is in the linear configuration, the middle nucleus
is at the origin and the $z$-axis passes through all three nuclei.
Let $(0,0,R_1)$ and $(0,0,R_2)$ be the coordinates of the upper and
lower nuclei (so $R_1>0$ and $R_2<0$). We consider the bending modes
by clamping the upper and lower nuclei to their fixed positions on
the $z$-axis and allowing the middle nucleus to move in the
perpendicular plane.  Let $(x,y,0)$ be the cartesian coordinates of
this middle nucleus, and let $(\tl{\rho},\phi)$ be the usual polar
coordinates associated with $(x,y)$ (see Figure
\ref{FigReferenceFrame}).

If $(x_1,x_2,\cdots,x_N)$ are the N three-dimensional electron
coordinates, the electronic Hamiltonian is
\begin{eqnarray*}
h(x,y)\, =\, -\, \f12\,\sum_{j=1}^N\, \Delta_{x_j}\, +\,
V(x,y;x_1,x_2,\cdots,x_N),
\end{eqnarray*}
where we have taken the electron mass to be 1, and the potential $V$
includes the repulsion forces between the nuclei, the attraction
forces between the nuclei and electrons, and the repulsion forces
between the electrons.  We think of $h(x,y)$ as having parametric
dependence on $(x,y)$ ({\it i.e.}\;it is a mapping from $\R^2$ to
the linear operators on the electronic Hilbert space), and we assume
it is a real symmetric operator. We assume that $V$ is a smooth
function in all variables.  Let $\ep^4$ be the ratio of the mass of
an electron to the mass of the middle nucleus. Then, the full
hamiltonian of this model is given by
\begin{eqnarray}
H(\ep)\, =\, -\, \f{\ep^4}2\, \Delta_{x,y}\, +\, h(x,y).
\label{TheHamiltonian}
\end{eqnarray}

Let $\mathcal{H}_{nuc}\, =\, L^2(\R^2,\,dx\,dy)$ and
$\mathcal{H}_{el}\, =\, L^2(\R^{3N})$, so that $H(\ep)$ acts on the
Hilbert space $\mathcal{H}_{nuc} \otimes \mathcal{H}_{el}$.  We
denote the inner product and norm on $\mH_{el}$ by $\langle\,
\cdot,\, \cdot\, \rangle_{el}$ and $\norm{\cdot}_{el}$ and similarly
on $\mH_{nuc}$ by  $\langle\, \cdot,\, \cdot\, \rangle_{nuc}$ and
$\norm{\cdot}_{nuc}$.

Let $L_z^{el}$ and $L_z^{nuc}\, =\, -i\,\f{\pa\, }{\pa \phi}$ be the
operators associated with the projections of the electronic and
nuclear angular momenta on the $z$-axis, respectively.  The operator
of total angular momentum about the $z$-axis is denoted by
$L_z^{TOT}\, =\, (I \otimes L_z^{el})\, +\, (L_z^{nuc} \otimes I)$.
We note that $H(\ep)$ commutes with $L_z^{TOT}$.  We consider the
electronic states when $(x,y)=(0,0)$. In this case the electronic
hamiltonian $h(0,0)$ commutes with $L_z^{el}$ since the nuclei are
in a linear arrangement.  So, for $\abs{l_z^{el}} \neq 0$ there are
two-fold degenerate electronic vectors $\psi_1,\ \psi_2 \in
\mathcal{H}_{el}$ satisfying $h(0,0)\, \psi_1\, =\,
E_0^{\abs{l_z^{el}}}\, \psi_1$ and $h(0,0)\, \psi_2\, =\,
E_0^{\abs{l_z^{el}}}\, \psi_2$, where $L_z^{el}\, \psi_1\, =\,
l_z^{el}\, \psi_1$ and $L_z^{el}\, \psi_2\, =\, -l_z^{el}\, \psi_2$.
Then, if the molecule is bent so that $(x,y)\neq (0,0)$, this
degeneracy splits since the nuclei are no longer in a linear
arrangement, and $h(x,y)$ no longer commutes with $L_z^{el}$ (see
\cite{JensenOsmann} for a discussion directly relating the breaking
of symmetry with the breaking of the degeneracy).  This is the
Renner-Teller effect.  As previously mentioned, the application of
the Born-Oppenheimer approximation is not straightforward in this
case.  There have been numerous papers related to the
Renner-Teller effect, few of which are relevant to our analysis
here.  We highlight one such paper by Brown and J{\o}rgensen
\cite{BrownJorgensen} for its completeness, and because it does
discuss effects beyond the leading order.  We encourage the reader
interested to learn the historical development and recent findings
of the theory to consult the review by Peri\'{c} and
Peyerimhoff \cite{nicereview}.

\begin{figure}[htb]
\begin{centering}
\includegraphics[width=0.27\textwidth, height=.2\textwidth]
{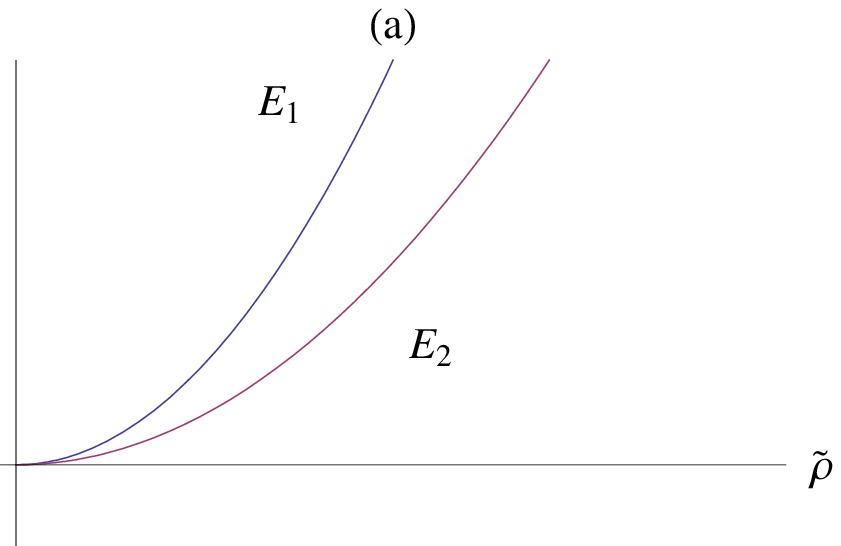}  \hspace{.5cm}
\includegraphics[width=0.27 \textwidth, height=.2\textwidth]
{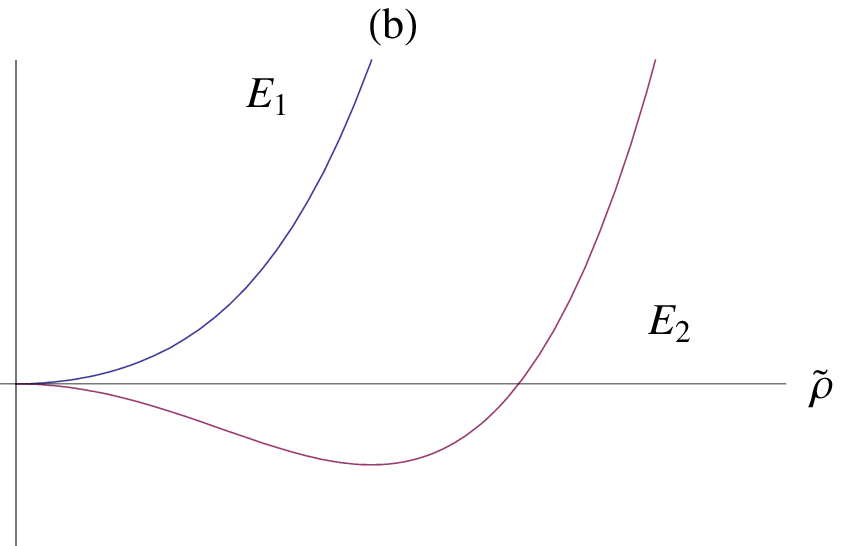}  \hspace{.5cm}
\includegraphics[width=0.27 \textwidth, height=.2\textwidth]
{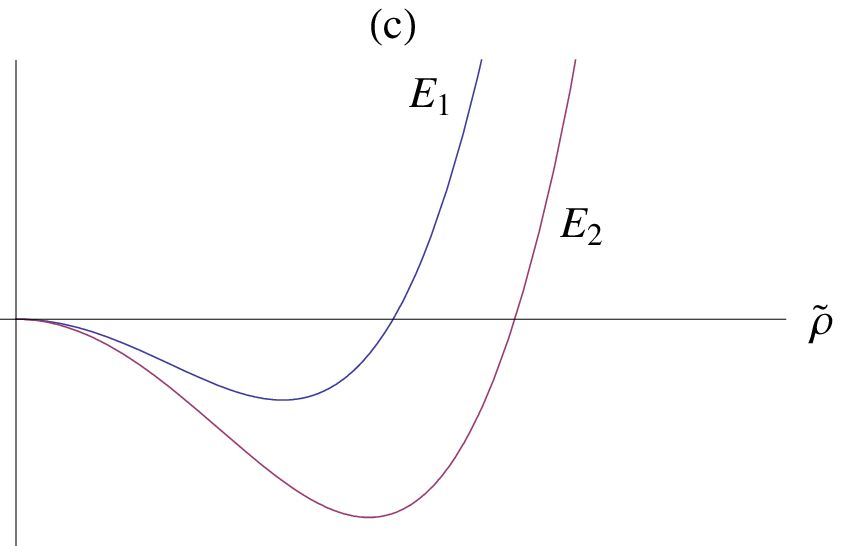} \caption{Potential energy surfaces of three
qualitatively different cases corresponding to an R-T pair of
electronic states.} \vskip .5cm \label{FigelecLevels}
\end{centering}
\end{figure}

Note that since changes in $\phi$ correspond to an overall molecular
rotation, the eigenvalues of $h(x,y)$ are independent of $\phi$.  Corresponding to the situation above where the electronic states at
$\tl{\rho}=0$ are linear combinations of eigenstates of $L_z^{el}$
with eigenvalues $l_z^{el},\,-l_z^{el}\, \neq\, 0$, consider
a pair of electronic eigenvalues $E_1(\tl{\rho})$ and
$E_2(\tl{\rho})$ of $h(x,y)$ that are degenerate at $\tl{\rho}=0$, but the degeneracy
breaks when $\tl{\rho}\neq 0$.  We refer to two such electronic
states as an R-T pair with value $\abs{l_z^{el}}$.  The eigenvalues of $h(x,y)$ provide the usual potential energy surfaces for the
nuclei, and there are several qualitatively different possibilities
where the Renner Teller effect is important.  See Figure \ref{FigelecLevels}.  
We refer to \cite{JensenOsmann,FoxandFriends} for further examples and
discussion of Renner-Teller surfaces.  We focus strictly on an R-T
pair of states corresponding to sketch (a) of Figure \ref{FigelecLevels},
where both surfaces have local minima at $\tl{\rho}=0$.  In this
case the optimal nuclear configuration, corresponding to both
electronic states of the R-T pair, is linear.  This was the
situation considered by Renner \cite{Renner} in 1934.

\vskip .5cm

Throughout this paper, we assume the following hypotheses:  There is
an R-T pair of states, with eigenvalues $E_1$ and $E_2$ both having minima at
$\tl{\rho}=0$.  We assume that for some neighborhood of
$\tl{\rho}=0$, $E_1$ and $E_2$ are isolated from the rest of the
spectrum and are $C^{\infty}$ in $(x,y)$.  This implies that
$E_1(\tl{\rho})$ and $E_2(\tl{\rho})$ have asymptotic expansions in
powers of $\tl{\rho}^2$.  We assume that splitting occurs at 2nd
order, that is, that $E_1$ and $E_2$ are asymptotic to
$\f{a\,+\,b}2\,\tl{\rho}^2$ and $\f{a\,-\,b}2\,\tl{\rho}^2$ for
small $\tl{\rho}$, respectively, for some $0 < b < a$ (we have taken $E_1(0)=E_2(0)=0$ for convenience).  Renner
\cite{Renner} argued that an R-T pair with value $\abs{l_z^{el}}=1$
will exhibit splitting at 2nd order, an R-T pair with value
$\abs{l_z^{el}}=2$ will exhibit splitting at 4th order, and in
general an R-T pair with value $\abs{l_z^{el}}=n$ will exhibit
splitting at order $2n$.  We instead assume 2nd order splitting
occurs and later prove that the R-T pair has value
$\abs{l_z^{el}}=1$, agreeing with Renner's argument.  We are now
ready to state our main theorem.

\begin{thm} \label{maintheorem}
Assume the hypotheses described above, in
particular that the potentials are smooth and there is an R-T pair
$E_1(\tl{\rho}),\ E_2(\tl{\rho})$ that are asymptotic to $\f{a+b}2
\tl{\rho}^2$ and $\f{a-b}2 \tl{\rho}^2$ respectively, where $0<b<a$.
Then for arbitrary $K$, there exist quasimode energies
$$
E_{\ep,K}\, =\, \sum_{k=0}^K\, \ep^k\, E^{(k)}
$$
and quasimodes
$$
\Phi_{\ep,K}\, =\, \sum_{k=0}^K\, \ep^k\, \Phi_\ep^{(k)}
$$
that satisfy
\begin{eqnarray*}
\norm{\,\lt(\,H(\ep)\, -\, E_{\ep,K}\,
\rt)\,\Phi_{\ep,K}\,}_{\mathcal{H}_{nuc} \otimes \mathcal{H}_{el}}\,
\leq\, C_K\, \ep^{K+1}\, \norm{\,\Phi_{\ep,K}\,}_{\mathcal{H}_{nuc}
\otimes \mathcal{H}_{el}}.
\end{eqnarray*}
The quasimodes are associated with the local wells of $E_1$ and
$E_2$ in a neighborhood of $\tl{\rho}=0$.

\end{thm}

\noindent Remarks:
\begin{enumerate}
\item  Quasimode estimates correspond to discrete eigenvalues of
$H(\ep)$ when $E_{\ep,K}$ lies below the essential spectrum as
characterized by the HVZ theorem \cite{RSV4}.

\item Since $[L_z^{TOT},H(\ep)]=0$, quasimodes can be
constructed to be eigenfunctions of $L_z^{TOT}$, with eigenvalues
$l_z^{TOT} \in \mb{Z}$.  The eigenstates of $H(\ep)$ corresponding
to $l_z^{TOT}=0$ are non-degenerate, while the eigenstates
corresponding to $\abs{l_z^{TOT}}\neq 0$ are two-fold degenerate. In
particular, there is an $l_z^{TOT}$ state and a $-l_z^{TOT}$ state,
with eigenfunctions that are complex conjugates of one another,
together forming a degenerate pair of states associated with $H(\ep)$.

\item The first two orders $E^{(0)}$ and $E^{(1)}$ are zero and the second
order $E^{(2)}$ is determined by the leading order eigenvalue
equation $\mb{H}_2\Psi=E^{(2)}\Psi$, on the Hilbert space\\
$L^2(\R^2,\,dX\,dY;\, \mb{C}^2)$, where
\begin{eqnarray*}
\mathbb{H}_2\, =\, \left(\begin{array}{cc} \displaystyle -\, \f12 \,
\Delta_{X,\,Y}\, +\, \f{a+b}2 \, X^2\, +\, \f{a-b}2 \, Y^2 &
\displaystyle b\, X\, Y
\\[5mm]
\displaystyle b\, X\, Y & \displaystyle -\, \f{1}2\, \Delta_{X,\,Y}\, +\, \f{a-b}2 \, X^2\, +\, \f{a+b}2 \, Y^2\\
    \end{array}\right).
\end{eqnarray*}
The higher order $E^{(k)}$ are determined through the perturbation
formulas presented in chapter \ref{QuasimodeConstruction}. All odd
order $E^{(k)}$ are zero (see Appendix).

\item The presence of the two levels $E_1(\tl{\rho})$ and
$E_2(\tl{\rho})$ gives rise to twice the number of vibrational
levels as usual in the following sense: If $b=0$, the upper and
lower component equations of the leading order equation which
determines $E^{(2)}$, are both two-dimensional harmonic oscillator
equations.  So, there will be two eigenfunctions, one associated
with the each of the upper and lower components, for each of the
usual eigenstates of the usual harmonic oscillator.  Then for small
$b$, this will give rise to two vibrational states via a
perturbative approach, for each of the usual harmonic oscillator
states.  This is shown in detail in section \ref{EfcnsEvals}.

\item For small $b$ and $\ep$, the ground state of $H(\ep)$ (meaning the lowest vibrational level
corresponding to the R-T pair we are considering) is degenerate,
corresponding to a pair of states which are eigenfunctions of
$L_z^{TOT}$ with eigenvalues $l_z^{TOT}=\pm 1$.  In section
\ref{EfcnsEvals}, we give plots which suggest that for approximately
$0.925a < b < a$ the ground state is non-degenerate, corresponding
to a state with $l_z^{TOT}=0$.

\end{enumerate}

The paper is organized as follows:  In section
\ref{QuasimodeConstruction}, we derive perturbation formulas to
construct the quasimodes that will enter in our main theorem.  In
section \ref{Properties}, we prove various properties of the leading
order Hamiltonian that are needed to prove the main theorem.  In
section \ref{EfcnsEvals}, we analyze the leading order eigenvalue
problem. Only some of the eigenvalues and eigenfunctions of the
leading order equation are solved for exactly.  In section
\ref{chapdegeneracy}, the degeneracy structure of the full
Hamiltonian $H(\ep)$ is discussed.  In section \ref{MainThmChap}, we
use the results of the previous chapters to prove the main theorem.

\section{The Construction of the
Quasimodes}\label{QuasimodeConstruction} \setcounter{equation}{0}
\setcounter{theorem}{0}

Before we begin the formal expansion, we first look at some
properties of the electronic eigenvectors and eigenvalues, construct
electronic basis vectors that are smooth in terms of the
nuclear coordinates, and derive the leading orders of the matrix
elements of the electronic Hamiltonian in this basis.

\subsection{The Two-Dimensional Electronic Basis Vectors}

For the N electrons, as well as the nuclei, we use the same fixed
reference frame previously described.  Let $(r_j,\,\theta_j,\,z_j)$
be the cylindrical coordinates of the $jth$ electron in this frame.
Suppose that for $\tl{\rho} > 0$, $\psi(x,y;\,
\theta_1,\theta_2,\cdots,\theta_N) : \R^2 \rightarrow
\mathcal{H}_{el}$ is an electronic eigenvector of $h(x,y)$.  We have
suppressed the dependence on $r_j$ and $z_j$ because it is
irrelevant to the discussion here.  The electronic eigenfunctions
are invariant with respect to a rotation of the entire molecule. So,
the eigenfunctions have the property
$$
\psi(x,\,y;\, \theta_1,\theta_2,\cdots,\theta_N)\, =\,
\psi(\tl{\rho},\,0;\,
\theta_1-\phi,\theta_2-\phi,\cdots,\theta_N-\phi)\,
$$
for $\tl{\rho} > 0$.  It follows that if $\psi(x,y)$ is continuous
at $\tl{\rho}=0$, then $\psi(0,0)$ has no $\theta_j$ dependence.  Since an
eigenvector corresponding to an R-T pair with positive
$\abs{l_z^{el}}$ value will have some $\theta_j$ dependence at
$\tl{\rho}=0$, we do not have well-defined continuous electronic
eigenfunctions of $h(x,y)$ in a neighborhood of $\tl{\rho}=0$, that
correspond to an R-T pair with value $\abs{l_z^{el}} > 0$.  We need basis vectors for the two-dimensional
eigenspace of $E_1(\tl{\rho})$ and $E_2(\tl{\rho})$ that are smooth
in $x$ and $y$.  The matrix elements of $h(x,y)$ in our electronic
basis determine the form of the leading order equations to follow.
We note that in deriving these matrix elements, we do not use the
matrix elements of $L_z^{el}$. Only second order splitting in $E_1$
and $E_2$ is needed, as well as the fact that our smooth basis
vectors are not eigenvectors of $h(x,y)$. This gives rise to
off-diagonal terms in the basis representation of $h(x,y)$. In this
sense, the unusual form of the leading order equations can be
thought of as a result of the discontinuity of the electronic
eigenvectors in the nuclear coordinates, {\it i.e.}\;there is no
smooth electronic basis that diagonalizes the electronic
hamiltonian.  We note that matrix elements we derive here, are related by an $(x,y)$-independent
unitary transformation to those given by Yarkony \cite{Yarkony}.  See also Worth and Cederbaum \cite{Worth} for a general
discussion of the topology and classification of different types of
intersections of potential surfaces.

We now describe our approach.  Choose any two normalized orthogonal electronic vectors $\psi_1$ and
$\psi_2$ that span the eigenvalue $0$ eigenspace of $h(0,0)$. Let
$P(x,\,y)$ denote the two dimensional projection onto the electronic
eigenspace associated to the two eigenvalues of $h(x,\,y)$. For
small $x$ and $y$, define
\begin{eqnarray}
\Psi_1(x,\,y)\ =\ \frac 1
{\sqrt{\,\langle\,\psi_1,\,P(x,\,y)\,\psi_1\rangle\,}}\
P(x,\,y)\,\psi_1. \label{basisvec1}
\end{eqnarray}
Let $P_1(x,\,y)$ denote the orthogonal projection onto this vector,
{\it i.e.},
$$
P_1(x,\,y)\ =\ \left|\,\Psi_1(x,\,y)\,\rangle\,
\langle\,\Psi_1(x,\,y)\,\right|.
$$
Next, define
$$
\chi(x,\,y)\ =\ \left(\,1\,-\,P_1(x,\,y)\,\right)\ P(x,\,y)\ \psi_2,
$$
and
\begin{eqnarray}
\Psi_2(x,\,y)\ =\ \frac 1{\sqrt{\,\langle\,\chi(x,\,y),\,
\chi(x,\,y)\,\rangle\,}}\ \chi(x,\,y). \label{basisvec2}
\end{eqnarray}
Then $\{\,\Psi_1(x,\,y),\,\Psi_2(x,\,y)\,\}$ is an orthonormal basis
for the range of $P(x,\,y)$.  From the formula \cite{RSV4}
$$
P(x,y)\, =\, \f{1}{2\pi i}\, \int_C\, (\lam-h(x,y))^{-1}\, d\lam,
$$
where $C$ is a closed path in the complex plane encircling
$E_1(\tl{\rho})$ and $E_2(\tl{\rho})$ but no other spectrum of
$h(x,y)$, we see that these vectors are smooth in $x$ and $y$, since
we have assumed that the potentials are smooth and hence the
resolvent of $h(x,y)$ is as well (recall we are only working in a
neighborhood of the origin $(x,y)=(0,0)$).  Note that we can arrange
for these vectors to be real, which we assume has been done.

\subsection{The Matrix Elements of the Electronic Hamiltonian}

The span of $\{\,\Psi_1(x,\,y),\,\Psi_2(x,\,y)\,\}$ is an invariant
subspace for $h(x,\,y)$. Using coordinates in this basis, the
restriction of $h(x,\,y)$ to this subspace is unitarily equivalent
to the real symmetric matrix
$$
\left(\,\begin{array}{cc}\vspace{2mm}
h_{11}(x,\,y)&h_{12}(x,\,y)\\
h_{21}(x,\,y)&h_{22}(x,\,y)
\end{array}\,\right),
$$
where
$$
h_{jk}(x,y)\ =\ \langle\,\Psi_j(x,\,y),\,
h(x,\,y)\,\Psi_k(x,\,y)\,\rangle.
$$
Again, since we have smooth potentials, $h_{ij}(x,y)$ can be
expanded in powers of $x$ and $y$.  Since we assume the degeneracy
splits at second order, the eigenvalues of this matrix are\\
$E_1(x,\,y)=\f{a\,+\,b}2\,(x^2+y^2)\,+\,O(\tl{\rho}^4)\ $ and $\
E_2(x,\,y)=\f{a\,-\,b}2\,(x^2+y^2)\,+\,O(\tl{\rho}^4)$.  Using these
expressions for the eigenvalues we show that up to an
$(x,y)$-independent unitary transformation, this matrix is
\begin{equation}
\left(\,\begin{array}{cc}\vspace{2mm} \displaystyle
\f{a+b}2 \ x^2\ +\ \f{a-b}2\, y^2 & \displaystyle \pm\, b\, x\, y \\
\displaystyle \pm\, b\, x\, y & \displaystyle \f{a-b}2\ x^2\ +\
\f{a+b}2\ y^2
\end{array}\,\right). \label{hmatrix}
\end{equation}

To show this, we consider a traceless,
real symmetric matrix
\begin{eqnarray}
\left(\,\begin{array}{cc}\vspace{2mm}
\tl{h}_{11}(x,\,y)&\tl{h}_{12}(x,\,y)\\
\tl{h}_{21}(x,\,y)& -\tl{h}_{11}(x,\,y)
\end{array}\,\right), \label{traceless}
\end{eqnarray}
with eigenvalues
$\tl{E}_{\pm}(x,\,y)=\pm\,\tl{\rho}^2\,+\,O(\tl{\rho}^4)$.  The form
in (\ref{hmatrix}) will follow from the analysis below.

Using
(\ref{traceless}), we have the characteristic equation
\begin{eqnarray}
\tl{E}_{\pm}^2\, +\, O(\tl{\rho}^6)=\, \tl{h}_{11}^2\, +\,
\tl{h}_{12}^2. \label{matrixeval}
\end{eqnarray}
By expanding in powers of $x$ and $y$ and equating orders in the
above equation, it can be easily shown that the constant and linear
terms of $\tl{h}_{11}$ and $\tl{h}_{12}$ must vanish.  We then
write,
\begin{eqnarray}
\left(\,\begin{array}{cc}\vspace{2mm}
\tl{h}_{11}(x,\,y)&\tl{h}_{12}(x,\,y)\\
\tl{h}_{21}(x,\,y)& -\tl{h}_{11}(x,\,y)
\end{array}\,\right)\, =\, A\, x^2\, +\, B\, y^2\, +\, C\, xy\, +\,
O(\tl{\rho}^3), \label{tracelessform}
\end{eqnarray}
where $A$, $B$, and $C$ are traceless 2 by 2 matrices with constant
entries.  We can apply a constant unitary transformation to
(\ref{tracelessform}) that diagonalizes $A$, which we assume has
been done. An obvious consequence of (\ref{matrixeval}) and
$\tl{E}_{\pm}(x,\,y)=\pm\,\tl{\rho}^2\,+\,O(\tl{\rho}^4)$ is that if
$A$ is diagonal, it must be
$$
A\, =\, \lt(\begin{array}{cc}\vspace{2mm}
1&0\\
0& -1
\end{array}\rt).
$$
We let
$$
B\, =\, \lt(\begin{array}{cc}\vspace{2mm}
b_{11}&b_{12}\\
b_{12} & -b_{11}
\end{array}\rt), \hskip .3cm \text{and} \hskip .3cm C\, =\, \lt(\begin{array}{cc}\vspace{2mm}
c_{11}& c_{12}\\
c_{12} & -c_{11}
\end{array}\rt),
$$
and use (\ref{matrixeval}) with
$\tl{E}_{\pm}(x,\,y)=\pm\,\tl{\rho}^2\,+\,O(\tl{\rho}^4)$ to solve
for $b_{ij}$ and $c_{ij}$ by equating the powers of $x$ and $y$.
This gives us four equations, the first equation comes from the
$y^4$ coefficients, the second comes from the $x^2y^2$ coefficients,
etc.
\begin{eqnarray}
y^4:\qquad 1 &=& b_{11}^2\, +\, b_{12}^2 \label{matrixform1}
\\
x^2y^2:\qquad  2 &=& 2\,b_{11}\, +\, c_{11}^2\, +\, c_{12}^2
\label{matixform2}
\\
xy^3:\qquad 0 &=& 2(b_{11}c_{11}\, +\, b_{12}c_{12})
\label{matrixform3}
\\
x^3y:\qquad 0 &=& 2c_{11} \label{matrixform4}
\end{eqnarray}
These equations have 3 solutions.  Two of the solutions are
\begin{eqnarray*}
(b_{11},b_{12},c_{11},c_{12})\, =\, (-1,0,0,\pm2),
\end{eqnarray*}
which give
\begin{eqnarray*}
\left(\,\begin{array}{cc}\vspace{2mm}
\tl{h}_{11}(x,\,y)&\tl{h}_{12}(x,\,y)\\
\tl{h}_{21}(x,\,y)& -\tl{h}_{11}(x,\,y)
\end{array}\,\right)\, =\, \left(\,\begin{array}{cc}\vspace{2mm}
x^2-y^2 & \pm\,2xy\\
\pm\,2xy & -(x^2-y^2)
\end{array}\,\right).
\end{eqnarray*}
These solutions give rise to (\ref{hmatrix}).  The only other
solution of equations (\ref{matrixform1})-(\ref{matrixform4}) is
\begin{eqnarray*}
(b_{11},b_{12},c_{11},c_{12})\, =\, (1,0,0,0),
\end{eqnarray*}
which gives rise to
$$
\left(\,\begin{array}{cc}\vspace{2mm}
h_{11}(x,\,y)&h_{12}(x,\,y)\\
h_{21}(x,\,y)&h_{22}(x,\,y)
\end{array}\,\right)\, =\, \left(\,\begin{array}{cc}\vspace{2mm} \displaystyle
\f{a+b}2 \ \tl{\rho}^2 & 0 \\
0 & \displaystyle \f{a-b}2\ \tl{\rho}^2
\end{array}\,\right).
$$
We do not consider this case.  Aside from being uninteresting, it implies that the
basis vectors are the eigenfunctions of $h(x,\,y)$ (at least to leading order).  We assume that the off diagonal terms in
(\ref{hmatrix}) are $bxy$, since the $-bxy$ case is related by the
trivial change of coordinates $y \mapsto -y$.

\subsection{The Formal Expansion}

To construct the quasimodes in theorem \ref{maintheorem}, we
introduce the scaled variables $(X,Y)=(x/\ep,y/\ep)$.  The intuition
of the Born-Oppenheimer approximation suggests that the adiabatic
effects will occur on the $(x,y)=(\ep X, \ep Y)$ scale, whereas the
semi-classical motion of the nuclei is determined on the $(X,Y)$
scale. In terms of the $(X,Y)$ variables, the Hamiltonian in
(\ref{TheHamiltonian}) is
\begin{eqnarray*}
H(\epsilon)\ =\ -\,\frac{\epsilon^2}2\,\Delta_{X,\,Y}\ +\
h(\epsilon\,X,\,\epsilon\,Y).\label{TheScaledHamiltonian}
\end{eqnarray*}
We define $\mH$ to be the Hilbert space $L^2(\R^2,\,dX\,dY;\,
\mb{C}^2)$ and we denote the inner product on this space by
$\langle\, \cdot,\, \cdot\, \rangle_{\mH}$.

We seek solutions to $H(\ep)\ \Psi(\ep,\,X,\,Y)=E(\ep)\
\Psi(\ep,\,X,\,Y)$.  The wave function $\Psi(\ep,\,X,\,Y)$ can be
written in terms of the orthonormal basis functions
$\{\,\Psi_1(x,\,y),\,\Psi_2(x,\,y)\,\}$ from (\ref{basisvec1}) and
(\ref{basisvec2}) as
\begin{eqnarray}\label{Psiepsilondefinition}
\hskip -10mm \Psi(\ep,\,X,\,Y)=f(\ep,\,X,\,Y)\
\Psi_1(\ep\,X,\,\ep\,Y)+g(\ep,\,X,\,Y)\ \Psi_2(\ep\, X,\,
\ep\,Y)+\psi_{\perp}(\ep,\,X,\,Y),
\end{eqnarray}
where $\langle\ \psi_{\perp},\ \Psi_i\ \rangle_{el} = 0$.

Substituting (\ref{Psiepsilondefinition}) in $H(\ep)\
\Psi(\ep,\,X,\,Y)=E(\ep)\ \Psi(\ep,\,X,\,Y)$ gives three equations;
one along $\Psi_1$, one along $\Psi_2$, and one in $span\{\Psi_1,\
\Psi_2\}^{\perp}$. We denote the projection on $span\{\Psi_1,\
\Psi_2\}^{\perp}$ by $P_{\perp}$.\\
Along $\Psi_1$:
\begin{eqnarray}
&& \hskip -15mm\displaystyle -\ \f{\ep^2}2\ \Delta_{X,\,Y}\ f\ +\
h_{11}\ f\ +\ h_{12}\ g\ -\ \f{\ep^2}2\ \langle\ \Psi_1,\
\Delta_{X,\,Y}\ \psi_{\perp}\ \rangle_{el}\ -\ \f{\ep^4}2\ f\
\langle\ \Psi_1, \Delta_{x,\,y}\ \Psi_1\ \rangle_{el}\ \notag
\\[3mm]
&&\hskip -15mm\displaystyle -\ \f{\ep^4}2\ g\ \langle\ \Psi_1,
\Delta_{x,\,y}\ \Psi_2\ \rangle_{el}\ -\ \ep^3\ \lt(\ \f{\partial
g}{\partial X}\ \langle\ \Psi_1,\f{\partial \Psi_2}{\partial x}\
\rangle_{el}\ +\ \f{\partial g}{\partial Y}\ \langle\
\Psi_1,\f{\partial \Psi_2}{\partial y}\ \rangle_{el}\ \rt)\notag
\\[3mm]
&&\hskip -15mm =\ E(\ep)\ f. \label{psi1}
\end{eqnarray}
Above we have used that $\langle\ \Psi_i,\f{\partial
\Psi_i}{\partial x}\ \rangle_{el}\ =\ 0$, which we know from
normalization and the fact that the electronic basis vectors were
chosen real. Along $\Psi_2$ we get a similar equation with $f
\leftrightarrow g,\ \Psi_1 \leftrightarrow \Psi_2 ,\ h_{11}
\leftrightarrow h_{22},\ h_{12} \leftrightarrow h_{21}.$\\
In $span\{\Psi_1,\ \Psi_2\}^{\perp}$:
\begin{eqnarray}
&& \hskip -12mm \displaystyle -\ \f{\ep^2}2\ P_{\perp}\lt[\
\Delta_{X,\,Y}\ \psi_{\perp}\ \rt]\ +\ \lt(h\ P_{\perp}\rt)\
\psi_{\perp}\ -\ \f{\ep^4}2\ f\ P_{\perp}\lt[\ \Delta_{x,\,y}\
\Psi_1\ \rt]\ -\ \f{\ep^4}2\ g\ P_{\perp}\lt[\ \Delta_{x,\,y}\
\Psi_2\ \rt] \nonumber
\\[3mm]
&& \hskip -12mm -\ \ep^3\ \lt(\ \f{\partial f}{\partial X}\
P_{\perp} \lt[\ \f{\partial \Psi_1}{\partial x}\ \rt]\ +\
\f{\partial f}{\partial Y}\ P_{\perp} \lt[\ \f{\partial
\Psi_1}{\partial y}\ \rt] +\ \f{\partial g}{\partial X}\ P_{\perp}\
\lt[\ \f{\partial \Psi_2}{\partial x}\ \rt]+\ \f{\partial
g}{\partial Y}\ P_{\perp}\ \lt[\ \f{\partial \Psi_2}{\partial y}\
\rt]\ \rt) \nonumber
\\[3mm]
&&\hskip -12mm =\ E(\ep)\ \psi_{\perp}. \label{perp1}
\end{eqnarray}
We adopt the following notation for simplicity:
\begin{eqnarray*}
T_{ij}(x,\,y)\ &=&\ \langle\ \Psi_i, \Delta_{x,\,y}\ \Psi_j\
\rangle_{el},\notag
\\[3mm]
A_{ij}(x,\,y)\ &=&\ \lt\langle\ \Psi_i, \f{\pa \Psi_j}{\pa x}
\rt\rangle_{el},\notag
\\[3mm]
B_{ij}(x,\,y)\ &=&\ \lt\langle\ \Psi_i, \f{\pa \Psi_j}{\pa y}
\rt\rangle_{el}. \label{BasisDerivInners}
\end{eqnarray*}
We have identities involving these quantities since $\{\, \Psi_1, \,
\Psi_2\, \}$ are orthonormal and real valued. For instance we know
the diagonal elements of $A$ and $B$ are zero and $A_{12}=-A_{21},\
B_{12}=-B_{21}$.

Now we expand all functions and operators with $\ep$ dependence. For
example, $\displaystyle f(\ep,\,X,\,Y)=\sum_{k=0}^\infty\ \ep^k\
f^{(k)}(X,\,Y)\ $. For functions and operators with exclusively
$(x,\,y)$ dependence, we know the form of the expansions.  For
example, $\displaystyle \Psi_1(x,\,y)\ =\ \Psi_1(\ep\,X,\,\ep\,Y)\
=\ \sum_{k=0}^\infty\ \ep^k\ \Psi_1^{(k)}(X,\,Y),$ where
$\displaystyle \Psi_1^{(k)}(X,\,Y)\ =\ \sum_{j=0}^k\ \f1{j!(k-j)!}\
\f{\pa^k \Psi_1}{\pa x^j \pa y^{k-j}}\,(0,0)\ X^k\ Y^{k-j}.$
Equations (\ref{psi1}) and (\ref{perp1}) become:
\begin{eqnarray}
&&\hskip -1.5cm \sum_{k=2}^\infty\ \ep^k\ \lt(-\f{1}2\rt)\
\Delta_{X,\,Y}\ f^{(k-2)}\ +\ \sum_{k=0}^\infty\ \ep^k\
\sum_{j=0}^k\ \lt[\,h_{11}^{(j)}\ f^{(k-j)}\ +\ h_{12}^{(j)}\
g^{(k-j)}\ \rt]\ \notag
\\[3mm]
&&\hskip -1.5cm +\ \sum_{k=2}^\infty\ \ep^k\ \sum_{j=2}^k\
\lt(-\f12\rt)\ \langle\ \Psi_1^{(j-2)},\ \Delta_{X,\,Y}\
\psi_{\perp}^{(k-j)}\ \rangle_{el}\ \notag
\\[3mm]
&&\hskip -1.5cm +\ \sum_{k=4}^\infty\ \ep^k\ \sum_{j=4}^k\
\lt(-\f12\rt)\ \lt[\,T_{11}^{(j-4)}\ f^{(k-j)}\ +\ T_{12}^{(j-4)}\
g^{(k-j)}\ \rt] \notag
\\[3mm]
&& \hskip -1.5cm +\ \sum_{k=3}^\infty\ \ep^k\ \sum_{j=3}^k\ \lt[\,-\
A_{12}^{(j-3)}\ \f{\partial \phantom{t}}{\partial X}\ -\
B_{12}^{(j-3)}\ \f{\partial \phantom{t}}{\partial Y}\ \rt]\
g^{(k-j)}\ =\ \sum_{k=0}^\infty\ \ep^k\ \sum_{j=0}^k E^{(j)}\ f^{(k-j)}
\label{psi2}
\end{eqnarray}
and
\begin{eqnarray}
&& \sum_{k=2}^\infty\ \ep^k\ \sum_{j=2}^k\ \lt(-\f12\rt)\
P_{\perp}^{(j-2)} \lt[\, \Delta_{X,\,Y}\ \psi_{\perp}^{(k-j)}\ \rt]\
+\ \sum_{k=0}^\infty\ \ep^k\ \sum_{j=0}^k\
\lt(\,h\,P_{\perp}\rt)^{(j)}\ \psi_{\perp}^{(k-j)} \notag
\\[3mm]
&& +\ \sum_{k=4}^\infty\ \ep^k\ \sum_{j=4}^k\ \sum_{l=4}^j\
\lt(-\f12\rt) \lt(\ P_{\perp}^{(j-l)} \lt[\, \lt(\Delta_{x,\,y}\
\Psi_1\rt)^{(l-4)}\ \rt]\ f^{(k-j)}\rt. \notag
\\[3mm]
&& \hskip 4.5cm \lt.+\ P_{\perp}^{(j-l)} \lt[\, \lt(\Delta_{x,\,y}\
\Psi_2\rt)^{(l-4)}\ \rt]\ g^{(k-j)}\ \rt) \notag
\\[3mm]
&& +\ \sum_{k=3}^\infty\ \ep^k\ \sum_{j=3}^k\ \sum_{l=3}^j\ \lt(\,-\
P_{\perp}^{(j-l)} \lt[\,\lt(\f{\pa \Psi_1}{\pa x}\rt)^{(l-3)}\ \rt]\
\f{\partial\phantom{t}}{\partial X}\rt. \notag
\\[3mm]
&& \hskip 4.5cm \lt.-\ P_{\perp}^{(j-l)} \lt[\,\lt(\f{\pa
\Psi_1}{\pa y}\rt)^{(l-3)}\ \rt]\ \f{\partial\phantom{t}}{\partial
Y}\ \rt)\ f^{(k-j)}\ \notag
\\[3mm]
&& +\ \sum_{k=3}^\infty\ \ep^k\ \sum_{j=3}^k\ \sum_{l=3}^j\ \lt(\,-\
P_{\perp}^{(j-l)} \lt[\,\lt(\f{\pa \Psi_2}{\pa x}\rt)^{(l-3)}\ \rt]\
\f{\partial\phantom{t}}{\partial X}\rt. \notag
\\[3mm]
&& \hskip 4.5cm \lt.-\ P_{\perp}^{(j-l)} \lt[\,\lt(\f{\pa
\Psi_2}{\pa y}\rt)^{(l-3)}\ \rt]\ \f{\partial\phantom{t}}{\partial
Y}\ \rt)\ g^{(k-j)} \notag
\\[3mm]
&& =\ \sum_{k=0}^\infty\ \ep^k\ \sum_{j=0}^k\ E^{(j)}\
\psi_{\perp}^{(k-j)}. \label{perp2}
\end{eqnarray}

We now collect terms at each order of $\ep$.  Recall there is an
equation along $\Psi_2$ analogous to (\ref{psi2}).  At each order,
we will combine these two similar equations into one matrix
equation.

\noindent {\bf Order 0}\quad The $\epsilon^0$ terms require
\begin{eqnarray}
\left(\,\begin{array}{cc}\vspace{2mm}
h_{11}^{(0)} & h_{12}^{(0)}\\
h_{21}^{(0)} & h_{22}^{(0)}
\end{array}\,\right)\
\left(\,\begin{array}{c}\vspace{2mm}
f^{(0)} \\
g^{(0)}
\end{array}\,\right)\
&=&\ E^{(0)}\ \left(\,\begin{array}{c}\vspace{2mm}
f^{(0)} \\
g^{(0)}
\end{array}\,\right), \label{order01}
\\[3mm]
\lt(\,h\,P_{\perp}\rt)^{(0)}\ \psi_{\perp}^{(0)}\ &=&\ E^{(0)}\
\psi_{\perp}^{(0)}. \label{order02}
\end{eqnarray}
The $h_{ij}(x,\,y)$ vanish until second order, so this forces
$E^{(0)}=0$ in (\ref{order01}), and consequently
$\psi_{\perp}^{(0)}=0$ after applying the reduced resolvent
$[\lt(\,h(x,y)\,P_{\perp}(x,y)\rt)^{(0)}\ ]_r^{-1}$ in
(\ref{order02}).

\noindent {\bf Order 1}\quad As above, the $\epsilon^1$ terms reduce
to \begin{eqnarray*} E^{(1)}\ \left(\,\begin{array}{c}\vspace{2mm}
f^{(0)} \\
g^{(0)}
\end{array}\,\right)\ &=&\ 0,
\\[3mm]
\lt(\,h\,P_{\perp}\rt)^{(0)}\ \psi_{\perp}^{(1)}\ &=&\ 0.
\end{eqnarray*}
So we get $E^{(1)}=0$ and $\psi_{\perp}^{(1)}=0$.

\noindent {\bf Order 2}\quad Using the known second order terms for
the $h_{ij}(x,y)$, the $\epsilon^2$ terms require
\begin{eqnarray*}
\mathbb{H}_2\ \left(\,\begin{array}{c}\vspace{2mm}
f^{(0)} \\
g^{(0)}
\end{array}\,\right)\ &=&\ E^{(2)}\ \left(\,\begin{array}{c}\vspace{2mm}
f^{(0)} \\
g^{(0)}
\end{array}\,\right),
\\[3mm]
\lt(\,h\,P_{\perp}\rt)^{(0)}\ \psi_{\perp}^{(2)}\ &=&\ 0,
\end{eqnarray*}
where
\begin{eqnarray*}
\mathbb{H}_2\, =\, \left(\begin{array}{cc} \displaystyle -\, \f12 \,
\Delta_{X,\,Y}\, +\, \f{a+b}2 \, X^2\, +\, \f{a-b}2 \, Y^2 &
\displaystyle b\, X\, Y
\\[5mm]
\displaystyle b\, X\, Y & \displaystyle -\, \f{1}2\, \Delta_{X,\,Y}\, +\, \f{a-b}2 \, X^2\, +\, \f{a+b}2 \, Y^2\\
    \end{array}\right). \label{H2matrix}
\end{eqnarray*}

Recall we have assumed the $+bxy$ case for the off diagonal entries.
By again applying the reduced resolvent in the last
equation we have $\psi_{\perp}^{(2)}=0$.  In chapter
\ref{Properties} we show that $\mathbb{H}_2$ is self-adjoint
(on the correct domain) and has purely discrete spectrum with
infinitely many eigenvalues for $a>b>0$.  We are only able to solve
for some of them exactly.  In chapter \ref{EfcnsEvals} we show that
there is at most a two-fold degeneracy in the eigenstates of
$\mb{H}_2$, but that no splitting occurs in the quasimode
eigenvalues, i.e., the degeneracy remains to all orders of $\ep$.
We can therefore proceed as if the eigenstates of $\mb{H}_2$ were
non-degenerate, since we can take any linear combination of
degenerate states for $f^{(0)}$ and $g^{(0)}$, and we know it will
lead to a valid quasimode and energy $E(\ep)$.  Fix $E^{(2)}, \
f^{(0)}$ and $g^{(0)}$ corresponding to one of the states of
$\mb{H}_2$.

\noindent {\bf Order 3}\quad The $\epsilon^3$ terms require
\begin{eqnarray}
\hskip -1.7cm \mathbb{H}_3\, \left(\,\begin{array}{c}\vspace{2mm}
f^{(0)} \\
g^{(0)}
\end{array}\right)\ +\ \lt(\, \mathbb{H}_2-E^{(2)}\, \rt)\,
\left(\,\begin{array}{c}\vspace{2mm}
f^{(1)} \\
g^{(1)}
\end{array}\right)\ =\ E^{(3)}\, \left(\,\begin{array}{c}\vspace{2mm}
f^{(0)} \\
g^{(0)}
\end{array}\right), \label{order31}
\end{eqnarray}
\begin{eqnarray}
\hskip -1.7cm \lt(\,h\,P_{\perp}\rt)^{(0)}\ \psi_{\perp}^{(3)}\ &=&\
\lt( P_{\perp}^{(0)} \lt[\,\lt(\f{\pa \Psi_1}{\pa x}\rt)^{(0)}\
\rt]\ \f{\partial\phantom{t}}{\partial X}\ +\ P_{\perp}^{(0)}
\lt[\,\lt(\f{\pa \Psi_1}{\pa y}\rt)^{(0)}\ \rt]\
\f{\partial\phantom{t}}{\partial Y}\ \rt)\ f^{(0)}\ \notag
\\[5mm]
\hskip -1.7cm && \hskip -1cm +\, \lt( P_{\perp}^{(0)}
\lt[\,\lt(\f{\pa \Psi_2}{\pa x}\rt)^{(0)}\ \rt]\
\f{\partial\phantom{t}}{\partial X}\ +\ P_{\perp}^{(0)}
\lt[\,\lt(\f{\pa \Psi_2}{\pa y}\rt)^{(0)}\ \rt]\
\f{\partial\phantom{t}}{\partial Y}\ \rt)\ g^{(0)}, \label{order32}
\end{eqnarray}
where
\begin{eqnarray*}
\mathbb{H}_3 &=& \left(
\begin{array}{cc}\vspace{2mm} \displaystyle h_{11}^{(3)} & \displaystyle h_{12}^{(3)}
\\
 \displaystyle h_{21}^{(3)} &
\displaystyle h_{22}^{(3)}\\
    \end{array}\right)\ +\ \left(\begin{array}{cc}\vspace{2mm} 0 &
     \displaystyle -\ A_{12}^{(0)}\ \f{\pa \phantom{t}}{\pa
    X}\ -\ B_{12}^{(0)}\ \f{\pa \phantom{t}}{\pa
    Y}
\\
 \displaystyle -\ A_{21}^{(0)}\
    \f{\pa \phantom{t}}{\pa X}\ -\ B_{21}^{(0)}\ \f{\pa \phantom{t}}{\pa
    Y} & 0 \\
    \end{array}\,\right).
\end{eqnarray*}
Since $\mathbb{H}_2$ is self-adjoint, we can take inner products of
both sides in (\ref{order31}) with $\displaystyle
\lt(\begin{array}{c} f^{(0)} \\ g^{(0)}
\end{array}
\rt)$ to obtain
$$
E^{(3)}\ =\ \lt\langle\ \displaystyle \lt(\begin{array}{c} f^{(0)}
\\
g^{(0)}
\end{array}
\rt),\, \mathbb{H}_3\, \lt(\begin{array}{c} \displaystyle f^{(0)}
\\
\displaystyle g^{(0)}
\end{array}
\rt)\ \rt\rangle_{\mH}.
$$
\vskip .3cm

In the appendix we argue that all of the odd order
$E^{(k)}$ are zero.  Let $Q_{\perp}$ be the projection in $\mH$ onto
the subspace perpendicular to the eigenspace of the eigenvalue
$E^{(2)}$ of $\mb{H}_2$.  Adopting ``intermediate normalization'' we
may choose the non-zero order wave functions perpendicular to the
eigenspace of $E^{(2)}$ (note that this will produce a
non-normalized quasimode), so that
\begin{eqnarray*}
\lt(\begin{array}{c} \displaystyle f^{(k)}
\\
\displaystyle g^{(k)}
\end{array}
\rt)\ =\ Q_{\perp}\, \lt(\begin{array}{c} \displaystyle f^{(k)}
\\
\displaystyle g^{(k)}
\end{array}
\rt),
\end{eqnarray*}
for $k \geq 1$.  Then from (\ref{order31}) we get
\begin{eqnarray}
\lt(\begin{array}{c} \displaystyle f^{(1)}
\\[1mm]
\displaystyle g^{(1)}
\end{array}\rt)&=&-\lt[\mathbb{H}_2-E^{(2)}\rt]^{-1}_r\ Q_\perp\ \mathbb{H}_3\, \lt(\begin{array}{c} \displaystyle
f^{(0)}
\\
\displaystyle g^{(0)}
\end{array}
\rt). \label{f1g1}
\end{eqnarray}
From (\ref{order32}) we have
\begin{eqnarray}
\psi_{\perp}^{(3)} &=& \lt[\lt(\,h\,P_{\perp}\rt)^{(0)}\rt]_r^{-1}\
\lt[\ \lt( P_{\perp}^{(0)} \lt[\,\lt(\f{\pa \Psi_1}{\pa
x}\rt)^{(0)}\ \rt]\ \f{\partial\phantom{t}}{\partial X}\ +\
P_{\perp}^{(0)} \lt[\,\lt(\f{\pa \Psi_1}{\pa y}\rt)^{(0)}\ \rt]\
\f{\partial\phantom{t}}{\partial Y}\ \rt)\ f^{(0)}\ \rt. \notag
\\[5mm]
&& \hskip .2cm \lt. +\ \lt( P_{\perp}^{(0)} \lt[\,\lt(\f{\pa
\Psi_2}{\pa x}\rt)^{(0)}\ \rt]\ \f{\partial\phantom{t}}{\partial X}\
+\ P_{\perp}^{(0)} \lt[\,\lt(\f{\pa \Psi_2}{\pa y}\rt)^{(0)}\ \rt]\
\f{\partial\phantom{t}}{\partial Y}\ \rt)\ g^{(0)}\ \rt].
\label{psiperp3}
\end{eqnarray}

\noindent{\bf Order 4}  The $\ep^4$ terms require
\begin{eqnarray}
\hskip -1cm \lt(\mathbb{H}_2-E^{(2)}\rt)\, \lt(\begin{array}{c}
\displaystyle f^{(2)}
\\[1mm]
\displaystyle g^{(2)}
\end{array}\rt)\, +\, \lt(\mathbb{H}_3-E^{(3)}\rt)\, \lt(\begin{array}{c} \displaystyle
f^{(1)}
\\[1mm]
\displaystyle g^{(1)} \end{array}\rt)\, +\,
\lt(\mathbb{H}_4-E^{(4)}\rt)\, \lt(\begin{array}{c} \displaystyle
f^{(0)}
\\[1mm]
\displaystyle g^{(0)} \end{array}\rt)\, =\, 0, \label{order41}
\end{eqnarray}
\begin{eqnarray}
&& \lt(\,h\ P_\perp\rt)^{(0)}\ \psi_\perp^{(4)}\ =\ -\
\lt(\,h\,P_{\perp}\rt)^{(1)}\ \psi_{\perp}^{(3)}\label{order42}
\\[3mm]
&& +\ \f12 \lt(\ P_{\perp}^{(0)} \lt[\, \lt(\Delta_{x,\,y}\
\Psi_1\rt)^{(0)}\ \rt]\ f^{(0)}\ +\ P_{\perp}^{(0)} \lt[\,
\lt(\Delta_{x,\,y}\ \Psi_2\rt)^{(0)}\ \rt]\ g^{(0)}\ \rt) \notag
\\[3mm]
&& +\ \sum_{j=3}^4\ \sum_{l=3}^j\ \lt(\,P_{\perp}^{(j-l)}
\lt[\,\lt(\f{\pa \Psi_1}{\pa x}\rt)^{(l-3)}\ \rt]\
\f{\partial\phantom{t}}{\partial X}\ +\ P_{\perp}^{(j-l)}
\lt[\,\lt(\f{\pa \Psi_1}{\pa y}\rt)^{(l-3)}\ \rt]\
\f{\partial\phantom{t}}{\partial Y}\ \rt)\ f^{(4-j)}\ \notag
\\[3mm]
&& +\ \sum_{j=3}^4\ \sum_{l=3}^j\ \lt(\, P_{\perp}^{(j-l)}
\lt[\,\lt(\f{\pa \Psi_2}{\pa x}\rt)^{(l-3)}\ \rt]\
\f{\partial\phantom{t}}{\partial X}\ +\ P_{\perp}^{(j-l)}
\lt[\,\lt(\f{\pa \Psi_2}{\pa y}\rt)^{(l-3)}\ \rt]\
\f{\partial\phantom{t}}{\partial Y}\ \rt)\ g^{(4-j)}, \notag
\end{eqnarray}
where
\begin{eqnarray*}
\mathbb{H}_4 &=& \lt(-\f12\rt)\ \left(
\begin{array}{cc}\vspace{2mm} \displaystyle T_{11}^{(0)} & \displaystyle T_{12}^{(0)}
\\
 \displaystyle T_{21}^{(0)} &
\displaystyle T_{22}^{(0)}\\
    \end{array}\right)\ +\ \left(
\begin{array}{cc}\vspace{2mm} \displaystyle h_{11}^{(4)} & \displaystyle h_{12}^{(4)}
\\
 \displaystyle h_{21}^{(4)} &
\displaystyle h_{22}^{(4)}\\
    \end{array}\right)\ +
\\[3mm]
&& \hskip 1cm    \left(\begin{array}{cc}\vspace{2mm} 0 &
     \displaystyle -\ A_{12}^{(1)}\ \f{\pa \phantom{t}}{\pa
    X}\ -\ B_{12}^{(1)}\ \f{\pa \phantom{t}}{\pa
    Y}
\\
 \displaystyle -\ A_{21}^{(1)}\
    \f{\pa \phantom{t}}{\pa X}\ -\ B_{21}^{(1)}\ \f{\pa \phantom{t}}{\pa
    Y} & 0 \\
    \end{array}\,\right).
\end{eqnarray*}

Using what we know through order 3, we can solve (\ref{order41}) and
(\ref{order42}).  From (\ref{order41}) we obtain:
\begin{eqnarray*}
E^{(4)}&=& \lt\langle\ \lt(\begin{array}{c} \displaystyle f^{(0)}
\\[1mm]
\displaystyle g^{(0)}
\end{array}\rt),\ \lt(\mathbb{H}_3-E^{(3)}\rt)\ \lt(\begin{array}{c}
\displaystyle f^{(1)}
\\[1mm]
\displaystyle g^{(1)} \end{array}\rt)\ \rt\rangle_{\mH}\ +\
\lt\langle\ \lt(\begin{array}{c}\vspace{1mm} f^{(0)} \\ g^{(0)}
\end{array}\rt),\ \mathbb{H}_4\ \lt(\begin{array}{c}\vspace{1mm} f^{(0)} \\ g^{(0)}
\end{array}\rt)\ \rt\rangle_{\mH} \label{E4solve}
\end{eqnarray*}
and
\begin{eqnarray*}
\hskip -1.5cm \lt(\begin{array}{c}\vspace{1mm} f^{(2)}
\\
g^{(2)}
\end{array}\rt)\ &=&\ -\ \lt[\mathbb{H}_2-E^{(2)}\rt]_r^{-1}\ Q_\perp\
\lt[\ \lt(\mathbb{H}_3-E^{(3)}\rt)\ \lt(\begin{array}{c}
\displaystyle f^{(1)}
\\[1mm]
\displaystyle g^{(1)} \end{array}\rt) \rt. +\ \lt. \mathbb{H}_4\
\lt(\begin{array}{c} \displaystyle f^{(0)}
\\[1mm]
\displaystyle g^{(0)} \end{array}\rt)\  \rt].
\end{eqnarray*}
From (\ref{order42}) we get
\begin{eqnarray*}
&&\psi_\perp^{(4)}\ =\ \lt[\lt(\,h\ P_\perp\rt)^{(0)} \rt]_r^{-1}\
\Bigg[ \ -\ \lt(\,h\,P_{\perp}\rt)^{(1)}\ \psi_{\perp}^{(3)}
\label{orderpsiperp4}
\\[2mm]
&& +\ \f12 \lt(\ P_{\perp}^{(0)} \lt[\, \lt(\Delta_{x,\,y}\
\Psi_1\rt)^{(0)}\ \rt]\ f^{(0)}\ +\ P_{\perp}^{(0)} \lt[\,
\lt(\Delta_{x,\,y}\ \Psi_2\rt)^{(0)}\ \rt]\ g^{(0)}\ \rt) \notag
\\[3mm]
&& +\ \sum_{j=3}^4\ \sum_{l=3}^j\ \lt(\,P_{\perp}^{(j-l)}
\lt[\,\lt(\f{\pa \Psi_1}{\pa x}\rt)^{(l-3)}\ \rt]\
\f{\partial\phantom{t}}{\partial X}\ +\ P_{\perp}^{(j-l)}
\lt[\,\lt(\f{\pa \Psi_1}{\pa y}\rt)^{(l-3)}\ \rt]\
\f{\partial\phantom{t}}{\partial Y}\ \rt)\ f^{(4-j)}\ \notag
\\[3mm]
&& \lt.+\ \sum_{j=3}^4\ \sum_{l=3}^j\ \lt(\, P_{\perp}^{(j-l)}
\lt[\,\lt(\f{\pa \Psi_2}{\pa x}\rt)^{(l-3)}\ \rt]\
\f{\partial\phantom{t}}{\partial X}\ +\ P_{\perp}^{(j-l)}
\lt[\,\lt(\f{\pa \Psi_2}{\pa y}\rt)^{(l-3)}\ \rt]\
\f{\partial\phantom{t}}{\partial Y}\ \rt)\ g^{(4-j)}\ \rt]. \notag
\end{eqnarray*}

\noindent {\bf Order \boldmath{$k \geq 5$}}\quad We now show that we can
proceed in this manner to any order of $\ep$ desired.  In chapter
\ref{Properties} we will show that all of the quantities involved
exist in the relevant Hilbert space.  If $k \geq 5$, the
$\epsilon^k$ terms require
\begin{eqnarray}
&& \hskip -1cm \lt(\mathbb{H}_2-E^{(2)}\rt)\ \lt(\begin{array}{c}
\displaystyle f^{(k-2)}
\\[1mm]
\displaystyle g^{(k-2)}
\end{array}\rt)\ +\ \lt(\mathbb{H}_3-E^{(3)}\rt)\ \lt(\begin{array}{c} \displaystyle
f^{(k-3)}
\\[1mm]
\displaystyle g^{(k-3)} \end{array}\rt)\ +\ \sum_{j=4}^{k-1}\
\lt(\mathbb{H}_j-E^{(j)}\rt)\ \lt(\begin{array}{c} \displaystyle
f^{(k-j)}
\\[1mm]
\displaystyle g^{(k-j)} \end{array}\rt) \notag
\\[3mm]
&&\hskip -1cm +\, \lt(\mathbb{H}_k-E^{(k)}\rt)\,
\lt(\begin{array}{c} \displaystyle f^{(0)}
\\[1mm]
\displaystyle g^{(0)} \end{array}\rt)\, +\, \sum_{j=2}^{k-3}\,
\lt(-\f12\rt)\, \lt(\begin{array}{c}\vspace{2mm} \displaystyle
\langle\ \Psi_1^{(j-2)},\ \Delta_{X,\,Y}\ \psi_\perp^{(k-j)}\
\rangle_{el}
\\
\displaystyle \langle\ \Psi_2^{(j-2)},\ \Delta_{X,\,Y}\
\psi_\perp^{(k-j)}\ \rangle_{el} \end{array}\rt)\, =\, 0,
\label{orderk1}
\end{eqnarray}
\begin{eqnarray}
&& \lt(\,h\ P_\perp\rt)^{(0)}\ \psi_\perp^{(k)}\ =\
\sum_{j=2}^{k-3}\ \f12\ P_{\perp}^{(j-2)} \lt[\, \Delta_{X,\,Y}\
\psi_{\perp}^{(k-j)}\ \rt]\ -\ \sum_{j=1}^{k-3}\
\lt(\,h\,P_{\perp}\rt)^{(j)}\ \psi_{\perp}^{(k-j)} \notag
\\[2mm]
&& +\ \sum_{j=4}^k\ \sum_{l=4}^j\ \f12 \lt(\ P_{\perp}^{(j-l)}
\lt[\, \lt(\Delta_{x,\,y}\ \Psi_1\rt)^{(l-4)}\ \rt]\ f^{(k-j)}\ +\
P_{\perp}^{(j-l)} \lt[\, \lt(\Delta_{x,\,y}\ \Psi_2\rt)^{(l-4)}\
\rt]\ g^{(k-j)}\ \rt) \notag
\\[3mm]
&& +\ \sum_{j=3}^k\ \sum_{l=3}^j\ \lt(\,P_{\perp}^{(j-l)}
\lt[\,\lt(\f{\pa \Psi_1}{\pa x}\rt)^{(l-3)}\ \rt]\
\f{\partial\phantom{t}}{\partial X}\ +\ P_{\perp}^{(j-l)}
\lt[\,\lt(\f{\pa \Psi_1}{\pa y}\rt)^{(l-3)}\ \rt]\
\f{\partial\phantom{t}}{\partial Y}\ \rt)\ f^{(k-j)}\ \notag
\\[3mm]
&& +\ \sum_{j=3}^k\ \sum_{l=3}^j\ \lt(\, P_{\perp}^{(j-l)}
\lt[\,\lt(\f{\pa \Psi_2}{\pa x}\rt)^{(l-3)}\ \rt]\
\f{\partial\phantom{t}}{\partial X}\ +\ P_{\perp}^{(j-l)}
\lt[\,\lt(\f{\pa \Psi_2}{\pa y}\rt)^{(l-3)}\ \rt]\
\f{\partial\phantom{t}}{\partial Y}\ \rt)\ g^{(k-j)} \notag
\\[3mm]
&& +\ \sum_{j=2}^{k-3}\ E^{(j)}\ \psi_{\perp}^{(k-j)},
\label{orderk2}
\end{eqnarray}
where
\begin{eqnarray*}
\mathbb{H}_j &=& \lt(-\f12\rt)\ \left(
\begin{array}{cc}\vspace{2mm} \displaystyle T_{11}^{(j-4)} & \displaystyle T_{12}^{(j-4)}
\\
 \displaystyle T_{21}^{(j-4)} &
\displaystyle T_{22}^{(j-4)}\\
    \end{array}\right)\ +\ \left(
\begin{array}{cc}\vspace{2mm} \displaystyle h_{11}^{(j)} & \displaystyle h_{12}^{(j)}
\\
 \displaystyle h_{21}^{(j)} &
\displaystyle h_{22}^{(j)}\\
    \end{array}\right)\ +
\\[3mm]
&& \hskip 1cm    \left(\begin{array}{cc}\vspace{2mm} 0 &
     \displaystyle -\ A_{12}^{(j-3)}\ \f{\pa \phantom{t}}{\pa
    X}\ -\ B_{12}^{(j-3)}\ \f{\pa \phantom{t}}{\pa
    Y}
\\
 \displaystyle -\ A_{21}^{(j-3)}\
    \f{\pa \phantom{t}}{\pa X}\ -\ B_{21}^{(j-3)}\ \f{\pa \phantom{t}}{\pa
    Y} & 0 \\
    \end{array}\,\right),
\end{eqnarray*}
for $j \geq 4$.

Following what we have seen through order 4, assume from previous
orders that
$$
\lt(\begin{array}{c}\vspace{1mm} f^{(j)} \\ g^{(j)}
\end{array}\rt) \text{ for } j=0,1,\ldots,k-3,\qquad E^{(j)} \text{ and } \psi_\perp^{(j)} \text{ for }
j=0,1,\ldots,k-1,
$$
are already determined.  Then, we can solve (\ref{orderk1}) and
(\ref{orderk2}) for $f^{(k-2)}$, $g^{(k-2)}$, $\psi_{\perp}^{(k)}$,
and $E^{(k)}$. From (\ref{orderk1}) we obtain:
\begin{eqnarray}
E^{(k)}&=& \sum_{j=3}^{k-1}\ \lt\langle\ \lt(\begin{array}{c}
\displaystyle f^{(0)}
\\[1mm]
\displaystyle g^{(0)}
\end{array}\rt),\ \lt(\mathbb{H}_j-E^{(j)}\rt)\ \lt(\begin{array}{c}
\displaystyle f^{(k-j)}
\\[1mm]
\displaystyle g^{(k-j)} \end{array}\rt)\ \rt\rangle_{\mH}\ +\
\lt\langle\ \lt(\begin{array}{c}\vspace{1mm} f^{(0)} \\ g^{(0)}
\end{array}\rt),\ \mathbb{H}_k\ \lt(\begin{array}{c}\vspace{1mm} f^{(0)} \\ g^{(0)}
\end{array}\rt)\ \rt\rangle_{\mH} \notag
\\[3mm]
&&\hskip 1cm -\ \f12\ \sum_{j=2}^{k-3}\ \lt\langle\
\lt(\begin{array}{c}\vspace{1mm} f^{(0)} \\ g^{(0)}
\end{array}\rt),\ \lt(\begin{array}{c}\vspace{1mm}
\langle\ \Psi_1^{(j-2)},\ \Delta_{X,\,Y}\ \psi_\perp^{(k-j)}\
\rangle_{el}
\\
\langle\ \Psi_2^{(j-2)},\ \Delta_{X,\,Y}\ \psi_\perp^{(k-j)}\
\rangle_{el}
\end{array}\rt)\ \rt\rangle_{\mH} \label{Ek}
\end{eqnarray}
and
\begin{eqnarray}
\hskip -2cm \lt(\begin{array}{c}\vspace{1mm} f^{(k-2)}
\\
g^{(k-2)}
\end{array}\rt)\ &=&\ -\ \lt[\mathbb{H}_2-E^{(2)}\rt]_r^{-1}\ Q_\perp\
\lt[\ \sum_{j=3}^{k-1}\ \lt(\mathbb{H}_j-E^{(j)}\rt)\
\lt(\begin{array}{c} \displaystyle f^{(k-j)}
\\[1mm]
\displaystyle g^{(k-j)} \end{array}\rt) \rt. \notag
\\[3mm]
\hskip -2cm && \hskip -.5cm +\ \lt. \mathbb{H}_k\
\lt(\begin{array}{c} \displaystyle f^{(0)}
\\[1mm]
\displaystyle g^{(0)} \end{array}\rt)\ -\ \f12\ \sum_{j=2}^{k-3}\
\lt(\begin{array}{c}\vspace{2mm} \displaystyle \langle\
\Psi_1^{(j-2)},\ \Delta_{X,\,Y}\ \psi_\perp^{(k-j)}\ \rangle_{el}
\\
\displaystyle \langle\ \Psi_2^{(j-2)},\ \Delta_{X,\,Y}\
\psi_\perp^{(k-j)}\ \rangle_{el} \end{array}\rt)\ \rt].
\label{orderfgk}
\end{eqnarray}
From (\ref{orderk2}) we get
\begin{eqnarray}
\psi_\perp^{(k)} &=& \lt[\lt(\,h\ P_\perp\rt)^{(0)} \rt]_r^{-1}\
\lt[\ \sum_{j=2}^{k-3}\ \f12\ P_{\perp}^{(j-2)} \lt[\,
\Delta_{X,\,Y}\ \psi_{\perp}^{(k-j)}\ \rt]\ -\ \sum_{j=1}^{k-3}\
\lt(\,h\,P_{\perp}\rt)^{(j)}\ \psi_{\perp}^{(k-j)} \rt. \notag
\\[2mm]
&& +\ \sum_{j=4}^k\ \sum_{l=4}^j\ \f12 \lt(\ P_{\perp}^{(j-l)}
\lt[\, \lt(\Delta_{x,\,y}\ \Psi_1\rt)^{(l-4)}\ \rt]\ f^{(k-j)}\ +\
P_{\perp}^{(j-l)} \lt[\, \lt(\Delta_{x,\,y}\ \Psi_2\rt)^{(l-4)}\
\rt]\ g^{(k-j)}\ \rt) \notag
\\[3mm]
&& +\ \sum_{j=3}^k\ \sum_{l=3}^j\ \lt(\,P_{\perp}^{(j-l)}
\lt[\,\lt(\f{\pa \Psi_1}{\pa x}\rt)^{(l-3)}\ \rt]\
\f{\partial\phantom{t}}{\partial X}\ +\ P_{\perp}^{(j-l)}
\lt[\,\lt(\f{\pa \Psi_1}{\pa y}\rt)^{(l-3)}\ \rt]\
\f{\partial\phantom{t}}{\partial Y}\ \rt)\ f^{(k-j)}\ \notag
\\[3mm]
&& +\ \sum_{j=3}^k\ \sum_{l=3}^j\ \lt(\, P_{\perp}^{(j-l)}
\lt[\,\lt(\f{\pa \Psi_2}{\pa x}\rt)^{(l-3)}\ \rt]\
\f{\partial\phantom{t}}{\partial X}\ +\ P_{\perp}^{(j-l)}
\lt[\,\lt(\f{\pa \Psi_2}{\pa y}\rt)^{(l-3)}\ \rt]\
\f{\partial\phantom{t}}{\partial Y}\ \rt)\ g^{(k-j)} \notag
\\[3mm]
&& \lt. +\ \sum_{j=2}^{k-3}\ E^{(j)}\ \psi_{\perp}^{(k-j)}\ \rt].
\label{orderpsiperpk}
\end{eqnarray}
So we can proceed in this manner to obtain $\Psi(\ep)$ and $E(\ep)$
up to any order in $\ep$.

\section{Properties of the Leading Order
Hamiltonian}\label{Properties} \setcounter{equation}{0}
\setcounter{theorem}{0}

We adopt the following notation throughout:

\begin{enumerate}

\item We let $I_2\, =\, \Id$. If $A$ is an operator on the Hilbert
space $L^2(\R^2,\,dX\,dY)$, then $A\,\Iten$, is the operator on
${\cal H}$ given by $\left(\begin{array}{cc} A & 0\\0 & A
\end{array}\right)$.

\item If $D(A)$ is the domain of the operator $A$ on the Hilbert
space $L^2(\R^2,\,dX\,dY)$, then\\
$D(A\,\Iten)\, =\, D(A)\, \oplus\, D(A)\, \subset\, {\cal H}$.

\end{enumerate}

In what follows, we prove various needed properties for the
expansion to all orders.
Let
$$
\mb{H}_2\ = -\ \f12\ \Delta_{X,\,Y}\, \Iten\ +\
\left(\begin{array}{cc} \displaystyle \f{a+b}2 \ X^2\ +\ \f{a-b}2 \
Y^2 & \displaystyle b\, X\, Y
\\[3mm]
\displaystyle b\, X\, Y & \displaystyle \ \ \f{a-b}2 \ X^2\ +\ \f{a+b}2 \ Y^2\\
    \end{array}\right).
$$
Note that if we let
$(\tilde{X},\,\tilde{Y})=(a^{1/4}\,X,\,a^{1/4}\,Y)$ and $\dsp
\tilde{b}=\f{b}{a}\, $, then
\begin{eqnarray}
\mb{H}_2\ &=&\ \sqrt{a}\ \lt[\, \lt(\ -\ \f12\
\Delta_{\tilde{X},\,\tilde{Y}}\ +\ \f12\ (\tilde{X}^2+\tilde{Y}^2)\
\rt)\ \Iten \rt. \notag
\\[3mm]
&&\lt. \hspace{3cm} +\ \tilde{b}\ \left(\begin{array}{cc}
\displaystyle \f12 \ \lt(\tilde{X}^2 - \tilde{Y}^2\rt) &
\displaystyle \tilde{X}\, \tilde{Y}
\\[3mm]
\displaystyle \tilde{X}\, \tilde{Y} & \displaystyle -\ \f12 \
\lt(\tilde{X}^2 -
\tilde{Y}^2\rt)\\
    \end{array}\right)\ \rt]. \label{rescaling}
\end{eqnarray}
We now use the Kato-Rellich Theorem \cite{RSV2} to prove
self-adjointness of $\mb{H}_2$.

\begin{thm}\label{theorem1}
If $\ a>b>0$, then $\mb{H}_2$ is self-adjoint on $D_{HO}\, \oplus\,
D_{HO}\, $, where $D_{HO}$ is the usual Harmonic oscillator domain
in $L^2(\R^2,dXdY)$, and essentially self-adjoint on
$\tilde{D}_{HO}\, \oplus\, \tilde{D}_{HO}\, $, where
$\tilde{D}_{HO}$ is any core for the usual Harmonic oscillator.
\end{thm}

\vskip 5mm \noindent {\bf Proof:}\quad

Define
$$
H_{HO}\ =\ \lt(\ -\ \f12\ \Delta_{X,\,Y}\ +\ \f12\ (X^2+Y^2)\ \rt)\
\Iten
$$
and
$$
V(\tilde{b})\ =\ \tilde{b}\ \left(\begin{array}{cc} \displaystyle
\f12 \ \lt(X^2 - Y^2\rt) & \displaystyle X\, Y
\\[3mm]
\displaystyle X\, Y & \displaystyle -\ \f12 \ \lt(X^2 -
Y^2\rt)\\
    \end{array}\right).
$$
We prove that for $0<\tilde{b}<1$, $V(\tilde{b})$ is relatively
bounded with respect to $H_{H0}$, with relative bound $\tilde{b}$.
The conclusion then follows from the Kato-Rellich theorem
\cite{RSV2} and (\ref{rescaling}).

For each fixed $X$ and $Y$, the eigenvalues of $V(b)$ are
$\pm\,\f{\tilde{b}}{2}\,(X^2+Y^2)$.  It follows that
$$
\lt\|\,V(\tilde{b})\,v\,\rt\|_e\ \le\ \tilde{b}\
\lt\|\lt(\,\f12\,(X^2+Y^2)\,\Iten\rt)\,v\,\rt\|_e,
$$
where $v\in {\mathbb{C}}^2$ is any two component vector, and we use
the usual Euclidean norm. This inequality implies the
$L^2({\mathbb{R}}^2,\,dX\,dY;\,\mathbb{C}^2)=\mH$ norm estimate
$$
\lt\|\,V(\tilde{b})\,\psi\,\rt\|_{\mH}\ \le\ \tilde{b}\
\lt\|\,\lt(\f12\,(X^2+Y^2)\,\Iten\rt)\,\psi\,\rt\|_{\mH},
$$
where $\psi(X,\,Y) \in \mH$ is a two-component vector-valued
function.

We now show that
\begin{eqnarray}
&& \hskip -2.5cm \|\,V(\tilde{b})\,\psi\,\|_\mH\ \le\ \tilde{b}\
\lt\|\,\lt(\f12\,(X^2+Y^2)\,\Iten\rt)\,\psi\,\rt\|_\mH\ \le\
\tilde{b}\ \|\,H_{HO}\,\psi\,\|_\mH\ +\ \tilde{b}\ \|\,\psi\,\|_\mH.
\label{relativebound}
\end{eqnarray}
for all $\psi \in D_{HO} \oplus D_{HO}\ $.  We have already shown the first inequality. The hard part is the
second estimate, which follows from
\begin{equation*}\label{GoodGrief}
\left\|\,\lt(\left(\,X^2\,+\,Y^2\,\rt)\,\Iten\,\rt) \psi\,\right\|
\quad\le\quad \left\|\,\lt(\lt(\,-\,\Delta_{X,Y}\,+\,
X^2\,+\,Y^2\,\rt)\,\Iten\rt)\,\psi\,\right\|\ +\ 2 \ \|\,\psi\,\|.
\end{equation*}
This easily follows from
\begin{equation}\label{MoreGrief}
\left\|\,\lt(\left(\,X^2\,+\,Y^2\,\rt)\,\Iten\rt)\,
\psi\,\right\|^{\,2} \quad\le\quad
\left\|\,\lt(\lt(\,-\,\Delta_{X,Y}\,+\,
X^2\,+\,Y^2\,\rt)\,\Iten\rt)\,\psi\,\right\|^{\,2}\ +\ 4\
\|\,\psi\,\|^{\,2}.
\end{equation}

Rather than proving this directly, let us first prove a simpler
relative bound estimate for the operators on $L^2(\R,dx)$.  We show
that for $\phi \in D\lt(-\, \frac{\partial^2}{\partial x^2}
\,+\,x^2\rt)$,
\begin{equation}\label{skunk}
\left\|\,x^2\,\phi\,\right\|^2\quad \le\quad \left\|\,\left(\,-\,
\frac{\partial^2}{\partial x^2}
\,+\,x^2\,\right)\,\phi\,\right\|^{\,2}\ +\ 2\ \|\,\phi\,\|^2.
\end{equation}
To prove this, let $p=-i\frac{\partial}{\partial x}$, and calculate
the commutators
$$
[x,\,p\,]\ =\ i\qquad\mbox{and}\qquad [x,\,p^2\,]\ =\ 2\,i\,p.
$$
We have
\begin{eqnarray}\nonumber
\left\|\,x^2\,\phi\,\right\|^2 &=&
\langle\,\phi,\,x^4\,\phi\,\rangle
\\[3mm]\nonumber
&=&\langle\,\phi,\,
\left((p^2+x^2)^2\,-\,x^2p^2\,-\,p^2x^2\,-\,p^4\right)\,
\phi\,\rangle
\\[3mm]\label{dogbreath}
&\leq& \left\|\,\left(p^2+x^2\right)\,\phi\,\right\|^2\ -\
\langle\,\phi,\, \left(x^2p^2+p^2x^2\right)\,\phi\,\rangle.
\end{eqnarray}
In this last expression, we use the commutators above to write
\begin{eqnarray}\nonumber
\langle\,\phi,\, \left(x^2p^2+p^2x^2\right)\,\phi\,\rangle
&=&\langle\,\phi,\, \left(\,xp^2x\,+\,x[x,\,p^2\,]\,
+\,xp^2x\,+\,[p^2,\,x\,]x\,\right)\,\phi\,\rangle
\\[3mm]\nonumber
&=&2\,\langle\,\phi,\,xp^2x\,\phi\,\rangle\ +\
2\,i\,\langle\,\phi,\,(xp\,-\,px)\,\phi\,\rangle
\\[3mm]\nonumber
&=&2\,\langle\,\phi,\,xp^2x\,\phi\,\rangle\ - 2\
\langle\,\phi,\,\phi\,\rangle.
\end{eqnarray}
In this last expression, the first inner product is the expectation
of a positive operator (since $xp^2x$ has the form $A^*A$ with
$A=px$). Using this and (\ref{dogbreath}), we see that
$$
\left\|\,x^2\,\phi\,\right\|^2\ \le\
\left\|\,\left(p^2+x^2\right)\,\phi\,\right\|^2\ +\ 2\
\|\,\phi\,\|^2,
$$
and (\ref{skunk}) is proved.

Now we simply mimic the proof of (\ref{skunk}) to prove
(\ref{MoreGrief}). We write
\begin{eqnarray}\nonumber
&& \hskip -.8cm \|\,\left(\,X^2\,+\,Y^2\right)\,\phi\,\|^2
\\[3mm]\nonumber
&& \hskip -.8cm =\ \lt\langle\,\phi,\,\left(\left(\,
-\,\Delta_{X,Y}\,+\,X^2\,+\,Y^2\right)^2 \,-\,\Delta_{X,Y}^2
\,+\,\Delta_{X,Y}\,\left(\,X^2\,+\,Y^2\right)
\,+\,\left(\,X^2\,+\,Y^2\right)\,\Delta_{X,Y}\, \right)
\phi\,\rt\rangle
\end{eqnarray}
The operator $\Delta_{X,Y}^2$ is positive. The operator
$-\,\Delta_X\,Y^2=-\,Y^2\,\Delta_X$ is also positive since it equals
$A^*A$ with $A=p_XY$. Similarly, $-\,\Delta_Y\,X^2=-\,X^2\,\Delta_Y$
is positive. By the commutator tricks we used above,
$-\,\Delta_X\,X^2-\,X^2\,\Delta_X$ and
$-\,\Delta_Y\,Y^2-\,Y^2\,\Delta_Y$ each are positive operators minus
twice the identity. Thus for all $\phi \in D_{HO}$,
\begin{eqnarray}\nonumber
&& \hspace{-1cm} \lt\langle\,\phi,\,\left(\left(\,
-\,\Delta_{X,Y}\,+\,X^2\,+\,Y^2\right)^2 \,-\,\Delta_{X,Y}^2
\,+\,\Delta_{X,Y}\,\left(\,X^2\,+\,Y^2\right)
\,+\,\left(\,X^2\,+\,Y^2\right)\,\Delta_{X,Y}\, \right)
\phi\,\rt\rangle
\\[3mm]\nonumber
&&\hspace{1cm} \le \lt\langle\,\phi,\,\left(\,
-\,\Delta_{X,Y}\,+\,X^2\,+\,Y^2\right)^2\,\phi\,\rt\rangle \ +\
4\,\lt\langle\,\phi,\,\phi\,\rt\rangle
\end{eqnarray}
and hence,
$$
\left\|\,\left(\,X^2\,+\,Y^2\right)\,
\phi\,\right\|^{\,2}\quad\le\quad
\left\|\,\left(\,-\,\Delta_{X,Y}\,+\,X^2\,+\,Y^2\right)\,
\phi\,\right\|^{\,2}\ +\ 4\ \|\,\phi\,\|^{\,2}\ .
$$
It follows that (\ref{MoreGrief}) holds for all $\psi \in D_{HO}
\oplus D_{HO}\ $.  This proves (\ref{relativebound}) and the theorem
follows.  $\square$

\vskip .5cm

Unless otherwise stated, it is assumed that by $\mb{H}_2$ we are
referring to this operator with domain $D(\mb{H}_2)\, =\, D_{H0}\,
\oplus\, D_{HO}$.  We now show that $\mb{H}_2$ has purely discrete
spectrum.

\vskip .5cm

\begin{thm}\label{theorem2}
If $\ a>b>0$,~ $\mb{H}_2$ has purely discrete spectrum, with
countably many eigenvalues $\{\mu_j(\mb{H}_2)\}_{j=1}^\infty$
satisfying
\begin{eqnarray*}\label{evalbounds}
N\sqrt{a-b}\, \leq\, \mu_{N(N-1)+1}(\mb{H}_2)\, \leq\,
\mu_{N(N-1)+2}(\mb{H}_2)\, \leq\, \ldots\, \leq\,
\mu_{N(N+1)}(\mb{H}_2)\, \leq\, N\sqrt{a+b},
\end{eqnarray*}
for $N=1,2,3,\ldots$
\end{thm}
\noindent {\bf Proof:}\quad

Let $(\rho,\phi)$ be the usual polar coordinates associated with
$(X,Y)$.  Define the unitary operators $U,\ W\,:\,{\cal
H}\,\rightarrow\,{\cal H}$ by (defined as multiplication operators
on ${\cal H}$):
\\
$$
U=\left(\begin{array}{cc} \displaystyle \cos(\phi) & \displaystyle -
\sin(\phi)
\\[3mm]
\displaystyle \sin(\phi) & \displaystyle \cos(\phi)\\
    \end{array}\right)
    \hskip 1cm \text{and} \hskip 1cm
    W=\f1{\sqrt{2}}\left(\begin{array}{cc} \displaystyle e^{i\phi} & \displaystyle
e^{-i\phi}
\\[3mm]
\displaystyle ie^{i\phi} & \displaystyle -ie^{-i\phi}\\
    \end{array}\right).
$$
\\
Define
$$
\mathbb{H}_0\ =\ U^{-1}\ \mathbb{H}_2\ U\ =\ \left(\begin{array}{cc}
\displaystyle - \f12 \ \Delta_{\rho,\phi}\ +\ \f1{2\rho^2}\ +\
\f{a+b}2 \ \rho^2 & \displaystyle \f1{\rho^2}\ \f{\pa
\phantom{t}}{\pa \phi}
\\[5mm]
\displaystyle -\ \f1{\rho^2}\ \f{\pa \phantom{t}}{\pa \phi} &
\displaystyle - \f12 \ \Delta_{\rho,\phi}\ +\ \f1{2\rho^2}\ +\
\f{a-b}2 \ \rho^2\\
    \end{array}\right),
$$
and
$$
\mathbb{H}_0^{\pm}\ =\ \left(\begin{array}{cc} \displaystyle - \f12
\ \Delta_{\rho,\phi}\ +\ \f1{2\rho^2}\ +\ \f{a \pm b}2 \ \rho^2 &
\displaystyle \f1{\rho^2}\ \f{\pa \phantom{t}}{\pa \phi}
\\[5mm]
\displaystyle -\ \f1{\rho^2}\ \f{\pa \phantom{t}}{\pa \phi} &
\displaystyle - \f12 \ \Delta_{\rho,\phi}\ +\ \f1{2\rho^2}\ +\
\f{a \pm b}2 \ \rho^2\\
    \end{array}\right),
$$
\\
and note that $\mathbb{H}_0^{-}\ \leq\ \mathbb{H}_0\ \leq\
\mathbb{H}_0^{+}.$  Now we define
\\
$$
\mathbb{H}_1^{\pm}\ =\ W^{-1}\ \mathbb{H}_0^{\pm}\ W\ =\
\left(\begin{array}{cc} \displaystyle - \f12 \ \Delta_{\rho,\phi}\
+\ \f{a \pm b}2 \ \rho^2 & \displaystyle 0
\\[5mm]
\displaystyle 0 & \displaystyle - \f12 \ \Delta_{\rho,\phi}\ +\
\f{a \pm b}2 \ \rho^2\\
    \end{array}\right).
$$

In the context of the min/max principle \cite{RSV4}, for all $n \in
\mb{N}$,
$$
\mu_n(\mb{H}_1^-)\ =\ \mu_n(\mb{H}_0^-)\ \leq\ \mu_n(\mb{H}_0)\ =\
\mu_n(\mb{H}_2)\ =\ \mu_n(\mb{H}_0)\ \leq\ \mu_n(\mb{H}_0^+)\ =\
\mu_n(\mb{H}_1^+).
$$
The operators $\mb{H}_1^{\pm}$ have purely discrete spectrum, with
$2N$-fold degenerate eigenvalues of  $N\sqrt{a \pm b}$ for
$N=1,2,\ldots\ $  So, $\mb{H}_2$ must have purely discrete spectrum
with eigenvalues $\mu_1(\mb{H}_2)\, \leq\, \mu_2(\mb{H}_2)\, \leq\,
\ldots\ $ satisfying the required bound.  $\square$

\vskip .5cm

To prove the quasimode can be expanded to any order in $\ep$, we
must show the terms arising at arbitrary order in the equations of
chapter \ref{QuasimodeConstruction} are in ${\cal H}$. This follows
from the propositions and lemmas we now prove.  A similar analysis
was needed in \cite{Hag/Joye1} and the proofs presented here are analogous to those found in
\cite{Hag/Joye1}.  For our purposes it must be shown
that the details can be extended to this situation on ${\cal H}$.

\vskip 1cm

\begin{lem}\label{lemma3}
Let $T(\al)$ be defined on a dense domain of a separable Hilbert
space $H$, and suppose that $T(\al)$ is an analytic family in the
sense of Kato for all $\, \al \in \mb{C}$, and self-adjoint for all
$\, \al \in \R$.  If, for all $\al \in \R$, $T(\al)$ has purely
discrete spectrum with eigenvalues accumulating at $\infty$, then
$T(\al)$ has purely discrete spectrum for all $\, \al \in \mb{C}$.
\end{lem}

\noindent {\bf Proof:}

First we note that if a self-adjoint operator has purely discrete
spectrum with eigenvalues accumulating at $\infty$, then it has
compact resolvent by Theorem XIII.64 of \cite{RSV4}. We also note
that for any closed operator $A$, $(A-\mu)^{-1}$ is compact for some
$\mu \in \rho(A)$ if and only if $(A-\mu)^{-1}$ is compact for all
$\mu \in \rho(A)$. This follows from the first resolvent formula.

We first show that if $T(\al)$ has compact resolvent for all $\al
\in \R$, then $T(\al)$ has compact resolvent for all $\al \in
\mb{C}$.  We then show that if a closed operator defined on a
separable Hilbert space has compact resolvent, then it must have
purely discrete spectrum.

Since $T(\al)$ is an entire analytic family, the resolvent
$R_{\al}(\lam)\, =\, (T(\al)\, -\, \lam)^{-1}$ is analytic in both
$\al$ and $\lam$ inside the set $R\, =\, \{\, (\,\al,\, \lam\, )\,
:\, \al \in \mb{C},\, \lam \in \rho(T(\al)) \,\}$.  From Theorem
XII.7 of \cite{RSV4}, R is open in both $\al$ and $\lam$. Let
$\mathfrak{B}(H)$ and $\mathfrak{C}(H)$ denote the bounded operators
and compact operators on the Hilbert space $H$ respectively. It
follows from the Hahn-Banach Theorem \cite{RSV1}, that for any $B\,
\in \mathfrak{B}(H) \setminus \mathfrak{C}(H)$, there exists $\ l_B
\in \mathfrak{B}(H)^\ast$ such that $l_B(B) \neq 0$, and $l_B=0$ on
$\mathfrak{C}(H)$.

Note that since $T(\al)$ is an analytic family, we know the
resolvent set is non-empty for all $\al \in \mb{C}$.  Define the set
$$
\Upsilon\, =\, \{\, \al \in \mb{C}\, :\, R_{\al}(\lam)\ \text{is
compact for all } \lam \in \rho(T(\al))\, \}.
$$
We show that $\Upsilon = \mb{C}$.

Let $B_s(z)$ denote an open disk in the complex plane of radius
$s>0$, centered at $z \in \mb{C}$.  Let $\lam_0 \in \rho(T(0))$.
Since the set $R$ is open, we know that there exists a disk
$B_{\delta}(0)$, such that $\lam_0 \in \rho(T(\al))$ for all $\al
\in B_{\delta}(0)$. Let $\ l \in \mathfrak{B}(H)^\ast$, such that $\
l\, $ is vanishing on $\mathfrak{C}(H)$. Then the function
$f(\al)=l(R_{\al}(\lam_0))$ defines an analytic map from
$B_{\delta}(0)$ into $\mb{C}$.  Since the resolvent is compact for
$\al \in \R$, we know $f(\al)=0$ for all $-\delta < \al < \delta$,
which implies $f(\al)=0$ for all $\al \in B_{\delta}(0)$.  Since $l$ was chosen arbitrarily in $\{\ l \in \mathfrak{B}(H)^\ast\ :\ 
l\, $ vanishes on $\mathfrak{C}(H)\ \}$, it
follows that $R_{\al}(\lam_0)$ is compact for all $\al \in
B_{\delta}(0)$ (if $R_{\al'}(\lam_0)$ was not compact for some $\al' \in
B_{\delta}(0)$, we could choose $l$ so that $l(R_{\al'}(\lam_0)) \neq 0$, contradicting $f(\al')=0$).  Hence $B_{\delta}(0) \subset \Upsilon$.

We now assume that $\Upsilon \neq \mb{C}$ and show this leads to a
contradiction.  Let\\ $r\, =\, \sup\{ \delta>0 \,:\, B_{\delta}(0)
\subset \Upsilon \}$.  Note that $0<r<\infty$ since we have assumed
$\Upsilon \neq \mb{C}$.  Then, there exists $\al_0$ with
$\abs{\al_0}=r$, such that every neighborhood of $\al_0$ contains a
point not in $\Upsilon$.  Let $\lam_0 \in \rho(T(\al_0))$ and choose
$\delta'>0$ small enough so that $\lam_0 \in \rho(T(\al))$ for all
$\al \in B_{\delta'}(\al_0)$. Choosing $l$ as before, we know $g(\al)\, =\, l(R_{\al}(\lam_0))$
is analytic on $B_{\delta'}(\al_0)$ and $g(\al)=0$ on $B_{r}(0) \cap
B_{\delta'}(\al_0)$.  So, $g(\al)=0$ on all of $B_{\delta'}(\al_0)$.
Again since $l$ was chosen arbitrarily, there exists an entire neighborhood of $\al_0$ in $\Upsilon$.  This is a contradiction, so $\Upsilon = \mb{C}$.

We now show that a closed operator with compact resolvent has purely
discrete spectrum.  Let $A$ be a closed operator.  Fix $\lam \in
\rho(A)$ and let $R(\lam)=(A-\lam)^{-1}$ be compact. Then the
spectrum of $R(\lam)$ is made up of at most countably many
eigenvalues of finite multiplicity that can only accumulate at 0
\textbf{\cite{RSV1}}. For $E \neq \lam$, we have
$$
A-E\, =\,
A-\lam-(E-\lam)=(E-\lam)(A-\lam)\lt(\f1{E-\lam}-R(\lam)\rt).
$$
From this we see that if $\f1{E-\lam} \in \rho(R(\lam))$, then $E
\in \rho(A)$.  So, if $E \in \sg(A)$, then $\f1{E-\lam} \in
\sg(R(\lam))$ and thus $\f1{E-\lam}$ is an isolated eigenvalue of
$R(\lam)$ with finite multiplicity.  Since,
$$
R(\lam)\Psi\, =\, \f1{E-\lam}\, \Psi\ \Leftrightarrow\ (E-\lam)\,
\Psi\, =\, (A-\lam)\, \Psi\ \Leftrightarrow\ A\Psi\, =\, E\, \Psi,
$$
it follows that $E$ is an isolated eigenvalue of $A$ with finite
multiplicity.  Therefore, $\sg(A)$ is made up of at most countably
many eigenvalues of finite multiplicity that can only accumulate at
infinity.  The conclusion of the Lemma follows.
 $\square$

\vskip .5cm

Before we prove Proposition \ref{prop4}, we consider a different
decomposition of $\mb{H}_2$.  We define $H_0$ and $V$ to be
$$
H_0\ =\ -\ \f12\ \Delta_{X,\,Y}\, \Iten\ \text{ and } \ V\ =\
\left(\begin{array}{cc} \displaystyle \f{a+b}2 \ X^2\ +\ \f{a-b}2 \
Y^2 & \displaystyle b X Y
\\[3mm]
\displaystyle b X Y & \displaystyle \ \ \f{a-b}2 \ X^2\ +\ \f{a+b}2 \ Y^2\\
    \end{array}\right),
    $$
\\[3mm]
so that $\mb{H}_2\, =\, H_0\, +\, V$.  Note that for any $X,\ Y$,
the eigenvalues of $V$ are $\f{a+b}2\ (X^2+Y^2)$ and $\f{a-b}2\
(X^2+Y^2)$.  So for $f,\ g \in L^2(\R^2)$,
\begin{eqnarray*}
\lt\langle\ \displaystyle \lt(\begin{array}{c} f
\\
g
\end{array}
\rt),\, V\, \lt(\begin{array}{c} \displaystyle f
\\
\displaystyle g
\end{array}
\rt)\ \rt\rangle\ &\geq&\ \f{a-b}2\ \int\ (X^2+Y^2)\ \lt(\abs{f}^2 + \abs{g}^2\rt)\
dX\
dY,
\end{eqnarray*}
and $V$ is a positive operator.

\vskip .5cm

\begin{prop}\label{prop4}
Let $\Psi = \lt(\begin{array}{c} \dsp f \\
\dsp g
\end{array}\rt) \in {\cal H}$ be a solution of
$\ \mb{H}_2\Psi=E\Psi$, with $E>0$.  Then,\\ $f,\ g \in
C^\infty(\R^2)$, $\nabla f,\ \nabla g \in L^2(\mb{R}^2)$, and for
any $\gm>0$,
$$
\dsp f,\ g \in D(e^{\gm\langle x \rangle}), \hskip 1cm \nabla f,\
\nabla g \in D(e^{\gm\langle x \rangle}), \hskip 1cm \Delta f,\
\Delta g \in D(e^{\gm\langle x \rangle}),
$$
where $\dsp \langle x \rangle\ =\ \sqrt{1+X^2+Y^2}.$
\end{prop}

\noindent {\bf Proof:}

Let $V_{11}\, =\, \f{a+b}2\, X^2\, +\, \f{a-b}2\, Y^2,\hskip .5cm
V_{12}\, =\, V_{21}\, =\, bXY,$ \hskip .5cm and \hskip .5cm
$V_{22}\, =\, \f{a-b}2\, X^2\, +\, \f{a+b}2\, Y^2$.\\
Then, $f,\ g$ satisfy the following pair of equations:
\begin{eqnarray}
\lt(-\, \Delta\, +\, V_{11}\rt)\, f\, +\, V_{12}\, g\, &=&\, E\, f
\label{1steqn}
\\
\lt(-\, \Delta\, +\, V_{22}\rt)\, g\, +\, V_{21}\, f\, &=&\, E\, g
\label{2ndeqn}
\end{eqnarray}
To show that $f,\ g \in C^\infty(\R^2)$, we follow the proof of
Theorem IX.26 of \cite{RSV2}.  Let $\Omega$ be a bounded open set in
$\R^2$. Since $f,\ g \in L^2(\R^2) = W_0$ and the $V_{ij} \in
C^\infty,$ we have $V_{11}f,\ V_{21}f,\ V_{12}g,\ V_{22}g \in
W_0(\Omega)$. It follows from (\ref{1steqn}) and (\ref{2ndeqn}) that
$\Delta f,\ \Delta g \in W_0(\Omega)$.  Then by the Lemma on pg. 52
of \cite{RSV2}, $f,\ g \in W_2(\Omega)$. Repeating the argument we
get $f,\ g \in W_m(\Omega)\ \forall\ m \in \mb{Z}$.  It follows from
Sobolev's Lemma that $f,\ g \in C^\infty$ on $\Omega$. Since
$\Omega$ was arbitrary $f,\ g \in C^\infty(\R^2).$

We now show $\nabla f,\ \nabla g \in L^2$.  We know $\Psi \in
D(\mb{H}_2)$.  Let $D(-\Delta)$ and $Q(-\Delta)$ be the domain of
self-adjointness and quadratic form domain of $-\Delta$
respectively.  Then
$$
D(\mb{H}_2)\, \subset\, D(-\Delta)\, \oplus\, D(-\Delta)\, \subset\,
Q(-\Delta)\, \oplus\, Q(-\Delta)\, =\, \lt\{\, \Psi\, =\,
\lt(\begin{array}{c} f
\\ g
\end{array}\rt)\ :\ \nabla f,\ \nabla g \in L^2(\R^2)\, \rt\}.
$$

We now use the Combes-Thomas argument (see theorem XIII.39 of
\cite{RSV4}) to prove that\\ $f,\ g \in D(e^{\gm\abs{X}})$. The
argument can be repeated for $D(e^{\gm\abs{Y}})$, and since
$$
e^{\gm\lt\langle x \rt\rangle}\ \leq\ e^\gm\
e^{\gm\lt(\abs{X}+\abs{Y}\rt)}\ \leq\ e^\gm\
e^{2\gm\max\lt\{\abs{X},\, \abs{Y}\rt\}}\ \leq\ e^\gm\
\lt(e^{2\gm\abs{X}} + e^{2\gm\abs{Y}}\rt),
$$
we then have $f,\ g \in D(e^{\gm\lt\langle x \rt\rangle})$.

For $\al \in \mb{R}$, consider the unitary group $W(\al)\, =\,
e^{i\al X}\, \Iten$ and the operator\\$\mb{H}_2(\al)\ =\
    W(\al)\mb{H}_2W(\al)^{-1}$.  We have\\
$$
\mb{H}_2(\al)\ =\ \mb{H}_2\ +\ \f{\al^2}2\ \Iten\ +\ i\, \al\,
\f{\pa \phantom{t}}{\pa X}\ \Iten\ .
$$\\
The operator $i\, \f{\pa \phantom{t}}{\pa X}$ is form bounded with
respect to $-\Delta$ with relative bound zero.  Since $V$ is
positive, it follows that $\, i\, \f{\pa \phantom{t}}{\pa X}\,
\Iten\, $ is form bounded with respect to $\mb{H}_2$ with relative
bound zero.  So, $\mb{H}_2(\al)$ is an entire analytic family in the
sense of Kato on $D(\mb{H}_2)$. Furthermore, since $\mb{H}_2(\al)$
is unitarily equivalent to $\mb{H}_2$ for $\alpha \in \R$, we know
that $\mb{H}_2(\al)$ is self-adjoint and
$\sg(\mb{H}_2)=\sg(\mb{H}_2(\al))$ for $\al \in \mb{R}$. Since
$\mb{H}_2$ has purely discrete spectrum, we know $\mb{H}_2(\al)$ has
purely discrete spectrum for $\al \in \mb{R}$. It follows from lemma
\ref{lemma3} that $\mb{H}_2(\al)$ has purely discrete spectrum
$\forall\ \al \in \mb{C}$.  Since $\mb{H}_2(\al)$ is an entire
analytic family in the sense of Kato, the eigenvalues are analytic
on $\mb{C}$ except possibly at isolated crossings \cite{RSV4}.
$W(\al)$ unitary implies that the eigenvalues are constant in a
neighborhood of the real axis and thus crossings will not be an
issue. Therefore, the eigenvalues are entire functions and constant
in $\al$.

Let $P(\al)$ be the projection onto the eigenspace corresponding to
the eigenvalue E of $\mb{H}_2(\al)$.  Then $P(\al)$ is entire in
$\al$ and has the form
$$
P(\al)\ =\ \f{-1}{2\pi i}\ \int_{\abs{\lam\, -\, E}\, =\, \ep}\
\lt(\, \mb{H}_2(\al)\ -\ \lam\, \rt)^{-1}\ d\lam\ .
$$
If $\al,\ \al_0 \in \mb{R}$,
\begin{eqnarray*}
P(\al+\al_0)\ &=&\ \f{-1}{2\pi i}\ \int_{\abs{\lam\, -\, E}\, =\,
\ep}\ \lt(\, \mb{H}_2(\al+\al_0)\ -\ \lam\, \rt)^{-1}\ d\lam
\\[3mm]
&=&\ \f{-1}{2\pi i}\ \int_{\abs{\lam\, -\, E}\, =\, \ep}\ W(\al_0)\
\lt(\, \mb{H}_2(\al)\ -\ \lam\, \rt)^{-1}\ W(\al_0)^{-1} d\lam
\\[3mm]
&=&\ W(\al_0)P(\al)W(\al_0)^{-1}\ .
\end{eqnarray*}
For $\al_0 \in \mb{R}$, the operator valued function $\, f(\al)\,
=\, W(\al_0)P(\al)W(\al_0)^{-1}\, -\, P(\al+\al_0)\, $ is entire in
$\al$. Since it vanishes $\forall\ \al \in \mb{R},$ it is zero
$\forall\ \al \in \mb{C}$.  So $P(\al+\al_0)\ =\
W(\al_0)P(\al)W(\al_0)^{-1}$, for $\al_0 \in \mb{R} \text{ and } \al
\in \mb{C}$.  The hypotheses of O'Connors lemma are satisfied
\cite{RSV4}.  So,
for the eigenvector $\Psi = \lt(\begin{array}{c} \dsp f \\
\dsp g
\end{array}\rt),$ we know $\Psi(\al)=W(\al)\Psi$ has an analytic continuation to all of
$\mb{C}$.  Therefore $f,\ g \in D(e^{\gm\abs{X}})$ for any $\gm>0$.

From this it now follows that $\Delta f,\ \Delta g \in
D(e^{\gm\lt\langle x \rt\rangle})$, for any $\gm>0$.  To see this,
consider
\begin{eqnarray*}
\lt\lvert\lt\lvert\ \lt(\begin{array}{c} e^{\gm\lt\langle x
\rt\rangle}\ \Delta\ f
\\[3mm]
e^{\gm\lt\langle x \rt\rangle}\ \Delta\ g \end{array}\rt)\
\rt\rvert\rt\rvert^2\ &=&\ 4\ \lt\lvert\lt\lvert\
\lt(\begin{array}{c} e^{\gm\lt\langle x \rt\rangle}\ \lt[\
\lt(V_{11}\ -\ E\ \rt)\ f\ +\ V_{12}\ g\ \rt]
\\[3mm]
e^{\gm\lt\langle x \rt\rangle}\ \lt[\ V_{21}\ f\ +\ \lt(V_{22}\ -\
E\ \rt)\ g\ \rt] \end{array}\rt)\ \rt\rvert\rt\rvert^2
\\[3mm]
&=&\ 4\ \lt(\ \lt\lvert\lt\lvert\ e^{\gm\lt\langle x \rt\rangle}\
\lt(V_{11}\ -\ E\ \rt)\ f\ +\ e^{\gm\lt\langle x \rt\rangle}\
V_{12}\ g\ \rt\rvert\rt\rvert_2^2\rt.
\\[3mm]
&& \hskip 3cm +\ \lt.\lt\lvert\lt\lvert\ e^{\gm\lt\langle x
\rt\rangle}\ V_{21}\ f\ +\ e^{\gm\lt\langle x \rt\rangle}\
\lt(V_{22}\ -\ E\ \rt)\ g\ \rt\rvert\rt\rvert_2^2\ \rt)
\end{eqnarray*}
Let $\beta>0$.  Then,
\begin{eqnarray*}
\int\ e^{2\gm\lt\langle x \rt\rangle}\ \abs{\ V_{21}\ f\ }^2\ dX\,
dY\ \leq\ \lt\lvert\lt\lvert\ V_{21}^2\ e^{-2\beta\lt\langle x
\rt\rangle\ }\rt\rvert\rt\rvert_\infty\ \lt\lvert\lt\lvert\
e^{2\lt\langle x \rt\rangle (\gm+\beta)}\ f\ \rt\rvert\rt\rvert_2^2\
<\ \infty
\end{eqnarray*}
So, $e^{\gm\lt\langle x \rt\rangle} V_{21}f \in L^2(\mb{R}^2)\,$ and
by similar arguments $\,e^{\gm\lt\langle x \rt\rangle}(V_{11}-E)f,\
\, e^{\gm\lt\langle x \rt\rangle }V_{12}g,\ \,e^{\gm\lt\langle x
\rt\rangle }(V_{22}-E)g$\\ $\in L^2(\mb{R}^2).$ Hence,
$$
\lt\lvert\lt\lvert\ \lt(\begin{array}{c} e^{\gm\lt\langle x
\rt\rangle}\ \Delta\ f
\\
e^{\gm\lt\langle x \rt\rangle}\ \Delta\ g \end{array}\rt)\
\rt\rvert\rt\rvert^2\ <\ \infty
$$
and $\Delta f,\ \Delta g \in D(e^{\gm\lt\langle x \rt\rangle})$.

For $\nabla f,\, \nabla g \in D(e^{\gm\lt\langle x \rt\rangle})$, we
apply Lemma 3.4 of \cite{Hag/Joye1}:\\[3mm]
Let $p \in C^1(\R^N)$ and suppose for some $C<\infty$, $\dsp
\lt\lvert \f{\nabla p(x)}{p(x)}\rt\rvert \leq 2C\ \ \forall\ x \in
\R^N.$  If\\[3mm]
$\dsp \int_{\R^N}\ (\abs{\, f\, }\, ^2\, +\, \abs{\, \Delta f\, }\,
^2)\, p\ dx < \infty$, then
\begin{eqnarray*}
&& \hskip -.7cm \lt(\, \int_{\R^N} \abs{\, \nabla f\, }\, ^2\, p\
dx\, \rt)^{1/2}\ \leq\ C\lt(\, \int_{\R^N} \abs{\, f\, }\, ^2\, p\
dx\rt)^{1/2}
\\[3mm]
&& \hskip 1.8cm +\ \lt[\, \lt(\, \int_{\R^N} \abs{\, f\, }\, ^2\, p\
dx\rt)^{1/2}\ \lt(\, \int_{\R^N} \abs{\, \Delta f\, }\, ^2\, p\
dx\rt)^{1/2}\ +\ C^2\int_{\R^N} \abs{\, f\, }\, ^2\, p\ dx\,
\rt]^{1/2}.
\end{eqnarray*}
We let $p\,(X,\,Y)\ =\ e^{2\gm\lt\langle x \rt\rangle}$.  Then $\dsp
\lt\lvert \f{\nabla p\,(X,\,Y)}{p\, (X,\,Y)}\rt\rvert \leq 2\gm\ \
\forall\ (X,\,Y) \in \R^2$.  We have already shown that for $f$ and
$g$, the right hand side in the lemma is finite for any $\gm>0$. So,
$\nabla f,\, \nabla g \in D(e^{\gm\lt\langle x \rt\rangle})$ for any
$\gm>0$.  $\square$

\vskip .5cm

\begin{cor}\label{cor5}
Let $R(\lam)=(\mb{H}_2-\lam)^{-1}$ for $\lam \in \rho(\mb{H}_2)$.
Let $P_E$ be the projection onto the eigenspace associated with $E$
and define $r(E)=[(\mb{H}_2-E)|_{Ran(I-P_E)}]^{-1}$, the reduced
resolvent at $E$. Then, $(e^{\gm\langle x
\rangle}\,\Iten)\,R(\lam)\,(e^{-\gm\langle x\rangle}\,\Iten)$ and
$(e^{\gm\langle x \rangle}\,\Iten)\,r(E)\,(e^{-\gm\langle
x\rangle}\Iten)$ are bounded on ${\cal H}$ for any $\gm>0$.  In
particular, if $\ \Psi \in D(e^{\gm \langle x \rangle}\,\Iten)$,
then $R(\lam)\, \Psi,\ r(E)\, \Psi \in D(e^{\gm \langle x
\rangle}\,\Iten)$.
\end{cor}

\noindent {\bf Note:}  See \cite{thesis} for a proof.

\vskip .5cm

\noindent We need the following lemma for proposition \ref{prop7}

\begin{lem}\label{lem6}
For fixed $t \in \R^n$, there exist $K>\tilde{K}>0$ and $S(t)>0$,
such that if $p \in \R^n$ satisfies $\sum_{j=1}^n p_j^2 \geq
S(t)^2$, then
$$
\tilde{K}\sum_{j=1}^n p_j^2 \leq \lt\lvert
\sum_{j=1}^n(p_j+it_j)^2\rt\rvert \leq K \sum_{j=1}^n p_j^2\ .
$$
Furthermore, $S(t)$ is uniformly bounded for $t$ in compact subsets
of $\R^n$.
\end{lem}

\noindent {\bf Proof:}

Let $\norm{t}\ =\ \lt[\sum_{j=1}^n\, t_j^{\,2}\rt]^{1/2},\
\norm{p\,}\ =\ \lt[\sum_{j=1}^n\, p_j^{\,2}\rt]^{1/2}$.  We prove
the Lemma with\\ $S(t)=1+4\norm{t},\ \tilde{K}=7/16$, and $K=17/16$.
We first show that for this choice of $S(t)$, $K=17/16$:
\begin{eqnarray*}
\lt\lvert\sum_{j=1}^n\,(p_j+it_j)^2\rt\rvert &\leq&
\sum_{j=1}^n\,\rvert p_j+it_j\lvert^2
\\[3mm]
&=& \sum_{j=1}^n\,(\,p_j^{\,2}+t_j^{\,2}\,)
\\[3mm]
&=& \norm{p\,}^2\,\lt[\, 1\, +\, \f{\norm{t}^2}{\norm{p\,}^2}\,\rt]
\\[3mm]
&\leq& \norm{p\,}^2\,\lt[\, 1\, +\,
\f{\norm{t}^2}{(1+4\norm{t})^2}\,\rt]
\\[3mm]
&\leq& \norm{p\,}^2\,\lt(\f{17}{16}\rt).
\end{eqnarray*}
In particular notice that this argument also shows
\begin{equation}
\norm{\,p\,}^{\,2}\, +\, \norm{\,t\,}^{\,2}\, \leq\, \f{17}{16}\,
\norm{\,p\,}^{\,2}. \label{upbound}
\end{equation}

Now we show that for this choice of $S(t)$, $\tilde{K}=7/16$:
\begin{eqnarray}
\lt\lvert\sum_{j=1}^n\,(p_j+it_j)^2\rt\rvert &=& \lt[\, \lt(\,
\sum_{j=1}^n\, (p_j^{\,2}-t_j^{\,2})\,\rt)^2\, +\, \lt(\,
2\sum_{j=1}^n\, p_j\, t_j\rt)^2 \,\rt]^{1/2} \notag
\\[3mm]
&\geq& \lt\lvert\ \ \ \lt\lvert\, \sum_{j=1}^n\,
(p_j^{\,2}-t_j^{\,2}) \,\rt\rvert\, -\, 2\lt\lvert\, \sum_{j=1}^n\,
p_j\, t_j \,\rt\rvert\ \ \ \rt\rvert \notag
\\[3mm]
&\geq& \sum_{j=1}^n\, (p_j^{\,2}-t_j^{\,2})\, -\, 2\,\sum_{j=1}^n\,
\abs{p_j\,}\,\abs{t_j\,} \notag
\\[3mm]
&\geq& \sum_{j=1}^n\, (p_j^{\,2}-t_j^{\,2})\, -\,
2\,\norm{p\,}\,\norm{t\,} \notag
\\[3mm]
&=& \norm{p\,}^2\,\lt(\, 1\, -\, \f{\norm{t\,}^2}{\norm{p\,}^2}\,
-\, 2\, \f{\norm{t\,}}{\norm{p\,}} \rt) \notag
\end{eqnarray}
\begin{eqnarray}
&\geq& \norm{p\,}^2\,\lt(\, 1\, -\,
\f{\norm{t\,}^2}{(1+4\norm{t})^2}\, -\, \f{2\,
\norm{t\,}}{1+4\norm{t}} \rt) \notag
\\[3mm]
&\geq& \norm{p\,}^2\,\lt(\f7{16}\rt). \label{lowbound}
\end{eqnarray}  $\square$

\vskip .5cm

\begin{prop}\label{prop7}
Let $\Psi = \lt(\begin{array}{c} \dsp f \\
\dsp g
\end{array}\rt)\, \in {\cal H}$ be a solution of
$\ \mb{H}_2\Psi=E\Psi$, with $E>0$.  Then, for any $\gm>0$, and any
$\al \in \mb{N}^2,\ D^{\al}f,\ D^{\al}g \in D(e^{\gm\langle
x\rangle})$, where $D^{\al}\, = \, \pa_{X}^{\,\al_1}\,
\pa_{Y}^{\,\al_2}$.
\end{prop}

\noindent {\bf Proof:}

We use a Paley-Wiener Theorem, Theorem IX.13 of \cite{RSV2}:

\noindent Let $\phi \in L^2(\R^n)$.  Then $e^{\gm\abs{x\,}}\,\phi \in
L^2(\R^n)$ for all $\gm <\gm'$ if and only if $\hat{\phi}$ has an
analytic continuation to the set $\{p\, :\, \abs{\mbox{Im}\,p\,}<
\gm' \}$ with the property that for each $t \in \R^n$ with
$\abs{t}<\gm',\ \hat{\phi}(\cdot + it) \in L^2(\R^n)$, and for any
$\gm <\gm',\ \sup_{\abs{t}\leq \gm}\,
\norm{\hat{\phi}(\cdot+it)\,}_2\, <\, \infty$.

If a function $\hat{\phi}$ satisfies the conditions in this theorem
we will say that $\hat{\phi}$ is ``P-W''.  Let $p_j\, =\, -i\,\pa_{x_j}\, $. We present the proof for general
$n$.  In our case we have $n=2$ with $x_1\, =\, X$ and $x_2\, =\,
Y$.

Proposition 4 shows that $\hat{f}$ and $\hat{g}$ are P-W for any
$\gm'>0$. In particular we know that $\hat{f},\ \hat{g}$ are
analytic everywhere. So the analyticity condition will be a
non-issue in the course of the proof.  $\widehat{\nabla f},\
\widehat{\nabla g},\ \widehat{\Delta f},\ \widehat{\Delta g}$ are
also P-W for any $\gm'>0$.  So $\ p\, \mapsto\, p_j\,\hat{f}(p),\ \
p\, \mapsto\, p_j\,\hat{g}(p),$ $p\, \mapsto\, \sum_{j=1}^n\,
p_j^{\,2}\,\hat{f}(p),\ $ and $\ p\, \mapsto\, \sum_{j=1}^n\,
p_j^{\,2}\,\hat{g}(p)\ $ are P-W for all $\gm'>0$.

Let $S(t)=1+4\norm{t\,}$ and $B_S$ be a ball of radius $S$ centered
at the origin.  Since $\sum_{j=1}^n\, p_j^{\,2}\,\hat{f}(p)\ $ is
P-W, with (\ref{lowbound}) we have
\begin{eqnarray}
\int_{\R^n\setminus B_{S(t)}}\ \norm{p\,}^4\,
\abs{\hat{f}(p+it)\,}^2\, dp\, &\leq& \lt(\f{16}{7}\rt)^2\,
\int_{\R^n\setminus B_{S(t)}}\
\lt\lvert\,\sum_{j=1}^n\,(p_j+it_j)^2\rt\rvert^{\,2}\abs{\hat{f}(p+it)\,}^2\,
dp \notag
\\[3mm]
&<& \infty \label{estimate1}
\end{eqnarray}
uniformly for $t$ in compact subsets of $\R^n$.  We only show
results involving $f$.  The same results hold with $f$ replaced by
$g$.

Note that since $S(t)$ and $\norm{\hat{f}(\cdot+it)\,}_2$ are
uniformly bounded for $t$ in compact subsets of $\R^n$, we only need
to prove estimates for $\norm{p\,} \geq S(t)$.  All of the integral
estimates that follow hold uniformly for $t$ in compact subsets of
$\R^n$.  From (\ref{upbound}) and (\ref{estimate1}) we have
\begin{eqnarray*}
&& \hskip -2cm \int_{\R^n\setminus B_{S(t)}}\
\abs{p_j+it_j\,}^{\,2}\,\abs{p_k+it_k\,}^{\,2}\,
\abs{\hat{f}(p+it)\,}^2\, dp
\\[3mm]
&& \leq\ \int_{\R^n\setminus B_{S(t)}}\ \lt(\, \norm{p\,}^{\,2}\,
+\, \norm{t\,}^{\,2}\,\rt)^2 \abs{\hat{f}(p+it)\,}^2\, dp
\\[3mm]
&& \leq\ \lt(\f{17}{16}\rt)^2\,\int_{\R^n\setminus B_{S(t)}}\
\norm{p\,}^4\, \abs{\hat{f}(p+it)\,}^2\, dp
\\[3mm]
&&< \infty\ .
\end{eqnarray*}
It follows that $\pa_{x_j}\, \pa_{x_k}\,f \in D(e^{\gm \langle x
\rangle})$ for any $\gm>0$.  Again the same will hold for $g$.

We now start an induction on the length $\abs{\al}$ in $D^{\al}f$
and $D^{\al}g$.  Assume that\\
$D^{\beta}f,\ D^{\beta}g \in D(e^{\gm \langle x \rangle})$ for any
$\gm>0$ and any $\abs{\beta}\leq m-1$.  It suffices to prove that
$D^{\al}f \in D(e^{\gm \langle x \rangle})$ for any $\gm>0$ and any
$\abs{\al\,}=m$.

Following the notation in the proof of Proposition 4, the eigenvalue
equation gives us
\begin{eqnarray*}
\Delta\, f\, &=&\ V_{12}\,g\, +\, (V_{11}\, -\, E)\, f\ ,
\label{eval1}
\\
\Delta\, g\, &=&\ V_{21}\,f\, +\, (V_{22}\, -\, E)\, g\ .
\label{eval2}
\end{eqnarray*}
where $V_{11},\ V_{12}=V_{21},$ and $V_{22}$ are polynomials in
$x_j$. Let $\abs{\al'\,}=m-2$.  Since the $V_{ij}$ are polynomial,
our induction hypothesis gives us $D^{\al'}\Delta\,f \in D(e^{\gm
\langle x \rangle})$ for any $\gm>0$.  It follows that for $j_k \in
\{1,2,\cdots,n\}$
\begin{eqnarray*}
\int_{\R^n\setminus B_{S(t)}}\, \abs{p_{j_1}+it_{j_1}\,}^{\,2}\,
\abs{p_{j_2}+it_{j_2}\,}^{\,2}\, \cdots\,
\abs{p_{j_{m-2}}+it_{j_{m-2}}\,}^{\,2}\,
\lt\lvert\,\sum_{j=1}^n\,(p_j+it_j)^2\rt\rvert^{\,2}
\abs{\hat{f}(p+it)\,}^2\, dp\, <\, \infty\ ,
\end{eqnarray*}
and from (\ref{lowbound}) we have
\begin{eqnarray*}
\int_{\R^n\setminus B_{S(t)}}\ \abs{p_{j_1}+it_{j_1}\,}^{\,2}\,
\abs{p_{j_2}+it_{j_2}\,}^{\,2}\, \cdots\,
\abs{p_{j_{m-2}}+it_{j_{m-2}}\,}^{\,2}\, \norm{p\,}^4\,
\abs{\hat{f}(p+it)\,}^2\, dp\, <\, \infty\ .
\end{eqnarray*}
Since the $j_k$ are arbitrary, we have
\begin{eqnarray*}
&&\hskip -.7cm \infty >
\\[3mm]
&&\hskip -.7cm \sum_{j_1,j_2,\cdots,j_{m-2}=1}^n\
\int_{\R^n\setminus B_{S(t)}}\ \abs{p_{j_1}+it_{j_1}\,}^{\,2}\,
\abs{p_{j_2}+it_{j_2}\,}^{\,2}\, \cdots\,
\abs{p_{j_{m-2}}+it_{j_{m-2}}\,}^{\,2}\, \norm{p\,}^4\,
\abs{\hat{f}(p+it)\,}^2\, dp
\\[3mm]
&&\hskip -.7cm =\, \sum_{j_1,j_2,\cdots,j_{m-2}=1}^n\
\int_{\R^n\setminus B_{S(t)}}\ (p_{j_1}^{\,2}+t_{j_1}^{\,2})\,
(p_{j_2}^{\,2}+t_{j_2}^{\,2})\, \cdots\,
(p_{j_{m-2}}^{\,2}+t_{j_{m-2}}^{\,2})\, \norm{p\,}^4\,
\abs{\hat{f}(p+it)\,}^2\, dp
\end{eqnarray*}
\begin{eqnarray*}
&& \hskip -4.7cm =\, \int_{\R^n\setminus B_{S(t)}}\
(\norm{p\,}^{\,2}\, +\, \norm{t}^{\,2})^{m-2}\, \norm{p\,}^4\,
\abs{\hat{f}(p+it)\,}^2\, dp
\\[3mm]
&&\hskip -4.7cm \geq\, \int_{\R^n\setminus B_{S(t)}}\
\norm{p\,}^{\,2(m-2)}\, \norm{p\,}^4\, \abs{\hat{f}(p+it)\,}^2\, dp
\\[3mm]
&& \hskip -4.7cm =\, \int_{\R^n\setminus B_{S(t)}}\
\norm{p\,}^{\,2m}\, \abs{\hat{f}(p+it)\,}^2\, dp
\end{eqnarray*}
Then using (\ref{upbound}), we have for any $j_k \in
\{1,2,\cdots,n\}$
\begin{eqnarray*}
&&\int_{\R^n\setminus B_{S(t)}}\ \abs{p_{j_1}+it_{j_1}\,}^{\,2}\,
\abs{p_{j_2}+it_{j_2}\,}^{\,2}\, \cdots\,
\abs{p_{j_{m}}+it_{j_{m}}\,}^{\,2}\, \abs{\hat{f}(p+it)\,}^2\, dp
\\[3mm]
&\leq& \int_{\R^n\setminus B_{S(t)}}\ (\norm{p\,}^{\,2}\, +\,
\norm{t}^{\,2})^{m}\, \abs{\hat{f}(p+it)\,}^2\, dp
\\[3mm]
&\leq& \lt(\f{17}{16}\rt)^m\, \int_{\R^n\setminus B_{S(t)}}\
\norm{p\,}^{\,2m}\, \abs{\hat{f}(p+it)\,}^2\, dp
\\[3mm]
&<& \infty\ .
\end{eqnarray*}
So, for arbitrary $j_k \in \{1,2,\cdots n\},\ p\, \mapsto\,
p_{j_1}\, p_{j_2}\, \cdots\, p_{j_m}\, \hat{f}(p)$ is P-W and it
follows that $D^{\al}f \in D(e^{\gm \langle x \rangle})$ for any
$\gm>0$ and any $\abs{\al}=m$.  The same argument will work with $f$
replaced by $g$ and the proposition is proved.
 $\square$

\vskip .5cm

\begin{lem}\label{lem8}
Let $\Psi\, =\, \lt(\begin{array}{c} f
\\ g
\end{array}\rt)
$, $R(\lam)=(\mb{H}_2-\lam)^{-1}$ for $\lam \in \rho(\mb{H}_2)$, and
$r(E)=(\mb{H}_2-E)_r^{-1}$ be the reduced resolvent at $E$. If $\
f,\ g \in C^\infty$ and $\ (D^{\al}\, \Iten)\, \Psi \in D(e^{\gm
\langle x \rangle}\, \Iten)$, for all $\ \al \in \mb{N}^2$ and any
$\ \gm >0$, then $\ (D^{\al}\, \Iten)\, R(\lam)\Psi,\ (D^{\al}\,
\Iten)\, r(E)\Psi \in D(e^{\gm \langle x \rangle}\, \Iten)$, for all
$\ \al \in \mb{N}^2$ and any $\gm > 0$.
\end{lem}

\noindent {\bf Proof:}

First note that for any $\gm_1>\gm_2>0$ and $j,k=0,1,2,\cdots$,
there exists $M>0$ such that
$$
\norm{e^{\gm_2 \langle x \rangle}\, X^j\,Y^k\, \phi \,}\, \leq\,
M\,\norm{\phi\,}\, +\, \norm{e^{\gm_1 \langle x \rangle}\, \phi \,}.
$$
This relative bound implies that if $\phi \in D(e^{\gm \langle x
\rangle})$ for all $\gm
> 0$, then $X^j\,Y^k\,\phi \in D(e^{\gm \langle x \rangle})$ for all
$\gm
> 0$, and arbitrary $j,k\, =\, 0,1,2,\cdots\ $.

By an argument similar to the one by which we obtained $f,\ g \in
C^\infty(\R^2)$ in the proof of Proposition \ref{prop4}, $R(\lam)$
and $r(E)$ map functions from $C^\infty(\R^2) \oplus C^\infty(\R^2)$
to\\ $C^\infty(\R^2) \oplus C^\infty(\R^2)$.

The following identity holds as long as the terms on the right hand
side are in\\ $L^2(\R^2)\, \oplus\, \L^2(\R^2)$:
\begin{eqnarray}
(\pa_X \Iten)\, R(\lam)\, \Phi\, =\, R(\lam)\, (\pa_X \Iten)\,
\Phi\, -\, R(\lam)\,[(\pa_X \Iten)(V)]\,R(\lam)\, \Phi,
\label{Resolvent Commutator}
\end{eqnarray}
where $\ [(\pa_X \Iten)(V)]\, =\, \lt(\begin{array}{cc} (a+b)\, X &
b\, Y
\\[2mm]
b\,Y & (a-b)\,X \end{array} \rt)$.
To see this, let $R(\lam)\,\Phi\, =\, \lt(\begin{array}{c} \psi_1 \\
\psi_2 \end{array}\rt)$ and we compute $[\,\pa_X \Iten,\
R(\lam)\,]$:
\begin{eqnarray*}
&&\hskip -.5cm \lt\{(\pa_X \Iten)\, R(\lam)\, -\, R(\lam)\,(\pa_X
\Iten)\rt\}\,\Phi
\\[4mm]
&&\hskip -.5cm = (\pa_X \Iten)\, \lt(\begin{array}{c} \psi_1 \\
\psi_2 \end{array}\rt)\, -\, R(\lam)\, (\pa_X \Iten)\, (\mb{H}_2\,
-\, \lam)\, \lt(\begin{array}{c} \psi_1 \\
\psi_2 \end{array}\rt)\,
\\[4mm]
&&\hskip -.5cm = \lt(\begin{array}{c} \pa_X\,\psi_1 \\
\pa_X\,\psi_2 \end{array}\rt)\ -\ R(\lam)\,\lt(\begin{array}{cc} \pa_X & 0 \\
0 & \pa_X \end{array}\rt)\, \lt[\ \lt(\begin{array}{c} (-\,\f12\,\Delta\,-\,\lam)\, \psi_1 \\
(-\,\f12\,\Delta\,-\,\lam)\, \psi_2 \end{array}\rt)\, +\,
\lt(\begin{array}{c} V_{11}\,\psi_1\,+\,V_{12}\,\psi_2 \\
V_{21}\,\psi_1\,+\,V_{22}\,\psi_2
\end{array}\rt)\ \rt]\,
\\[4mm]
&&\hskip -.5cm = \lt(\begin{array}{c} \pa_X\,\psi_1 \\
\pa_X\,\psi_2 \end{array}\rt)\ -\ R(\lam)\, \lt[\ \lt(\begin{array}{c} (-\,\f12\,\Delta\,-\,\lam)\, \pa_X\,\psi_1 \\
(-\,\f12\,\Delta\,-\,\lam)\, \pa_X\,\psi_2 \end{array}\rt)\, +\,
\lt(\begin{array}{c} \pa_X(V_{11})\,\psi_1\,+\,\pa_X(V_{12})\,\psi_2 \\
\pa_X(V_{21})\,\psi_1\,+\,\pa_X(V_{22})\,\psi_2
\end{array}\rt) \rt.
\\[4mm]
&& \hskip 7cm \lt.+\, \lt(\begin{array}{c} V_{11}\,\pa_X(\psi_1)\,+\,V_{12}\,\pa_X(\psi_2) \\
V_{21}\,\pa_X(\psi_1)\,+\,V_{22}\,\pa_X(\psi_2)
\end{array}\rt)\ \rt]\,
\\[4mm]
&&\hskip -.5cm = \lt(\begin{array}{c} \pa_X\,\psi_1 \\
\pa_X\,\psi_2 \end{array}\rt)\ -\ R(\lam)\, \lt[\ (\mb{H}_2-\lam)\, \lt(\begin{array}{c} \pa_X\,\psi_1 \\
\pa_X\,\psi_2 \end{array}\rt)\, +\, [(\pa_X \Iten)(V)]\, \lt(\begin{array}{c} \psi_1 \\
\psi_2 \end{array}\rt)\, \rt]
\\[4mm]
&&\hskip -.5cm =\ -\ R(\lam)\, [(\pa_X \Iten)(V)]\, \lt(\begin{array}{c} \psi_1 \\
\psi_2 \end{array}\rt)
\\[4mm]
&&\hskip -.5cm =\ -\ R(\lam)\, [(\pa_X \Iten)(V)]\, R(\lam)\, \Phi.
\end{eqnarray*}
Clearly (\ref{Resolvent Commutator}) holds with $X$ replaced by $Y$.

From the hypotheses on $\Psi$ and Corollary \ref{cor5}, we know that
for all $\gm>0$, $R(\lam)\, (\pa_X \Iten)\, \Psi \in D(e^{\gm
\langle x \rangle}\Iten)\, \subset\, L^2(\R^2)\, \oplus\,
L^2(\R^2)$.  From Corollary \ref{cor5}  and the note above, we know
that $R(\lam)\,[(\pa_X \Iten)(V)]\,R(\lam)\, \Psi\, \in D(e^{\gm
\langle x \rangle}\Iten)\, \subset\, L^2(\R^2)\, \oplus\, L^2(\R^2)$
for all $\gm>0$.  From this we see that (\ref{Resolvent Commutator})
holds when applied to $\Psi$ and therefore $\, (\pa_X \Iten)\,
R(\lam)\, \Phi \in D(e^{\gm \langle x \rangle}\Iten)$ for all
$\gm>0$. Similarly, $\, (\pa_Y \Iten)\, R(\lam)\, \Phi \in D(e^{\gm
\langle x \rangle}\Iten)$ for all $\gm>0$.

By applying (\ref{Resolvent Commutator}) repeatedly, we see that
$(D^{\al} \Iten)\, R(\lam)\, \Psi$ is a linear combination of terms
of the form
$$
R(\lam)\, [(D^{\al_1} \Iten)(V)]\, R(\lam)\, [(D^{\al_2}
\Iten)(V)]\, \cdots\, R(\lam)\, [(D^{\al_{m-1}} \Iten)(V)]\,
R(\lam)\,(D^{\al_{m}} \Iten)\,\Psi,
$$
where $\sum_{j=1}^m\, \abs{\al_j\,}\, =\, \abs{\al\,}$.  Since the
$[(D^{\al_j}\Iten)(V)]$ are matrices with polynomial entries, we use
Corollary \ref{cor5} and the note above to obtain $(D^{\al} \Iten)\,
R(\lam)\, \Psi \in D(e^{\gm \langle x \rangle}\Iten)$ for all $\al
\in \mb{N}^{\,2}$ and $\gm > 0$.  The conclusion involving $(D^{\al}
\Iten)\, r(E)\, \Psi$ follows by writing the reduced resolvent in the form (see Theorem XII.5 in
\cite{RSV4})
\begin{eqnarray}
r(E)\, =\, \f1{2\pi i}\, \int_{\abs{\lam-E}\, =\, \Gamma\, >\, 0}\,
R(\lam)\, \f1{\lam-E}\,d\lam\ . \notag
\end{eqnarray}

$\square$

\vskip .5cm

\begin{thm}\label{thm9}
For $k
\geq 2$, let $\Psi^{(k-2)}=\lt(\begin{array}{c} f^{(k-2)} \\
g^{(k-2)}
\end{array}\rt)$, and $\psi_{\perp}^{(k)}$ be determined by the
perturbation formulas of chapter \ref{QuasimodeConstruction}. Then,
$f^{(k-2)},\ g^{(k-2)},\ \psi_{\perp}^{(k)}
 \in C^\infty(\R^2)$,
$f^{(k-2)},\ g^{(k-2)},\ \norm{\psi_{\perp}^{(k)}}_{el}
 \in L^2(\R^2)$ and $f^{(k-2)},\ g^{(k-2)},\ \norm{\psi_{\perp}^{(k)}}_{el} \in D(e^{\gm \langle x
\rangle}\,)$, for any $\gm >0$.  In addition, $D^{\al}\,f^{(k-2)}$,
$D^{\al}\, g^{(k-2)}$, $\norm{D^{\al}\,\psi_{\perp}^{(k)}}_{el} \in
D(e^{\gm \langle x \rangle})\,$ for all $\ \al \in \mb{N}^2$ and any
$\gm > 0$.
\end{thm}

\noindent {\bf Proof:}

We refer to a function in $D(e^{\gm \langle x \rangle})\,$ (or
$D(e^{\gm \langle x \rangle} \Iten)\,$) for any $\gm > 0$, as
exponentially decaying with arbitrary $\gm$. We first note that from
the proof of lemma \ref{lem8}, multiplication by polynomials in $X$
and $Y$ preserves exponential decay with arbitrary $\gm$.

Since $\Psi^{(0)}=\lt(\begin{array}{c} f^{(0)} \\
g^{(0)}
\end{array}\rt)$ is determined at second order in $\ep$ as an eigenfunction
of $\mb{H}_2$, we already know from propositions \ref{prop4} and
\ref{prop7} that $\Psi^{(0)}$ satisfies the conclusion.

The $f^{(1)}$ and $g^{(1)}$ given by equation (\ref{f1g1}) are
determined by $\mb{H}_3$ followed by a projection $Q_{\perp}$, and
reduced resolvent $\mb{H}_2-E^{(2)}$, acting on $\Psi^{(0)}$. By
corollary \ref{cor5} we know that the reduced resolvent preserves
exponential decay with arbitrary $\gm$.  The projection $Q_{\perp}$
was the projection in $\mH$ onto the subspace perpendicular to the
eigenspace of the eigenvalue $E^{(2)}$ of $\mb{H}_2$.  From
proposition \ref{prop4}, we know that the eigenvectors of $\mb{H}_2$
have exponential decay with arbitrary $\gm$, and so it follows that
$Q_{\perp}$ will preserve exponential decay with arbitrary $\gm$.
 Since the matrix entries of $\mb{H}_3$ only contain polynomials and
derivatives in $X$ and $Y$, we know from lemma \ref{lem8} that
$(\mb{H}_3\, \Psi^{(0)})$ will have exponential decay with arbitrary
$\gm$.  It follows that $\Psi^{(1)}$ will have exponential decay
with arbitrary $\gm$.  By a similar argument, $\Psi^{(1)} \in
C^\infty \oplus C^\infty$.  From the definition of $Q_{\perp}$ along
with proposition \ref{prop7}, we see that all of the derivatives of
$Q_{\perp}\,\mb{H}_3\,\Psi^{(0)}$ are exponentially decaying with
arbitrary $\gm$.  It then follows from lemma \ref{lem8} that all of
the derivatives of $\Psi^{(1)}$ are exponentially decaying with
arbitrary $\gm$.

Recall that
$\psi_{\perp}^{(0)}=\psi_{\perp}^{(1)}=\psi_{\perp}^{(2)}=0$.  From
equation (\ref{psiperp3}) we know that
\begin{eqnarray*}
&&\norm{\psi_{\perp}^{(3)}}_{el} \leq\\[3mm] && \hskip -.2cm \lt\lvert\lt\lvert\
\lt[\lt(\,h\,P_{\perp}\rt)^{(0)}\rt]_r^{-1}\ P_{\perp}^{(0)}
\lt(\f{\pa \Psi_1}{\pa x}\rt)^{(0)}\ \rt\rvert\rt\rvert_{el}\
\lt\lvert\,\f{\partial f^{(0)}}{\partial X}\rt\rvert\notag
\ +\ \lt\lvert\lt\lvert\
\lt[\lt(\,h\,P_{\perp}\rt)^{(0)}\rt]_r^{-1}\ P_{\perp}^{(0)}
\lt(\f{\pa \Psi_1}{\pa y}\rt)^{(0)}\ \rt\rvert\rt\rvert_{el}\
\lt\lvert\,\f{\partial f^{(0)}}{\partial Y}\rt\rvert \notag
\\[5mm]
&& \hskip -.7cm +\ \lt\lvert\lt\lvert\
\lt[\lt(\,h\,P_{\perp}\rt)^{(0)}\rt]_r^{-1}\ P_{\perp}^{(0)}
\lt(\f{\pa \Psi_2}{\pa x}\rt)^{(0)}\ \rt\rvert\rt\rvert_{el}\
\lt\lvert\,\f{\partial f^{(0)}}{\partial X}\rt\rvert \notag
\ +\ \lt\lvert\lt\lvert\
\lt[\lt(\,h\,P_{\perp}\rt)^{(0)}\rt]_r^{-1}\ P_{\perp}^{(0)}
\lt(\f{\pa \Psi_2}{\pa y}\rt)^{(0)}\ \rt\rvert\rt\rvert_{el}\
\lt\lvert\,\f{\partial f^{(0)}}{\partial Y}\rt\rvert. \notag
\end{eqnarray*}
By assumption, $\lt(\f{\pa \Psi_1}{\pa x}\rt)^{(0)} \in \mH_{el}$,
and $\lt[\lt(\,h\,P_{\perp}\rt)^{(0)}\rt]_r^{-1}$ and
$P_{\perp}^{(0)}$ are bounded operators on $\mH_{el}$.  So we have
\begin{eqnarray}
&&\norm{\psi_{\perp}^{(3)}}_{el}\ \leq\ A\ \lt\lvert\,\f{\partial
f^{(0)}}{\partial X}\rt\rvert\ +\ B\ \lt\lvert\,\f{\partial
f^{(0)}}{\partial Y}\rt\rvert, \notag
\end{eqnarray}
for some positive real numbers $A$ and $B$ and
$\norm{\psi_{\perp}^{(3)}}_{el}$ is exponentially decaying for
arbitrary $\gm$ by proposition \ref{prop7}.  Also,
$\psi_{\perp}^{(3)} \in C^\infty(\R^2)$ since its $(X,Y)$ dependence
comes strictly from derivatives of $f^{(0)}$ and $g^{(0)}$.  By a
similar argument, we see that $\norm{\,D^{\al}\,
\psi_{\perp}^{(3)}}_{el}$ is exponentially decaying with arbitrary
$\gm$, from proposition \ref{prop7}.

One can now use induction on $k$ to show the conclusion.  For the
induction hypothesis, assume that $\Psi^{(k-3)}$ and
$\psi_{\perp}^{(k-1)}$ given by the perturbation formulas of chapter
\ref{QuasimodeConstruction} satisfy the conclusions.  Using
equations (\ref{orderfgk}) and (\ref{orderpsiperpk}) to determine
$\Psi^{(k-2)}$ and $\psi_{\perp}^{(k)}$, the conclusion follows
from the propositions and lemmas previously proved. $\square$

\section{The Eigenstates of the Leading Order
Hamiltonian}\label{EfcnsEvals} \setcounter{equation}{0}
\setcounter{theorem}{0}

\noindent We adopt the following notation throughout:

\begin{enumerate}

\item The operator of nuclear angular momentum about the
z-axis is denoted by\\ $\dsp L_z^{nuc}=-i\f{\pa\ }{\pa \phi}$.  The
operator of total electronic angular momentum about the z-axis is
denoted by $\dsp L_z^{el}$.  The operator of total angular momentum
about the z-axis is denoted by $L_z^{TOT}\, =\, (L_z^{nuc} \otimes
I)\, +\, (I \otimes L_z^{el})$.

\item We let $L_n^k(x)$ be the associated Laguerre polynomials, as defined in \cite{Messiah}.

\end{enumerate}

The first non-vanishing terms in our perturbation expansion are
$E^{(2)},\ f^{(0)}(X,\,Y),\ g^{(0)}(X,\,Y)$ arising from the
eigenvalue equation
\begin{eqnarray*}
\mathbb{H}_2\ \left(\,\begin{array}{c}\vspace{2mm}
f^{(0)} \\
g^{(0)}
\end{array}\,\right)\ &=&\ E^{(2)}\ \left(\,\begin{array}{c}\vspace{2mm}
f^{(0)} \\
g^{(0)}
\end{array}\,\right),
\end{eqnarray*}
where
$$
\mb{H}_2\ = -\ \f12\ \Delta_{X,\,Y}\, \Iten\ +\
\left(\begin{array}{cc} \displaystyle \f{a+b}2 \ X^2\ +\ \f{a-b}2 \
Y^2 & \displaystyle b\, X\, Y
\\[3mm]
\displaystyle b\, X\, Y & \displaystyle \ \ \f{a-b}2 \ X^2\ +\ \f{a+b}2 \ Y^2\\
    \end{array}\right).
$$

Let $(\rho,\phi)$ be the usual polar coordinates associated with
$(X,Y)$.  Define the unitary operators $U,\ Z\,:\,{\cal
H}\,\rightarrow\,{\cal H}$ by:
$$
U=\left(\begin{array}{cc} \displaystyle \cos(\phi) & \displaystyle -
\sin(\phi)
\\[3mm]
\displaystyle \sin(\phi) & \displaystyle \cos(\phi)\\
    \end{array}\right)
\hskip 1cm    \text{ and } \hskip 1cm
    Z=\f1{\sqrt{2}}\left(\begin{array}{cc} \displaystyle 1 & \displaystyle
1
\\[2mm]
\displaystyle i & \displaystyle -i\\
    \end{array}\right).
$$
Let $r=a^{1/4}\,\rho$, $\ \dsp \tilde{b}=\f{b}{a}\, $, and
\begin{eqnarray*}
H_U &=& \f1{\sqrt{a}}\ U^{-1}\, \mathbb{H}_2\, U
\\[2mm]
&=&\,\lt(\lt( - \f12\, \f{\pa^2 \phantom{t}}{\pa r^2}\, -\,
\f1{2r}\, \f{\pa \phantom{t}}{\pa r}\, +\, \f12 \, r^2\, +\,
\f{(L_z^{nuc})^2+1}{2r^2} \rt)\, \Iten \rt)\, +\,
\left(\begin{array}{cc} \displaystyle \f{\tl{b}}2\,r^2 &
\displaystyle \f{i}{r^2}\,L_z^{nuc}
\\[5mm]
\displaystyle -\,\f{i}{r^2}\,L_z^{nuc} & \displaystyle
-\,\f{\tl{b}}{2}\,r^2
    \end{array}\right)
    \\[2mm]
    H_{UZ} &=& \f1{\sqrt{a}}\ (UZ)^{-1}\, \mathbb{H}_2\, (UZ)
    \\[2mm]
    &=&\, \lt(\lt( -
\f12\, \f{\pa^2 \phantom{t}}{\pa r^2}\, -\, \f1{2r}\, \f{\pa
\phantom{t}}{\pa r}\, +\, \f12 \, r^2 \rt)\, \Iten \rt)\, +\,
\left(\begin{array}{cc} \displaystyle \f{(L_z^{nuc}-1)^2}{2r^2} &
\displaystyle \f{\tl{b}}2 \, r^2
\\[7mm]
\displaystyle \f{\tl{b}}2 \, r^2 & \displaystyle
\f{(L_z^{nuc}+1)^2}{2r^2}
    \end{array}\right).
\end{eqnarray*}
Both $H_U$ and $H_{UZ}$ commute with $L_z^{nuc}\, \Iten$. So, we
search for eigenfunctions of these operators of the form $\dsp \lt(
\begin{array}{c} e^{\pm i\abs{l}\phi}\, \psi_1(r) \\[3mm] e^{\pm i \abs{l}\phi}\, \psi_2(r)
\end{array}\rt),\, \ \abs{l}=0,1,2,\cdots\
$  We warn the reader that although $l$ arises here as an eigenvalue of
$L_z^{nuc}$, at this point we should not associate any physical
meaning to $l$.  Here we are dealing with the operators $H_U$ and
$H_{UZ}$, which are related to $\mb{H}_2$ by the operations of $U$
and $Z$.  The physical meaning of $l$ will become apparent in
theorem \ref{thm5.1}.

We note that $(U^{-1}\, \mathbb{H}_2\, U)\Psi\, =\, E\,\Psi$ was the
leading order equation obtained by Renner \cite{Renner}, which is
unitarily equivalent to our leading order equation $\mb{H}_2\,\Psi\,
=\, E\,\Psi$.  Renner showed that some of the eigenvalues can be
solved for exactly, and used regular perturbation theory up to
second order to approximate the other eigenvalues.  These equations
have been studied by several other authors, for instance
\cite{BrownJorgensen,Herzbook}.  We repeat some of Renner's
results here, but we calculate the perturbation series to much
higher orders, demonstrating that many of the series are diverging inside the region of
interest. We also illustrate that there is likely a crossing
involving the ground state eigenvalue of $\mb{H}_2$ near $b \approx
0.925a$. The ground state appears to be degenerate for $0 < b <
0.925 a$ and non-degenerate for $0.925 a < b < a$.

\subsection{The Exactly Solvable \boldmath{$l=0$}
States}\label{sectionl0states}

The $l=0$ states (no angular dependence) are exactly solvable.  In
this case $H_U$ reduces to
$$
H_U^{[l=0]}\, =\, \left(\begin{array}{cc} \displaystyle - \f12\,
\f{\pa^2 \phantom{t}}{\pa r^2}\, -\, \f1{2r}\, \f{\pa
\phantom{t}}{\pa r}\, +\, \f{1+\tl{b}}2 \, r^2\, +\, \f{1}{2r^2}\, &
\displaystyle 0
\\[5mm]
\displaystyle 0 & \displaystyle - \f12\, \f{\pa^2 \phantom{t}}{\pa
r^2}\, -\, \f1{2r}\, \f{\pa \phantom{t}}{\pa r}\, +\, \f{1-\tl{b}}2
\, r^2\, +\, \f{1}{2r^2}
    \end{array}\right).
$$
We recognize that the component equations are of the same form as
the radial equation for angular momentum 1 states of the two
dimensional Isotropic Harmonic Oscillator.  From the first component
equation, the eigenvalues and eigenfunctions (non-normalized) are
\begin{eqnarray*}
E_{N_+}=(2N_++2)\sqrt{1+\tl{b}}\ , \hskip .5cm \lt(
\begin{array}{c} \tl{r_+}\, L_{N_+}^{1}(\tl{r_+}^2)\, e^{-\tl{r_+}^2/2} \\[3mm]
0
\end{array}\rt), \hskip .5cm N_+=0,1,2,\cdots \label{L0states1}
\end{eqnarray*}
where $\tl{r_+}\, =\, (1+\tl{b})^{1/4}\,r$.  From the second
component equation, the eigenvalues and eigenfunctions
(non-normalized) are
\begin{eqnarray*}
E_{N_-}=(2N_{-}+2)\sqrt{1-\tl{b}}\ , \hskip .5cm \lt(
\begin{array}{c} 0 \\[3mm]
\tl{r_-}\, L_{N_-}^{1}(\tl{r_-}^2)\, e^{-\tl{r_-}^2/2}
\end{array}\rt), \hskip .5cm N_-=0,1,2,\cdots \label{L0states2}
\end{eqnarray*}
where $\tl{r_-}\, =\, (1-\tl{b})^{1/4}\,r$.

Since $\mb{H}_2$ is unitarily equivalent to $\sqrt{a}\,H_U$, we see
these states give rise to eigenvalues and eigenfunctions of
$\mb{H}_2$ given by
\begin{eqnarray}
E_{N_-}&=&(2N_{-}+2)\,\sqrt{a-b} \notag
\\[3mm]
\Psi_{N_-}^{[l=0]}(\rho,\phi) &=& \lt(
\begin{array}{c} -\,\tl{r_-}\,\sin(\phi)\, L_{N_-}^{1}(\tl{r_-}^2)\, e^{-\tl{r_-}^2/2}
\\[3mm]
\tl{r_-}\,\cos(\phi)\, L_{N_-}^{1}(\tl{r_-}^2)\,
e^{-\tl{r_-}^2/2}\end{array}\rt),\ \ {N_-}=0,1,2,\cdots
\label{L0states3}
\end{eqnarray}
where $\tl{r_-}\, =\, (a-b)^{1/4}\,\rho$, and
\begin{eqnarray}
E_{N_+} &=& (2N_++2)\,\sqrt{a+b} \notag
\\[3mm]
\Psi_{N_+}^{[l=0]}(\rho,\phi) &=& \lt(
\begin{array}{c} \tl{r_+}\,\cos(\phi)\, L_{N_+}^{1}(\tl{r_+}^2)\, e^{-\tl{r_+}^2/2}
\\[3mm]
\tl{r_+}\,\sin(\phi)\, L_{N_+}^{1}(\tl{r_+}^2)\,
e^{-\tl{r_+}^2/2}\end{array}\rt),\ \ N_+=0,1,2,\cdots
\label{L0states4}
\end{eqnarray}
where $\tl{r_+}\, =\, (a+b)^{1/4}\,\rho$.

\subsection{The Perturbation Calculation For the \boldmath{$l\neq 0$}
States}\label{sectionlneq0states}

In this case, $H_{UZ}$ reduces to
\begin{eqnarray*}
&& H_{UZ}^{[\pm\abs{l}]}\ =\ \left(\begin{array}{cc} \displaystyle - \f12\, \f{\pa^2
\phantom{t}}{\pa r^2}\, -\, \f1{2r}\, \f{\pa \phantom{t}}{\pa r}\,
+\, \f12 \, r^2\, +\, \f{(\abs{l}\mp1)^2}{2r^2} & 0
\\[7mm]
0 & \displaystyle - \f12\, \f{\pa^2 \phantom{t}}{\pa r^2}\, -\,
\f1{2r}\, \f{\pa \phantom{t}}{\pa r}\, +\, \f12 \, r^2\, +\,
\f{(\abs{l}\pm1)^2}{2r^2}
    \end{array}\right) \notag
    \\[5mm]
    &&\hskip 2.5cm +\ \f{\tl{b}}2\, r^2\, \lt(\begin{array}{cc} 0 & 1 \\ 1 & 0
    \end{array}\rt).
\end{eqnarray*}
Denote the eigenfunctions of $H_{UZ}^{[\pm\abs{l}]}$ by $\dsp \lt(
\begin{array}{c} f^{[\pm \abs{l}]}(r) \\[3mm] g^{[\pm \abs{l}]}(r)
\end{array}\rt)$.  It is clear that if $\dsp \lt(
\begin{array}{c} f^{[\abs{l}]}(r) \\[3mm] g^{[\abs{l}]}(r)
\end{array}\rt)$ is an\\[3mm] eigenfunction of $H_{UZ}^{[\abs{l}]}$ with eigenvalue $E$, then $\dsp \lt(
\begin{array}{c} f^{[-\abs{l}]}(r) \\[3mm] g^{[- \abs{l}]}(r)
\end{array}\rt)\, =\, \dsp \lt(
\begin{array}{c} g^{[\abs{l}]}(r) \\[3mm] f^{[\abs{l}]}(r)
\end{array}\rt)$ is an eigenfunction of $H_{UZ}^{[-\abs{l}]}$ with eigenvalue
$E$.  So we only need to find the eigenfunctions and eigenvalues of
the $H_{UZ}^{[\abs{l}]}$.

We have not been able to solve for the eigenvalues and
eigenfunctions in this case exactly.  We use regular perturbation
theory with perturbation parameter $\tl{b}$, letting
$H_{UZ}^{[\abs{l}]}=H_0^{[\abs{l}]}+\tl{b}\tl{V}$, where
\begin{eqnarray*}
H_0^{[\abs{l}]} &=& \left(\begin{array}{cc} \displaystyle -\ \f12
\f{\pa^2 \phantom{t}}{\pa r^2}\ -\ \f1{2r} \f{\pa \phantom{t}}{\pa
r} + \f12 r^2 + \f{(\abs{l}-1)^2}{2r^2} & 0
\\[7mm]
0 & \displaystyle -\ \f12 \f{\pa^2 \phantom{t}}{\pa r^2}\ -\ \f1{2r}
\f{\pa \phantom{t}}{\pa r} + \f12 r^2 + \f{(\abs{l}+1)^2}{2r^2}
\end{array}\rt) \notag
\\[5mm]
\tl{V}\, &=&\, \f{1}2\, r^2\, \lt(\begin{array}{cc} 0 & 1 \\ 1 & 0
    \end{array}\rt).
\end{eqnarray*}
One can show using the relative bound found in equation
(\ref{relativebound}), that $\tl{V}$ is relatively bounded with respect to $H_0^{[\abs{l}]}$ on $\mH$.
So, we know that in terms of $\tl{b}$, $H_{UZ}^{[\abs{l}]}$ is an
analytic family of type A for small $\tl{b}$ \cite{RSV4}. Therefore,
the eigenvalues and eigenfunctions will be analytic functions of
$\tl{b}$ in a neighborhood of $\tl{b}=0$.

We expand the eigenvalues and eigenfunctions of $H_{UZ}^{[\abs{l}]}$
in a series in $\tl{b}$:
\begin{eqnarray}
E^{N,\abs{l}}(\tl{b})\, =\, \sum_{k=0}^\infty\, E_k^{N,\abs{l}}\,
\tl{b}^k, \hskip 2cm \Psi^{N,\abs{l}}(\tl{b})\, =\,
\sum_{k=0}^\infty\, \Psi_k^{N,\abs{l}}\, \tl{b}^k \label{pertseries}
\end{eqnarray}
and solve for the coefficients $E_k^{N,\abs{l}},\
\Psi_k^{N,\abs{l}}$ recursively.  Here $N$ indexes the energy levels
of $H_{UZ}^{[\abs{l}]}$ for fixed $\abs{l}$.  Again from the two-dimensional isotropic oscillator, the eigenfunctions of
$H_0^{[\abs{l}]}$ are known exactly. The lowest state is
non-degenerate, with eigenvalue and eigenfunction given by
\begin{eqnarray}
E_0^{0,\abs{l}}=\abs{l}, \hskip .5cm \Psi_0^{0,\abs{l}}\, =\, \lt(
\begin{array}{c} r^{\abs{l}-1}\, e^{-r^2/2} \\[3mm]
0
\end{array}\rt). \label{H0statesground}
\end{eqnarray}
The rest of the states are two-fold degenerate, with eigenvalues and
eigenfunctions given by
\begin{eqnarray}
\Psi_{0,\,up}^{N,\abs{l}}\ &=&\, \lt(
\begin{array}{c} r^{\abs{l}-1}\, L_N^{\abs{l}-1}(r^2)\,e^{-r^2/2} \\[3mm]
0
\end{array}\rt), \qquad E_0^{N,\abs{l}}\ =\ 2N+\abs{l},
\notag
\\[3mm]
\Psi_{0,\,dwn}^{N,\abs{l}}\, &=&\, \lt(
\begin{array}{c} 0 \\[3mm]
r^{\abs{l}+1}\, L_{N-1}^{\abs{l}+1}(r^2)\,e^{-r^2/2}
\end{array}\rt), \qquad \ N=1,2,\cdots \label{H0statesexcited}
\end{eqnarray}
The functions $\{e^{i\,l\,\phi}\,r^{\abs{l}}\,
L_K^{\abs{l}}(r^2)\, e^{-r^2/2}\}_{\substack{\phantom{t} \\
l\in \mb{Z}
\\ K=0,1,2,\cdots}}$ form a basis for $L^2(\R^2)$ by theorem
XIII.64 of \cite{RSV4}.  Then for fixed $l\in\mb{Z}$, the functions
$\{r^{\abs{l}}\, L_K^{\abs{l}}(r^2)\, e^{-r^2/2}\}_{K=0,1,2,\cdots}$
form a basis for the projection of $L^2(\R^2)$ onto $r$-dependent
multiples of $e^{i\,l\,\phi}$.  We can then use the following
orthonormal basis for the perturbation expansion:
\begin{eqnarray*}
\lt\{ \lt(
\begin{array}{c} B_{N,\abs{l}-1}\,r^{\abs{l}-1}\, L_N^{\abs{l}-1}(r^2)\,e^{-r^2/2} \\[3mm]
0
\end{array}\rt),\ \lt(
\begin{array}{c} 0 \\[3mm]
B_{N,\abs{l}+1}\,r^{\abs{l}+1}\, L_{N}^{\abs{l}+1}(r^2)\,e^{-r^2/2}
\end{array}\rt) \rt\}_{N=0,1,2,\cdots}
\end{eqnarray*}
where the $B_{N,\abs{l}}$ are constants of normalization.  The matrix
elements of the perturbation $\tl{V}$ in this basis can be obtained
explicitly \cite{thesis}.

\subsubsection{The Non-Degenerate Perturbation
Calculation}\label{subsectionNonDegeneratePerturbation}

\begin{figure}[htb]
\begin{centering}
\leavevmode
\includegraphics[width=0.45\textwidth]
{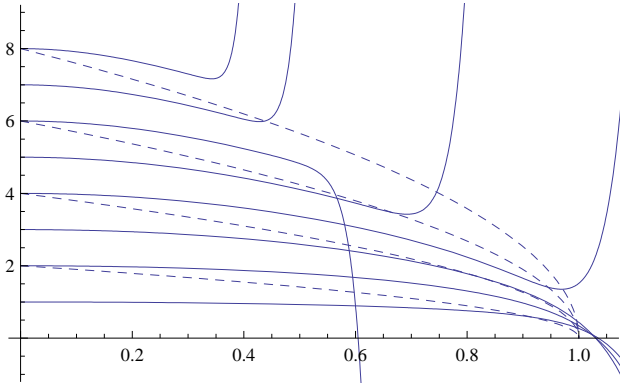}  \vspace{-.1in} \caption{A plot of the
perturbation series versus $\tl{b}$, of the non-degenerate, lowest
lying eigenvalue of $H_{UZ}^{[\abs{l}]}$ up to order 28, for
$\abs{l}=1,2,\cdots, 8$.  The dashed curves are
$(2N+2)\sqrt{1-\tl{b}}$, for $N=0,1,2,3$, which are the $l=0$ states
that we have solved for exactly.} \vskip .5cm
\label{Fignondegenlevels}
\end{centering}
\end{figure}

Recall from (\ref{H0statesground}), for fixed $\abs{l}\neq 0$, the
lowest lying eigenvalue of $H_0^{[\abs{l}]}$ is
$E_0^{0,\abs{l}}=\abs{l}$ (non-degenerate).  Since $H_{UZ}^{\abs{l}}$ is an analytic family, we use non-dengenerate, regular perturbation theory.  Using the
$Mathematica$ software package, we easily computed the exact
perturbation coefficients up to 28th order for the non-degenerate,
lowest lying eigenvalue $E^{0,\abs{l}}$, for several values of
$\abs{l}$ (See Figure \ref{Fignondegenlevels}).

Recall that we are concerned with the case where $0 < b < a$, so
that $0 < \tl{b}=\f{b}{a} <1$.  The functions
$E^{0,\abs{l}}(\tl{b})$ likely do not exist as eigenvalues of
$H_{UZ}^{[\abs{l}]}$ if $\tl{b} \geq 1$.  Seemingly the radii of
convergence of the series are smaller as $\abs{l}$ increases.  It
appears that for $\abs{l}=1,2,$ and $3$ the radius of convergence is
likely close to 1 (if not larger).  For $\abs{l} > 4$, the series
are behaving erratically for values of $\tl{b} < 1$. The $\abs{l}=4$
case appears to be borderline, with radius of convergence possibly
only slightly smaller than 1.  This divergent behavior was seen even
from the low order coefficients for the larger values of $\abs{l}$.
The singularities are likely caused by avoided crossings between two
states with the same value of $\abs{l}$, as suggested in
\cite{Renner, BrownJorgensen, Herzbook}.

We highlight the crossing between the $\abs{l}=1$ state and the
lowest lying $l=0$ state near $\tl{b}=0.925$.  Recall that for
$l\neq 0$, the eigenvalue $E^{J,\abs{l}}$ of $H_{UZ}^{[\abs{l}]}$ is
also an eigenvalue of $H_{UZ}^{[-\abs{l}]}$.  Together these states
correspond to a degenerate eigenvalue of the original operator
$\mb{H}_2$. The $l=0$ states are all non-degenerate for $b>0$.  So,
this crossing implies that the ground state of $\mb{H}_2$ is
degenerate for approximately $0 < b < 0.925a$ and non-degenerate for
$0.925 a < b < 1$.

\subsubsection{The Degenerate Perturbation Calculation}\label{subsectionDegenereatePerturbation}

Recall from (\ref{H0statesground}), that only the ground state of
$H_0^{[\abs{l}]}$ is non-degenerate if $\abs{l}\neq 0$.  In the
perturbation calculation described in section
\ref{subsectionNonDegeneratePerturbation}, we used regular
non-degenerate perturbation theory to obtain the perturbation
coefficients for these eigenvalues.  From (\ref{H0statesexcited}),
we have that for fixed $\abs{l}\neq 0$, $H_0^{[\abs{l}]}$ also has
two-fold degenerate eigenvalues of $E_0^{N,\abs{l}}=2N+\abs{l}$ for
$N=1,2,\cdots$  So we must use degenerate perturbation theory to
calculate the perturbation coefficients of these eigenvalues. Recall
from (\ref{H0statesexcited}), the degenerate pair of eigenfunctions
corresponding to $E_0^{N,\abs{l}}$ are given by
$\Psi_{0,up}^{N,\abs{l}}$ and $\Psi_{0,dwn}^{N,\abs{l}}$.

\begin{figure}[htb]
\begin{centering}
\leavevmode
\includegraphics[width=0.45\textwidth]
{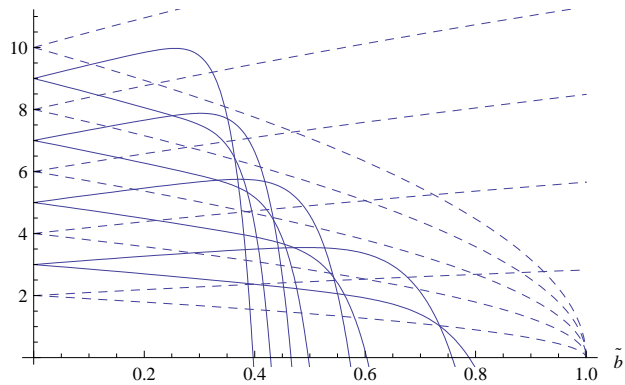}  \vspace{-.1in} \caption{A plot of the
perturbation series versus $\tl{b}$ up to order 12, of the first 8
eigenvalues of $H_{UZ}^{[\abs{l}]}$ that are degenerate at zeroth
order, for $\abs{l}=1$. The dashed curves are
$(2N+2)\sqrt{1-\tl{b}}$ and $(2N+2)\sqrt{1+\tl{b}}$, for
$N=0,1,2,3,4$, which are the $l=0$ states that we have solved for
exactly.} \vskip .5cm \label{FigdegenlevelsL1}
\end{centering}
\end{figure}

\begin{figure}[htb]
\begin{centering}
\leavevmode
\includegraphics[width=0.45\textwidth]
{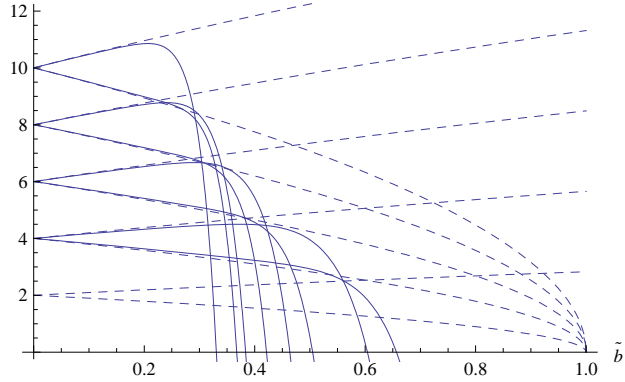}  \vspace{-.1in} \caption{A plot of the
perturbation series versus $\tl{b}$ up to order 12, of the first 8
eigenvalues of $H_{UZ}^{[\abs{l}]}$ that are degenerate at zeroth
order, for $\abs{l}=2$. The dashed curves are
$(2N+2)\sqrt{1-\tl{b}}$ and $(2N+2)\sqrt{1+\tl{b}}$, for
$N=0,1,2,3,4$, which are the $l=0$ states that we have solved for
exactly.} \vskip .5cm \label{FigdegenlevelsL2}
\end{centering}
\end{figure}

Employing degenerate perturbation theory in the usual manner, we
find there is splitting that occurs at first order (we omit the details).
Armed with the proper linear combinations we can then proceed as in
the non-degenerate case.  Using the $Mathematica$ software package,
we easily computed the exact perturbation coefficients up to 12th
order for the first few eigenvalues $E^{N,\abs{l}}$ that are
degenerate at zeroth order, for several values of $\abs{l}$ (See
Figures \ref{FigdegenlevelsL1}, \ref{FigdegenlevelsL2}).

While the splitting is nicely illustrated, we see that all of the
series likely have radii of convergence well below 1.  The radius of
convergence appears to decrease as $\abs{l}$ or $N$ increase. The
divergent behavior was seen even at low orders of the perturbation
coefficients.

\begin{figure}[htb]
\begin{centering}
\leavevmode
\includegraphics[width=0.5\textwidth]
{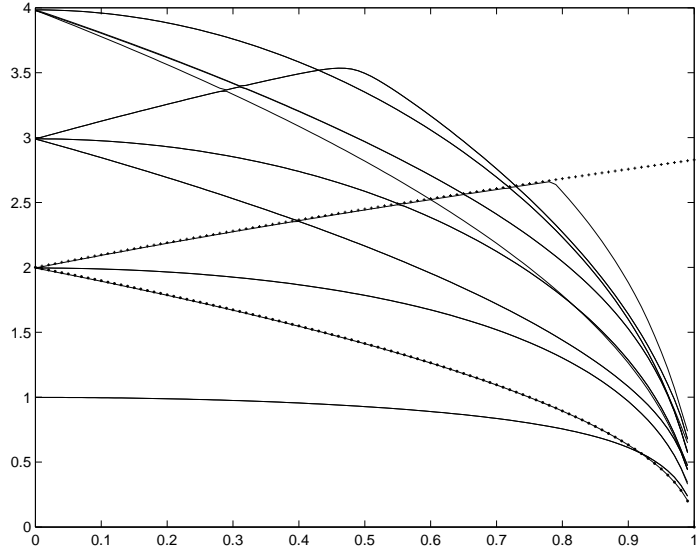}  \vspace{-.1in} \caption{A plot of lowest 17
eigenvalues of $\mb{H}_2$ as a function of $\tl{b}$, on $0 < \tl{b}
<1$, as approximated by a finite difference scheme.  The dotted
curves are $2\sqrt{1-\tl{b}}$ and $2\sqrt{1+\tl{b}}$, which are the
lowest of the $l=0$ states that we have solved for exactly.}
\label{Figmatlablevels}
\end{centering}
\end{figure}

We also used an elementary finite difference scheme to approximate
the eigenvalues at several values of $\tl{b}$, for $0<\tl{b}<1$. The
results are given in Figure \ref{Figmatlablevels}.  The plot was
generated by approximating the lowest lying 17 eigenvalues for a
fixed $\tl{b}$ value, then the value of $\tl{b}$ was changed and the
lowest 17 eigenvalues were calculated again.  This was repeated at
steps of $\Delta \tl{b}=.01$ from $0\leq\tl{b}<.99$.  Recall that
the $l=0$ states were exactly solvable.  For comparison, the exact
values of the lowest lying $l=0$ states were plotted as dotted
curves.  We see that the finite difference scheme approximates these
eigenvalues so well that the dotted curve are hardly distinguishable
from the finite difference approximation of these eigenvalues.  Near
$\tl{b}=0$, the 17 eigenvalues that are being approximated can be
identified by their values at $\tl{b}=0$:
\begin{enumerate}

\item The curve that has value 1 at $\tl{b}=0$ is actually two overlapping eigenvalues of $\mb{H}_2$ corresponding to the degenerate
pair of lowest lying $\abs{l}=1$ states, one for $l=1$ and $l=-1$.

\item There are three curves that have value 2 at $\tl{b}=0$.  Two of the curves are the
non-degenerate $l=0$ states (one increases with $\tl{b}$ and one
decreases with $\tl{b}$).  The other curve is two overlapping
eigenvalues corresponding to the degenerate pair of lowest lying
$\abs{l}=2$ states, one for $l=2$ and $l=-2$.  These curves together
account for four eigenvalues of $\mb{H}_2$.

\item There are three curves that have value 3 at $\tl{b}=0$.  Two of
the curves are overlapping degenerate $\abs{l}=1$ states, (one
degenerate pair increases with $\tl{b}$ and one degenerate pair
decreases with $\tl{b}$).  The other curve an overlapping degenerate
pair of lowest lying $\abs{l}=2$ states.  These curves together
account for six eigenvalues of $\mb{H}_2$.

\item There are three curves that have value 4 at $\tl{b}=0$.  One
of the curves is a non-degenerate $l=0$ state, one is a degenerate
pair of $\abs{l}=2$ states, and one is a degenerate pair of
$\abs{l}=4$ states. Together these curves account for five
eigenvalues of $\mb{H}_2$.

\end{enumerate}

This plot supports the claim that a crossing occurs involving the
ground state near $\tl{b}=0.925$. While the finite difference scheme
is crude, we are inclined to trust the qualitative features of the
results considering the lowest of the exactly solvable $l=0$
eigenvalues were so well approximated, even near $\tl{b}=1$ as seen
in the figure.  We note that as $\tl{b}$ increases from zero,
avoided crossings involving states with the same value of $\abs{l}$
occur, as well as crossings involving states with different values
of $\abs{l}$.  When the uppermost curve is involved with such a
phenomenon it will appear to change behavior suddenly without
reason, but this is only because we can only see the lowest 17
eigenvalues at each $\tl{b}$.

\section{Degeneracy of the Quasimode Energies}\label{chapdegeneracy}
\setcounter{equation}{0} \setcounter{theorem}{0}

The eigenfunctions of $\mb{H}_2$ provide the zeroth order states for
the quasimode expansion.  Recall that if  $\dsp \lt(\begin{array}{c}
f^{(0)} \\ g^{(0)} \end{array}\rt)\,$ is an eigenfunction of
$\mb{H}_2$, we have derived perturbation formulas in chapter
\ref{QuasimodeConstruction} that determine the functions
$f^{(k)}(X,\,Y),\ g^{(k)}(X,\,Y),$ and $\psi_{\perp}^{(k)}(X,\,Y)$
that enter in equation (\ref{Psiepsilondefinition}) as the
asymptotic series
\begin{eqnarray*}
&& \hskip -.4cm \Phi_{\ep}\, =\,
\\[3mm]
&& \hskip -.4cm \Psi_1(\ep\,X,\,\ep\,Y)\, \sum_{k=0}^\infty\,
f^{(k)}(X,\,Y)\, \ep^k\, +\, \Psi_2(\ep\,X,\,\ep\,Y)\,
\sum_{k=0}^\infty\, g^{(k)}(X,\,Y)\, \ep^k\, +\, \sum_{k=0}^\infty\,
\psi_{\perp}^{(k)}(X,\,Y)\, \ep^k
\end{eqnarray*}
\begin{eqnarray*}
&& \hskip -.4cm =\, \sum_{k=0}^\infty\, \ep^k\, \lt(\,
\sum_{j=0}^k\, \lt(\, \Psi_1^{(j)}(X,\,Y)\, f^{(k-j)}(X,\,Y)\, +\,
\Psi_2^{(j)}(X,\,Y)\, g^{(k-j)}(X,\,Y)\,\rt)\, +\,
\psi_{\perp}^{(k)}(X,\,Y)\, \rt),
\\[3mm]
&& \hskip -.4cm =:\, \sum_{k=0}^\infty\, \ep^k\, \Phi_k,
\end{eqnarray*}
where $\lt\{\Psi_1(\ep\,X,\,\ep\,Y),\ \Psi_2(\ep\,X,\,\ep\,Y)\rt\}$
is the electronic eigenfunction basis.  The $f^{(k)}$ and $g^{(k)}$
have no electronic dependence (they are scalar functions) and
$\psi_{\perp}^{(k)}$ has both electronic and nuclear dependence.

Recall that for $\abs{l} \neq 0$, if $\dsp \lt(
\begin{array}{c} f^{[\abs{l}]}(r) \\[3mm] g^{[\abs{l}]}(r)
\end{array}\rt)$ is an eigenfunction of $H_{UZ}^{[\abs{l}]}$ with eigenvalue $E$, then\\[3mm]
$\dsp \lt(
\begin{array}{c} f^{[-\abs{l}]}(r) \\[3mm] g^{[- \abs{l}]}(r)
\end{array}\rt)\, =\, \dsp \lt(
\begin{array}{c} g^{[\abs{l}]}(r) \\[3mm] f^{[\abs{l}]}(r)
\end{array}\rt)$ is an eigenfunction of $H_{UZ}^{[-\abs{l}]}$ with eigenvalue
$E$.  So if $\abs{l} \neq 0$, we have two-fold degenerate
eigenfunctions of $\mb{H}_2$ of the form (recall $r=a^{1/4}\rho$)
\begin{eqnarray}
\lt(\begin{array}{c} F^{(0)} \\[3mm] G^{(0)} \end{array}\rt)\, &=:&\, U\,Z\, \lt(\begin{array}{c} e^{i\abs{l}\phi}\, f^{[\abs{l}]}(r) \\[3mm]
e^{i\abs{l}\phi}\, g^{[\abs{l}]}(r) \end{array}\rt) \notag
\\[3mm]
&=&\, \f1{\sqrt{2}}\, \lt(\begin{array}{c} e^{i(\abs{l}-1)\phi}\,
f^{[\abs{l}]}(r)\, +\, e^{i(\abs{l}+1)\phi}\, g^{[\abs{l}]}(r)
\\[3mm]
i\, \lt(\, e^{i(\abs{l}-1)\phi}\, f^{[\abs{l}]}(r)\, -\,
e^{i(\abs{l}+1)\phi}\, g^{[\abs{l}]}(r)\, \rt)
\end{array}\rt) \label{plusL}
\end{eqnarray}
and
\begin{eqnarray}
U\,Z\, \lt(\begin{array}{c} e^{-i\abs{l}\phi}\, g^{[\abs{l}]}(r) \\[3mm]
e^{-i\abs{l}\phi}\, f^{[\abs{l}]}(r) \end{array}\rt)\, &=&\,
\f1{\sqrt{2}}\, \lt(\begin{array}{c} e^{-i(\abs{l}-1)\phi}\,
f^{[\abs{l}]}(r)\, +\, e^{-i(\abs{l}+1)\phi}\, g^{[\abs{l}]}(r)
\\[3mm]
-i\, \lt(\, e^{-i(\abs{l}-1)\phi}\, f^{[\abs{l}]}(r)\, -\,
e^{-i(\abs{l}+1)\phi}\, g^{[\abs{l}]}(r)\, \rt)
\end{array}\rt) \notag
\\[3mm]
&=&\, \lt(\begin{array}{c} \overline{F^{(0)}} \\[3mm] \overline{G^{(0)}}
\end{array}\rt)\, \label{minusL}
\end{eqnarray}
By taking appropriate linear combinations, these degenerate zeroth
order functions lead to two orthogonal quasimodes using the
perturbation formulas of chapter \ref{QuasimodeConstruction},
possibly degenerate (no splitting) or non-degenerate (splitting).

We adopt the following nomenclature:  We refer to the eigenfunctions
of $\mb{H}_2$ that arise from the eigenfunctions of
$H_{UZ}^{[\abs{l}]}$, where $\abs{l} \neq 0$, as $+\abs{l}$ states.
We refer to the eigenfunctions of $\mb{H}_2$ that arise from the
eigenfunctions of $H_{UZ}^{[-\abs{l}]}$, where $\abs{l} \neq 0$, as
$-\abs{l}$ states.  We refer to the eigenfunctions of $\mb{H}_2$
that arise from the eigenfunctions of $H_{U}^{[\abs{l}=0]}$ as
$\abs{l}=0$ states.

\begin{thm}\label{thm5.1}
Let $L_z^{TOT}$ be the operator of total angular momentum around the
$z$-axis and $0 < \tl{b} <1$. Then:
\begin{enumerate}
\item For $l\neq 0$, each $\,+\abs{l}$ state generates a quasimode $\Phi_{\ep}^A$ of $H(\ep)$ that satisfies $L_z^{TOT}\,
\Phi_{\ep}^A\, =\, \abs{l}\, \Phi_{\ep}^A$.  The corresponding
degenerate $\,-\abs{l}$ state generates a quasimode $\Phi_{\ep}^B$
that satisfies $\Phi_{\ep}^B\, =\, \overline{\Phi}_{\ep}^A$ and
$L_z^{TOT}\, \Phi_{\ep}^B\, =\, -\abs{l}\, \Phi_{\ep}^B$.  The
$\Phi_{\ep}^A$ and $\Phi_{\ep}^B$ quasimodes are orthogonal, and
asymptotic to two-fold degenerate eigenfunctions of $H(\ep)$.  We
see that linear combinations of the these two-fold degenerate $\,\pm
\abs{l}$ states also generate valid quasimodes.

\item Each $\abs{l}=0$ state generates a quasimode that is asymptotic to
a non-degenerate eigenfunction of $H(\ep)$.
\end{enumerate}
In either case, the zeroth order of the electronic eigenfunction
basis vectors $\Psi_1(0,0)$ and $\Psi_2(0,0)$ are linear
combinations of eigenfunctions of $L_z^{el}$ with eigenvalues $\pm
1$.
\end{thm}

\noindent {\bf Remark:}

The physical meaning of $l$ is now apparent.  It corresponds to the
total angular momentum about the z-axis of the wave function being
approximated.  From the proof to follow, it will be clear that the
zeroth order $\Phi_0$ of a quasimode, can be constructed to satisfy
$L_z^{TOT}\, \Phi_0\, =\, l_z^{TOT}\, \Phi_0$.  In this case it is a
linear combination of two states of the form
$$
\Xi_{l_z^{TOT}-1}(\vec{r}_{nuc})\, \tl{\Psi}_+(\vec{r}_{el})\quad
\text{and} \quad \Xi_{l_z^{TOT}+1}(\vec{r}_{nuc})\,
\tl{\Psi}_-(\vec{r}_{el}),
$$
where
\begin{eqnarray*}
&&L_z^{el}\, \tl{\Psi}_+\, =\, \tl{\Psi}_+, \hskip 4.6cm L_z^{el}\,
\tl{\Psi}_-\, =\, -\,\tl{\Psi}_-,\\[3mm] &&L_z^{nuc}\, \Xi_{l_z^{TOT}-1}\,
=\, (l_z^{TOT}-1)\, \Xi_{l_z^{TOT}-1}, \qquad L_z^{nuc}\,
\Xi_{l_z^{TOT}+1}\, =\, (l_z^{TOT}+1)\, \Xi_{l_z^{TOT}+1}.
\end{eqnarray*}

\noindent {\bf Proof:}

Since $[H(\ep),\ L_z^{TOT}]=0$, we know that the true eigenfunctions
$\Psi(\ep)$ of $H(\ep)$ can be constructed to satisfy $L_z^{TOT}\,
\Psi(\ep)\, =\, l_z^{TOT}\, \Psi(\ep)$, at each $\ep$ in a
neighborhood of 0, for some $l_z^{TOT} \in \mb{Z}$.  This implies
that $L_z^{TOT}\, \overline{\Psi(\ep)}\, =\, -\, l_z^{TOT}\,
\overline{\Psi(\ep)}$, since $\overline{L_z^{TOT}\, \Psi(\ep)}\, =\,
-\, L_z^{TOT}\, \overline{\Psi(\ep)}$.  We can therefore arrange so
that the asymptotic series
$\Phi_{\ep}=\sum_{k=0}^{\infty}\ep^k\Phi_{k}$ satisfies $L_z^{TOT}\,
\Phi_{\ep}\, =\, l_z^{TOT}\, \Phi_{\ep}$ at each order of $\ep$.  We
then know that each order $\Phi_k$ of the quasimode, and its complex
conjugate, are eigenfunctions of $L_z^{TOT}$ with eigenvalues
$l_z^{TOT}$ and $-l_z^{TOT}$ respectively.

\noindent We now separate into two cases:

\noindent {\bf Case 1: $\abs{l} \neq 0$}

In this case, we have degenerate zeroth order states of the form in
(\ref{plusL}) and (\ref{minusL}).  Regardless of whether splitting
occurs, assume that we depart from zeroth order with a
correct linear combination $\dsp \lt(\begin{array}{c} f^{(0)} \\
g^{(0)} \end{array}\rt)\, =\, \al\, \lt(\begin{array}{c} F^{(0)} \\
G^{(0)} \end{array}\rt)\,
+\, \beta\, \lt(\begin{array}{c} \overline{F^{(0)}} \\
\overline{G^{(0)}} \end{array}\rt)\,$, so that this leads to a valid
quasimode, which satisfies $L_z^{TOT}\, \Phi_{\ep}\, =\, l_z^{TOT}\,
\Phi_{\ep}$.  Then the zeroth order $\Phi_0$ must also be an
eigenfunction of $L_z^{TOT}$ with eigenvalue $l_z^{TOT}$. The
$\Phi_0$ function is given by
\begin{eqnarray}
\Phi_0\, &=&\, \Psi_1(0,0)\, f^{(0)}\, +\, \Psi_2(0,0)\, g^{(0)}
\notag
\\[3mm]
&=&\, \Psi_1(0,0)\, \lt(\al\, F^{(0)}\, +\, \beta\,
\overline{F^{(0)}}\rt)\, +\, \Psi_2(0,0)\, \lt(\al\, G^{(0)}\, +\,
\beta\, \overline{G^{(0)}}\rt) \notag
\\[3mm]
&=&\, \al\, \lt[\, e^{i(\abs{l}-1)\phi}\, f^{[\abs{l}]}\,
\lt(\Psi_1(0,0)\, +\, i\,\Psi_2(0,0)\rt)\, +\,
e^{i(\abs{l}+1)\phi}\, g^{[\abs{l}]}\, \lt(\Psi_1(0,0)\, -\,
i\,\Psi_2(0,0)\rt)\, \rt] \notag
\\[3mm]
&+& \beta\, \lt[\, e^{-i(\abs{l}-1)\phi}\, f^{[\abs{l}]}\,
\lt(\Psi_1(0,0)\, -\, i\,\Psi_2(0,0)\rt)\, +\,
e^{-i(\abs{l}+1)\phi}\, g^{[\abs{l}]}\, \lt(\Psi_1(0,0)\, +\,
i\,\Psi_2(0,0)\rt)\, \rt]. \notag
\end{eqnarray}
We now plug this into the equation $L_z^{TOT}\,\Phi_0\, -\,
l_z^{TOT}\, \Phi_0\, =\, 0$, and for $\abs{l} \geq 2$, we project
along $e^{i(\abs{l}-1)\phi}\,f^{[\abs{l}]}$,
$e^{i(\abs{l}+1)\phi}\,g^{[\abs{l}]}$,
$e^{-i(\abs{l}-1)\phi}\,f^{[\abs{l}]}$, and
$e^{-i(\abs{l}+1)\phi}\,g^{[\abs{l}]}$, and obtain the following
four equations:
\begin{eqnarray}
L_z^{el}\, \lt(\Psi_1(0,0)\, +\, i\,\Psi_2(0,0)\rt)\, =\,
(l_z^{TOT}-\abs{l}+1)\, \lt(\Psi_1(0,0)\, +\, i\,\Psi_2(0,0)\rt)
\label{firstLequation}
\\[3mm]
L_z^{el}\, \lt(\Psi_1(0,0)\, -\, i\,\Psi_2(0,0)\rt)\, =\,
(l_z^{TOT}-\abs{l}-1)\, \lt(\Psi_1(0,0)\, -\, i\,\Psi_2(0,0)\rt)
\label{secondLequation}
\\[3mm]
L_z^{el}\, \lt(\Psi_1(0,0)\, -\, i\,\Psi_2(0,0)\rt)\, =\,
(l_z^{TOT}+\abs{l}-1)\, \lt(\Psi_1(0,0)\, -\, i\,\Psi_2(0,0)\rt)
\label{thirdLequation}
\\[3mm]
L_z^{el}\, \lt(\Psi_1(0,0)\, +\, i\,\Psi_2(0,0)\rt)\, =\,
(l_z^{TOT}+\abs{l}+1)\, \lt(\Psi_1(0,0)\, +\, i\,\Psi_2(0,0)\rt)
\label{fourthLequation}
\end{eqnarray}
Equations (\ref{firstLequation}) and (\ref{secondLequation}) hold as
long as $\al \neq 0$ and equations (\ref{thirdLequation}) and
(\ref{fourthLequation}) hold as long as $\beta \neq 0$.  By
combining (\ref{firstLequation}) and (\ref{fourthLequation}) we
obtain $l_z^{TOT}-\abs{l}+1=l_z^{TOT}+\abs{l}+1$ which contradicts
our assumption that $\abs{l} \neq 0$.  So, either $\al=0$ or
$\beta=0$.  Assume that $\beta=0$ and take $\al=1$, so that
equations (\ref{firstLequation}) and (\ref{secondLequation}) still
hold. Since $L_z^{TOT}\, \Phi_0\, =\, l_z^{TOT}\, \Phi_0$, we know
that $L_z^{TOT}\, \overline{\Phi_0}\, =\, -l_z^{TOT}\,
\overline{\Phi_0}$.  Using this equation and projecting along
$e^{-i(\abs{l}-1)\phi}\,f^{[\abs{l}]}$ and
$e^{-i(\abs{l}+1)\phi}\,g^{[\abs{l}]}$, we obtain equations similar
to (\ref{thirdLequation}) and (\ref{fourthLequation}), but with
$l_z^{TOT}$ replaced by $-l_z^{TOT}$:
\begin{eqnarray}
L_z^{el}\, \lt(\Psi_1(0,0)\, -\, i\,\Psi_2(0,0)\rt)\, =\,
(-l_z^{TOT}+\abs{l}-1)\, \lt(\Psi_1(0,0)\, -\, i\,\Psi_2(0,0)\rt)
\label{thirdLequationprime}
\\[3mm]
L_z^{el}\, \lt(\Psi_1(0,0)\, +\, i\,\Psi_2(0,0)\rt)\, =\,
(-l_z^{TOT}+\abs{l}+1)\, \lt(\Psi_1(0,0)\, +\, i\,\Psi_2(0,0)\rt).
\label{fourthLequationprime}
\end{eqnarray}
By combining (\ref{firstLequation}) and (\ref{fourthLequationprime})
we obtain $l_z^{TOT}=\abs{l}$ and these equations now reduce to
\begin{eqnarray}
L_z^{el}\, \lt(\Psi_1(0,0)\, -\, i\,\Psi_2(0,0)\rt)\, &=&\, -\,
\lt(\Psi_1(0,0)\, -\, i\,\Psi_2(0,0)\rt) \label{Lelecminus1}
\\[3mm]
L_z^{el}\, \lt(\Psi_1(0,0)\, +\, i\,\Psi_2(0,0)\rt)\, &=&\,
\Psi_1(0,0)\, +\, i\,\Psi_2(0,0). \label{Lelecplus1}
\end{eqnarray}

By repeating the argument with $\al=0,\ \beta=1$, we would instead
find $l_z^{TOT}=-\abs{l}$.  From this analysis we see that $\al=1,\
\beta=0$ and $\al=0,\ \beta=1$ are correct linear combinations that
will generate two orthogonal quasimodes $\Phi_{\ep}^A$ and
$\Phi_{\ep}^B$ respectively.  These quasimodes satisfy $L_z^{TOT}\,
\Phi_{\ep}^A\, =\, \abs{l} \Phi_{\ep}^A$ and $L_z^{TOT}\,
\Phi_{\ep}^B\, =\, -\abs{l}\, \Phi_{\ep}^B$ and that they are
asymptotic to eigenfunctions of $H(\ep)$.  We note that
$\Phi_0^B=\overline{\Phi}_0^A$.  Since $H(\ep)$ commutes with
complex conjugation, we have that $\overline{\Phi}_{\ep}^A$ is also
asymptotic to an eigenfunction with the same eigenvalue as
$\Phi_{\ep}^A$. Since quasimodes are determined by their zeroth
order eigenfunctions through the perturbation formulas of chapter
\ref{QuasimodeConstruction}, this imples that
$\Phi_{\ep}^B=\overline{\Phi}_{\ep}^A$ since
$\Phi_0^B=\overline{\Phi}_0^A$.  So, the $\Phi_{\ep}^A$ and
$\Phi_{\ep}^B$ correspond to a degenerate pair and we see that no
splitting occurs in the perturbation expansion. As a result, any
linear combination would be a correct one. The $\Phi_{\ep}^A$ and
$\Phi_{\ep}^B=\overline{\Phi}_{\ep}^A$ generated by the combinations
$\al=1,\ \beta=0$ and $\al=0,\ \beta=1$ respectively, are the
quasimodes that satisfy $L_z^{TOT}\, \Phi_{\ep}^A\, =\, \abs{l}\,
\Phi_{\ep}^A$ and $L_z^{TOT}\, \Phi_{\ep}^B\, =\, -\abs{l}\,
\Phi_{\ep}^B$.

If $\abs{l}=1$, we take projections of $L_z^{TOT}\,\Phi_0\, -\,
l_z^{TOT}\, \Phi_0\, =\, 0$ along
$f^{[1]}$, $e^{2i\phi}\,g^{[1]}$, and $e^{-2i\phi}\,g^{[\abs{1}]}$,
and obtain three equations.  By proceeding in a similar manner to the analysis in the $\abs{l}\geq
2$ case above, we would obtain $l_z^{TOT}=1$ if $\al =1,\ \beta=0$
and $l_z^{TOT}=-1$ if $\al=0,\ \beta=1$.  In either case, we would
obtain (\ref{Lelecminus1}) and (\ref{Lelecplus1}) and the desired
results follow as in the $\abs{l}\geq 2$ case above.

\noindent {\bf Case 2: $\abs{l} = 0$}

If $\abs{l} = 0$, we have non-degenerate eigenfunctions of
$\mb{H}_2$ of the form in equations (\ref{L0states3}) or
(\ref{L0states4}).  In any case, we see that $\Phi_0$ is real.  Then from
$L_z^{TOT}\, \Phi_0\, =\, l_z^{TOT}\, \Phi_0$ and $L_z^{TOT}\,
\overline{\Phi}_0\, =\, -l_z^{TOT}\, \overline{\Phi}_0$, it is clear
that $l_z^{TOT}=0$ in this case. By plugging into $L_z^{TOT}\,
\Phi_0\, =\, 0$ and taking projections along $e^{i\phi}$ and
$e^{-i\phi}$ in a manner similar to the $\abs{l} \neq 0$ case, we
obtain the same relations for the electronic basis vectors at zeroth
order given by equations (\ref{Lelecminus1}) and (\ref{Lelecplus1}).
$\square$

\vskip .5cm

\begin{cor}\label{Cor5.2}
For all $\tl{b}$ in some interval $(0,\delta)$, if $\ep$ is
sufficiently small, the ground state of $H(\ep)$ (corresponding to
the R-T pair of states we are considering) is degenerate.
\end{cor}

\noindent{\bf Proof:}

From our perturbation analysis, we have that for all $\tl{b}$ in
some interval $(0,\delta)$, the ground state of $\mb{H}_2$ is
degenerate, arising from the $l=\pm1$ states of $H_{UZ}$.  The
previous theorem tells us that these generate a degenerate pair of
quasimodes $\Phi_{\ep}^A$ and $\Phi_{\ep}^B$ that satisfy
$L_z^{TOT}\, \Phi_{\ep}^A\, =\, \Phi_{\ep}^A$, $\Phi_{\ep}^B\, =\,
\overline{\Phi}_{\ep}^A$ and $L_z^{TOT}\, \Phi_{\ep}^B\, =\, -\,
\Phi_{\ep}^B$.  If the quasimode energy lies below the essential
spectrum, then this will correspond to the lowest lying eigenvalue
of $H(\ep)$ corresponding to the R-T pair.  $\square$

\vskip .5cm

Our perturbation calculations suggest that there is a crossing
involving this eigenvalue with the lowest lying $l=0$ eigenvalue,
somewhere near $\tl{b}=0.925$.  The ground state seemingly
corresponds to these $l=\pm1$ states for $0 < \tl{b} < 0.925$ and
corresponds to the non-degenerate, lowest lying $l=0$ state for
$0.925 < \tl{b} < 1$.  We now prove that the ground state of
$\mb{H}_2$ cannot arise from any other $\abs{l}$ states.

\vskip .5cm

\begin{prop}\label{prop5.3}
Let $0 < \tl{b} <1$.  Then the ground state of $H_{UZ}$ is either an
$\abs{l}=1$ state or an $l=0$ state.
\end{prop}

\noindent {\bf Proof:}

As previously mentioned, since $[H_{UZ},L_z^{nuc}]=0$ we assume
$\dsp \Psi\, =\,\lt(
\begin{array}{c} e^{\pm i\abs{l}\phi}\, \psi_1(r) \\[3mm] e^{\pm i \abs{l}\phi}\, \psi_2(r)
\end{array}\rt)$ and then $H_{UZ}$ can be written in the form
\begin{eqnarray}
H_{UZ}^{[\pm \abs{l}]}\, =\, H_{UZ}^{[0]}\, +\, \f1{2r^2}\,
\left(\begin{array}{cc} \displaystyle \abs{l}^2 \mp 2\abs{l} & 0
\\[3mm]
0 & \abs{l}^2 \pm 2\abs{l}
    \end{array}\right). \label{NoLBiggerthan2}
\end{eqnarray}

We now show that $H_{UZ}^{[0]}$ must have an eigenvalue below the
eigenvalues of $H_{UZ}^{[\pm \abs{l}]}$ if $\abs{l} \geq 2$.
Recall that $\sg\lt(H_{UZ}^{[\abs{l}]}\rt)\, =\,
\sg\lt(H_{UZ}^{[-\abs{l}]}\rt)$, so we only consider
$H_{UZ}^{[\abs{l}]}$.
 Since $\f1{2r^2}\, \lt(\abs{l}^2 \pm 2\abs{l}\rt)
> 0$ for $\abs{l}>2$, we know from (\ref{NoLBiggerthan2}) that $H_{UZ}^{[\abs{l}]}\, >
H_{UZ}^{[0]}$ for all $\abs{l}>2$.  It easily follows that the
lowest eigenvalue of $H_{UZ}^{[0]}$ must lie below the eigenvalues
of $H_{UZ}^{[\pm \abs{l}]}$ for all $\abs{l} >2$.

The presence of the off-diagonal terms in $H_{UZ}^{[\abs{l}]}$ when
$\tl{b} \neq 0$, implies that both components of the eigenvectors
must be non-vanishing.  Let $\Psi$ be the eigenvector corresponding
to the lowest lying eigenvalue of $H_{UZ}^{[2]}$.  Then,
\begin{eqnarray*}
\lt\langle\, \Psi,\, H_{UZ}^{[2]}\, \Psi\, \rt\rangle\, &=&\,
\lt\langle\, \Psi,\, H_{UZ}^{[0]}\, \Psi\, \rt\rangle\, +\,
\lt\langle\, \Psi,\, \f1{2r^2}\, \left(\begin{array}{cc}
\displaystyle 0 & 0
\\[3mm]
0 & 8
    \end{array}\right)\, \Psi\, \rt\rangle \notag
\\[3mm]
&>&\, \lt\langle\, \Psi,\, H_{UZ}^{[0]}\, \Psi\, \rt\rangle
\ \geq\, \inf\, \sg\lt(H_{UZ}^{[0]}\rt). \label{NoL2groundstate}
\end{eqnarray*}
We see that $H_{UZ}^{[0]}$ has at least one eigenvalue below the
eigenvalues of $H_{UZ}^{[2]}$.  So, the ground state of $H_{UZ}$
must correspond to the ground state of $H_{UZ}^{[0]}$ or
$H_{UZ}^{[\pm 1]}$.  $\square$

\section{Proof of the Main Theorem}\label{MainThmChap}
\setcounter{equation}{0} \setcounter{theorem}{0}

Here we use the quasimode expansion constructed in section
\ref{QuasimodeConstruction} to sketch the proof of theorem
\ref{maintheorem}.  Our candidates for the approximate wave function
and energy in the theorem are
\begin{eqnarray}
\Phi_{\ep,K}\, =\, F(\ep \rho)\, \Psi_{\ep,K},\ \qquad \text{where} \notag
\end{eqnarray}
\begin{eqnarray}
\Psi_{\ep,K}\, =\, \lt(\, \sum_{j=0}^{K-2}\, \ep^j\, \lt(\Psi_1(\ep
X,\ep Y)\, f^{(j)}(X,Y)\, +\, \Psi_2(\ep X,\ep Y)\,
g^{(j)}(X,Y)\rt)\, +\, \sum_{j=0}^K\, \ep^j\,
\psi_{\perp}^{(j)}(X,Y)\rt) \notag
\end{eqnarray}
and $E_{\ep,K}\, =\, \sum_{j=0}^{K}\, \ep^j\, E^{(j)}$, where the
$f^{(j)},\ g^{(j)},\ \psi_{\perp}^{(j)},$ and $E^{(j)}$ are
determined by the perturbation formulas in section
\ref{QuasimodeConstruction}.  The cut-off function $F(\ep \rho)$ is
needed to restrict the analysis to a neighborhood of the local
minimum of the electronic eigenvalues $E_1(\tl{\rho})$ and
$E_2(\tl{\rho})$ at $\tl{\rho}=0$, where $E_1$ and $E_2$ are
isolated from the rest of the spectrum of $h(\ep X, \ep Y)$ and also
where the functions and operators that we have expanded into powers
of $\ep$ (such as $\Psi_1(\ep X, \ep Y)$) have asymptotic expansions
(recall that $(x,y)=(\ep X,\,\ep Y)$ and $\tl{\rho}=\ep \rho$).  We
require that the cut-off function $F(\tl{\rho})\,:\,\R^2\rightarrow
[0,1]$ be smooth in both variables $x$ and $y$.  It has support in
some neighborhood where $\tl{\rho} < S$. Also, $F(\tl{\rho})=1$ for
$\tl{\rho} \leq R$, where $0<R<S$. So, the derivatives of
$F(\tl{\rho})$ with respect to $x$ and $y$ vanish outside the region
$R \leq \tl{\rho} \leq S$.

We have
\begin{eqnarray}
(H(\ep)-E_{\ep, K})\, \Phi_{\ep,K}\, =\, F(\ep \rho)\,
(H(\ep)-E_{\ep, K})\, \Psi_{\ep,K}\, -\, \f{\ep^2}{2}\, \lt[\,
\Delta_{X,Y}\, ,\, F(\ep \rho)\, \rt]\, \Psi_{\ep,K}.
\label{boundedequation}
\end{eqnarray}
To prove the theorem, one can first show the norm of $\Phi_{\ep,K}$
is asymptotic to $\ep$.  It then suffices to prove that both terms
on the right hand side of (\ref{boundedequation}) are finite linear
combinations of the form $\ep^{J}\, G$, where
$\norm{G}_{\mH_{nuc}\otimes\mH_{el}}<\infty$ and $J \geq K+2$.
Recall from chapter \ref{QuasimodeConstruction} that $P_{\perp}$ was
the projection in $\mH_{el}$ onto $\lt\{\Psi_1,
\Psi_2\rt\}^{\perp}$. We can write
\begin{eqnarray}
(H(\ep)-E_{\ep, K})\, \Psi_{\ep,K}\, =\, \chi_1(\ep,X,Y)\,
\Psi_1(\ep X,\ep Y)\, +\, \chi_2(\ep,X,Y)\, \Psi_2(\ep X,\ep Y)\,
+\, \chi_{\perp}(\ep,X,Y), \notag
\end{eqnarray}
where $\chi_{\perp}\, =\, P_{\perp}\lt[(H(\ep)-E_{\ep, K})\,
\Psi_{\ep,K} \rt]$ (so $\chi_1$ and $\chi_2$ have no electronic
dependence, but $\chi_{\perp}$ does have electronic dependence).

The analysis regarding $\chi_2$ and $\chi_{\perp}$ is similar to
that of $\chi_1$ and will be omitted.  Using (\ref{psi1}) with our
definition of $\Psi_{\ep,K}$, we have
\begin{eqnarray}
&& \hskip -.5cm \chi_1\, =\, -\ \f{\ep^2}2\ \Delta_{X,\,Y}\
\sum_{j=0}^{K-2}\,\ep^j\, f^{(j)}\ +\ h_{11}\
\sum_{j=0}^{K-2}\,\ep^j\,f^{(j)}\ +\ h_{12}\
\sum_{j=0}^{K-2}\,\ep^j\,g^{(j)}\notag\\[3mm] &&\hskip -.5cm -\ \f{\ep^2}2\,
\langle\, \Psi_1,\, \Delta_{X,\,Y}\, \psi_{\perp}\, \rangle_{el}\ -\
\f{\ep^4}2\, \sum_{j=0}^{K-2}\,\ep^j\,f^{(j)}\, \langle\, \Psi_1,\,
\Delta_{x,\,y}\, \Psi_1\, \rangle_{el}\ -\ \f{\ep^4}2\,
\sum_{j=0}^{K-2}\,\ep^j\,g^{(j)}\, \langle\, \Psi_1,
\Delta_{x,\,y}\, \Psi_2\, \rangle_{el} \notag
\\[3mm]
&& \hskip -.5cm -\ \ep^3\, \lt(\, \sum_{j=0}^{K-2}\,\ep^j\,
\f{\partial g^{(j)}}{\partial X}\, \langle\, \Psi_1,\,\f{\partial
\Psi_2}{\partial x}\, \rangle_{el}\, +\,
\sum_{j=0}^{K-2}\,\ep^j\,\f{\partial g^{(j)}}{\partial Y}\,
\langle\, \Psi_1,\,\f{\partial \Psi_2}{\partial y}\, \rangle_{el}\,
\rt)\ -\ \sum_{j=0}^{K}\,\ep^j\,E^{(j)}\,
\sum_{l=0}^{K-2}\,\ep^l\,f^{(l)}. \notag
\end{eqnarray}

To show $\norm{F(\ep \rho)\,\chi_1\,\Psi_1}_{\mH_{nuc}\otimes
\mH_{el}}\, \leq\, C\,\ep^{K+2}$, we can consider the terms in the
above equation separately and use the triangle inequality. Analogous
to equations (\ref{psi2}) and (\ref{perp2}), we expand all functions
with $(\ep X, \ep Y)$ dependence into powers of $\ep$, however, we
truncate the series here and add an error term. For example, we can
write $h_{11}(\ep X, \ep Y)\, =\, \sum_{j=0}^{K}\, \ep^j\,
h_{11}^{(j)}\, +\, \ep^{K+1}\, h_{11}^{err}(X,Y)$, where we know
$h_{11}^{err}(X,Y)$ is in $C^\infty(X,Y)$ and is bounded by a
polynomial in $X$ and $Y$ of order $K+1$ on $\supp(F(\ep \rho))$. If
we do this, we know all terms of order $\ep^j$, for $j \leq K$, will
cancel in the above equations, since the terms of $f$, $g$, $E$, and
$\psi_{\perp}$ were chosen using the perturbation formulas. We show
how to deal with the $h_{11}$ term arising in $F(\ep
\rho)\,\chi_1\,\Psi_1$ only, the rest of the terms are handled
similarly.  Considering only expressions of order $\ep^{K+1}$ or
higher, this term can be written
\begin{eqnarray}
h_{11}(\ep X,\ep Y)\ \sum_{j=0}^{K-2}\,\ep^j\, f^{(j)}\
&=& \sum_{\substack{0 \leq j \leq K-2 \\[1mm] 0 \leq l \leq K \\[1mm] j+l \geq
K+1 }}\, \ep^{l+j}\, h_{11}^{(l)}\, f^{(j)}\, +\,
\sum_{j=0}^{K-2}\,\ep^{K+j+1}\, h_{11}^{err}(X,Y)\, f^{(j)}. \notag
\end{eqnarray}
Then using the results of section \ref{Properties}, in particular
that $f^{(j)} \in D(e^{\gm \langle x \rangle})$, one can show that
\begin{eqnarray}
&&\lt\lvert\lt\lvert\, F(\ep \rho)\,h_{11}(\ep X,\ep Y)\,
\sum_{j=0}^{K-2}\,\ep^j\, f^{(j)}\,\Psi_1(\ep X, \ep
Y)\,\rt\rvert\rt\rvert_{\mH_{nuc}\otimes \mH_{el}} \
\leq\ \sum_{\substack{0 \leq j \leq K-2 \\[1mm] 0 \leq l \leq K \\[1mm] j+l \geq
K+1 }}\, \ep^{l+j+1}\, C_{l,j}\ +\ \sum_{j=0}^{K-2}\,\ep^{K+j+2}\,
D_{l,j}, \notag
\end{eqnarray}
where $C_m$, $D_m$, $C_{l,j}$, $D_{l,j}\ <\ \infty$.  We see that
this term is indeed of order greater or equal to $O(\ep^{K+2})$.
All of the terms of $\chi_1$ and $\chi_{\perp}$ can be handled in a
similar fashion using the results of theorem \ref{thm9}.

The term involving the derivatives of $F$ in equation
(\ref{boundedequation}) are handled using theorem \ref{thm9} as
well.  The derivatives of $F$ are supported away from the origin and
the terms of $\Psi_{\ep,K}$ are exponentially decaying.  We consider
the terms involving derivatives with respect to $X$.  Let $M$ be
larger than $\sup_{\scriptsize{\R^2}}\,\abs{\f{\pa^2 F}{\pa x^2}}\,$
and $\sup_{\scriptsize{\R^2}}\,\abs{\f{\pa F}{\pa x}}$. Then, using
theorem \ref{thm9}, one can easily show that
\begin{eqnarray*}
&&\hskip -1cm \norm{\lt[\, \f{\pa^2\, }{\pa X^2}\, ,\, F(\ep \rho)\,
\rt]\, \Psi_{\ep,K}}_{\mH_{nuc}\otimes\mH_{el}} \notag
\\[3mm]
&&\leq M\,e^{-\gm\sqrt{1+R^2/\ep^2}}\,\lt(\ep^2\,\norm{e^{\gm\langle
x \rangle}\, \Psi_{\ep,K}}_{\mH_{nuc}\otimes\mH_{el}}\ +\
2\,\ep\,\norm{e^{\gm\langle x \rangle}\, \f{\pa \Psi_{\ep,K}}{\pa
X}}_{\mH_{nuc}\otimes\mH_{el}}\rt)\notag
\\[3mm]
&&\leq O(\ep^\infty).
\end{eqnarray*}
The conclusion of the theorem follows.  $\square$

\vskip .5cm

\renewcommand{\theequation}{A-\arabic{equation}}
\setcounter{equation}{0}  
\section*{Appendix}  

We now argue that the odd terms in the $E(\ep)$ series must be zero.
See \cite{thesis} for a detailed proof which utilizes the perturbation
formulas derived in chapter \ref{QuasimodeConstruction}.

The Hamiltonian of interest in terms of the scaled nuclear
coordinates $(X,Y)=(x/\ep,y/\ep)$ is given by
$$
H(\ep)\, =\, \f{-\ep^2}{2}\, \Delta_{X,Y}\, +\, h(\ep X,\ep Y),
$$
where $h(x,y)$ is the electronic hamiltonian that also contains the
nuclear repulsion terms.  If $E(\ep)$ is an eigenvalue of $H(\ep)$,
then $E(-\ep)$ is an eigenvalue of
$$
H(-\ep)\, =\, \f{-\ep^2}{2}\, \Delta_{X,Y}\, +\, h(-\ep X,-\ep Y).
$$
Under the unitary change $\tl{X}=-X$, $\tl{Y}=-Y$, we see that
$H(-\ep)$ becomes
$$
H(-\ep)\, =\, \f{-\ep^2}{2}\, \Delta_{\tl{X},\tl{Y}}\, +\, h(\ep
\tl{X},\ep \tl{Y}).
$$
It is clear that $H(\ep)$ and $H(-\ep)$ share the same eigenvalues.
This does not immediately imply $E(\ep)=E(-\ep)$, since there could
be a pair of eigenvalues related by $E_A(\ep)=E_B(-\ep)$. However,
this would imply that $E_A^{(2)}=E_B^{(2)}$, and then theorem \ref{thm5.1}
implies $E_A(\ep)=E_B(\ep)$.  Therefore, $E(\ep)=E(-\ep)$, and as a result the odd terms in the expansion
must vanish.

\newpage

\noindent {\Large\textbf{Acknowledgement}}

\vskip .5cm

\noindent  It is a pleasure to thank Professor George A. Hagedorn for his advise and
many useful comments.\\

\noindent This research was supported in part by National Science
Foundation Grant DMS--0600944 while at Virginia Polytechnic Institute and State University, and also by the Institute for Mathematics and its Applications at the University of Minnesota, with funds provided by the National Science Foundation.

%
%

%




\bibliographystyle{plain}

\end{document}